\documentclass[useAMS,usenatbib]{mn2e}

\usepackage{url,graphicx,rotating}

\newcommand{\SSp}{\emph{SAGE-Spec}}
\newcommand{\UBcol}{U$-$B}
\newcommand{\BVcol}{B$-$V}
\newcommand{\VRcol}{V$-$R}
\newcommand{\VKcol}{V$-$K}
\newcommand{\JKcol}{V$-$K}

\def\aap{A\&A}                % Astronomy and Astrophysics
\def\aaps{A\&AS}              % Astronomy and Astrophysics, Supplement
\def\aj{AJ}                   % Astronomical Journal
\def\ao{Applied Optics}       % Applied Optics
\def\apj{ApJ}                 % Astrophysical Journal
\def\apjl{ApJ}                % Astrophysical Journal, Letters
\def\apjs{ApJS}               % Astrophysical Journal, Supplement
\def\apss{Ap\&SS}             % Astrophysics and Space Science
\def\araa{ARA\&A}             % Annual Review of Astronomy and Astrophysics
\def\mnras{MNRAS}             % Monthly Notices of the RAS
\def\nat{Nature}              % Nature
\def\pasj{PASJ}               % Publ. of the Astronomical Society of Japan
\def\pasp{PASP}               % Publ. of the Astron. Society of the Pacific
\title[SAGE-Spec: Point source classification I.]{The SAGE-Spec
  Spitzer Legacy program: The life-cycle of \\ dust and gas in the
  Large Magellanic Cloud.\\Point source classification I.}

\author[P.~M.~Woods et al.]{Paul M.~Woods$^{1}$\thanks{E-mail: Paul.Woods@manchester.ac.uk},
J.~M.~Oliveira$^{2}$,
F.~Kemper$^{1}$, 
J.~Th.~van Loon$^{2}$,
B.~A.~Sargent$^{3}$,\newauthor
M.~Matsuura$^{4,5}$,
R.~Szczerba$^{6}$,
K.~Volk$^{3}$, 
A.~A.~Zijlstra$^{1}$,
G.~C.~Sloan$^{7}$,
E.~Lagadec$^{8,1}$,\newauthor
I.~McDonald$^{1}$,
O.~Jones$^{1}$,
V.~Gorjian$^{9}$,
K.~E.~Kraemer$^{10}$,
C.~Gielen$^{11}$, 
M.~Meixner$^{3}$,\newauthor 
R.~D.~Blum$^{12}$, 
M.~Sewi\l o$^{3}$, 
D.~Riebel$^{13}$,
B.~Shiao$^{3}$,
C.-H.~R.~Chen$^{14}$,
M.~L.~Boyer$^{3}$,\newauthor 
R.~Indebetouw$^{14,15}$, 
V.~Antoniou$^{16}$, 
J.-P.~Bernard$^{17}$, 
M.~Cohen$^{18}$, 
C.~Dijkstra$^{19}$,\newauthor
M.~Galametz$^{20}$,
F.~Galliano$^{20}$,
Karl D.~Gordon$^{3}$, 
J.~Harris$^{20}$, 
S.~Hony$^{20}$, 
J.~L.~Hora$^{21}$, \newauthor
A.~Kawamura$^{22}$, 
B.~Lawton$^{3}$,
J.~M.~Leisenring$^{14}$, 
S.~Madden$^{20}$, 
M.~Marengo$^{16,21}$, \newauthor
C.~McGuire$^{1}$,
A.~J.~Mulia$^{23}$,
B.~O'Halloran$^{24}$,
K.~Olsen$^{17}$, 
R.~Paladini$^{25}$, 
D.~Paradis$^{25}$, \newauthor
W.~T.~Reach$^{25}$, 
D.~Rubin$^{20}$, 
K.~Sandstrom$^{26,18}$,
I.~Soszy{\'n}ski$^{27}$,
A.~K.~Speck$^{23}$, \newauthor
S.~Srinivasan$^{28,13}$,
A.~G.~G.~M.~Tielens$^{29}$, 
E.~van Aarle$^{11}$, 
S.~D.~Van Dyk$^{25}$, \newauthor
H.~Van Winckel$^{11}$, 
Uma~P.~Vijh$^{30}$, 
B.~Whitney$^{31}$,
A.~N.~Wilkins$^{7}$
\\$^{1}${Jodrell Bank Centre for Astrophysics, Alan Turing Building, School of Physics and Astronomy, The University of Manchester,}
\\\phantom{$^{1}$}{Oxford Road, Manchester, M13 9PL, UK.}
\\$^{2}${School of Physical \& Geographical Sciences, Lennard-Jones Laboratories,
Keele University, Staffordshire, ST5 5BG, UK}
\\$^{3}${Space Telescope Science Institute, 3700 San Martin Drive, Baltimore, MD 21218}
\\$^{4}${Institute of Origins, Department of Physics and Astronomy, University
College London, Gower Street, London, WC1E 6BT, UK
}
\\$^{5}${Institute of Origins, Mullard Space Science Laboratory, University College
London, Holmbury St. Mary, Dorking, Surrey, RH5 6NT, UK
}
\\$^{6}${N.~Copernicus Astronomical Center,
Rabianska 8,  87-100 Torun, Poland
}
\\$^{7}${Department of Astronomy, Cornell University, Ithaca, NY 14853}
\\$^{8}${European Southern Observatory, Karl-Schwarzschild-Stra\ss e 2, D-85748 Garching bei M\"unchen, Germany}
\\$^{9}${JPL/Caltech, MS 169-506, 4800 Oak grove Dr., Pasadena, CA 91109}
\\$^{10}${US Air Force Research Laboratory, Space Vehicles Directorate, 29 Randolph Road, Hanscom AFB, MA 01731}
\\$^{11}${Instituut voor Sterrenkunde, Katholieke Universiteit Leuven, Celestijnenlaan 200D, 3001 Leuven, Belgium}
\\$^{12}${NOAO, 950 North Cherry Avenue, Tucson, AZ 85719}
\\$^{13}${Department of Physics and Astronomy, Johns Hopkins University, Homewood Campus, Baltimore, MD 21218}
\\$^{14}${Department of Astronomy, University of Virginia, P.O. Box 400325, Charlottesville, VA 22904}
\\$^{15}${National Radio Astronomy Observatory,
520 Edgemont Road,
Charlottesville, VA 22903}
\\$^{16}${Department of Physics \& Astronomy, Iowa State University, Ames, IA 50011, USA}
\\$^{17}${Centre d'\'Etude Spatiale des Rayonnements,
9 Av.~du Colonel Roche, BP 44346,
31028 Toulouse C{\'e}dex 4, France}
\\$^{18}${Radio Astronomy Laboratory, University of California at Berkeley, 601 Campbell Hall, Berkeley, CA 94720-3411}
\\$^{19}${Passiebloemweg 31, 1338 TT Almere, The Netherlands}
\\$^{20}${Laboratoire AIM, CEA/DSM - CNRS - Universit{\'e} Paris Diderot DAPNIA/Service d'Astrophysique B\^{a}t.~709, CEA-Saclay}
\\\phantom{$^{20}$}{F-91191 Gif-sur-Yvette C{\'e}dex, France}
\\$^{20}${Steward Observatory, University of Arizona, 933 North Cherry Avenue, Tucson, AZ 85721}
\\$^{21}${Harvard-Smithsonian Center for Astrophysics, 60 Garden Street, MS 65, Cambridge, MA 02138-1516}
\\$^{22}${Department of Astrophysics, Nagoya University, Chikusa-Ku, Nagoya 464-01, Japan}
\\$^{23}${Physics \& Astronomy Department,
 University of Missouri,
 Columbia, MO 65211}
\\$^{24}${Astrophysics Group,
Imperial College London,
Blackett Laboratory,
Prince Consort Road,
London, SW7 2AZ,UK}
\\$^{25}${Spitzer Science Center, California Institute of Technology, MS 220-6, Pasadena, CA 91125}
\\$^{26}${Max Planck Institut f{\"u}r Astronomie, D-69117 Heidelberg, Germany}
\\$^{27}${Warsaw University Observatory, Al. Ujazdowskie 4, 00-478 Warszawa, Poland}
\\$^{28}${Institut d'Astrophysique de Paris, 98 bis, Boulevard Arago, Paris 75014, France}
\\$^{29}${Leiden Observatory,
P.O. Box 9513,
NL-2300 RA Leiden,
The Netherlands
}
\\$^{30}${Ritter Astrophysical Research Center, University of Toledo, Toledo OH 43606}
\\$^{31}${Space Science Institute, 4750 Walnut Street, Suite 205, Boulder, CO 80301}
}

\begin{document}

\date{}

\pagerange{\pageref{firstpage}--\pageref{lastpage}} \pubyear{2010}

\maketitle

\label{firstpage}

\begin{abstract}
We present the classification of 197 point sources observed with the
\emph{Infrared Spectrograph} in the \SSp\ Legacy program on the
\emph{Spitzer Space Telescope}.  We introduce a decision-tree method
of object classification based on infrared spectral features,
continuum and spectral energy distribution shape, bolometric
luminosity, cluster membership, and variability information, which is
used to classify the \SSp\ sample of point sources. The decision tree
has a broad application to mid-infrared spectroscopic surveys, where
supporting photometry and variability information are available.  We
use these classifications to make deductions about the stellar
populations of the Large Magellanic Cloud and the success of
photometric classification methods. We find 90 asymptotic giant branch
(AGB) stars, 29 young stellar objects, 23 post-AGB objects, 19 red
supergiants, eight stellar photospheres, seven background galaxies,
seven planetary nebulae, two H{\sc ii} regions and 12 other objects,
seven of which remain unclassified.
\end{abstract}


\section{Introduction}

The \emph{SAGE-LMC} program \citep{mei06}, the vanguard of the
\emph{Surveying the Agents of Galaxy Evolution} (\emph{SAGE})
collaboration, is a \emph{Spitzer Space Telescope} Legacy Program
which took a photometric inventory of the Large Magellanic Cloud (LMC)
using the \emph{Infrared Array Camera} \citep[IRAC;][]{faz04} and
\emph{Multi-Band Imaging Photometer for Spitzer}
\citep[MIPS;][]{rie04} instruments on board the \emph{Spitzer Space
  Telescope} \citep{wer04}.  Observations were taken over two epochs
separated by three months. The survey detected some 6.5 million point
sources at 3.6, 4.5, 5.8, 8.0\,$\mu$m (IRAC) and 24, 70, 160\,$\mu$m
(MIPS). To follow up on the \emph{SAGE-LMC} program, the \SSp\ project
\citep{kem10}, obtained 196 staring-mode pointings using
\emph{Spitzer}'s \emph{Infrared Spectrograph} \citep[IRS;][]{hou04}
(5.2-38\,$\mu$m) of positions selected from the \emph{SAGE-LMC}
catalog.  In addition, several other \emph{Spitzer} programs have
targeted objects in the LMC with the IRS; see Table 4 of \citet{kem10}
for an overview of these.

Characterization of the point sources observed in the \emph{SAGE-LMC}
survey, \SSp\ survey, and IRS data archive, builds an inventory of
dusty sources and their inter-relation in the LMC. One of the major
science outputs of \SSp\ will be to relate the \emph{SAGE-LMC}
photometry to the spectral characteristics for different types of
objects in the LMC, and ultimately, to classify all the photometric
point sources in the LMC. This paper makes the first steps towards
this goal. We discuss the method of classification into various source
types in \S2. In \S3 we have executed the classification method on the
point sources observed in the \SSp\ program. This classification,
combined with the photometric classification of the 250 brightest
infrared sources in the LMC \citep{kas08,buc09} and a large sample of
LMC young stellar objects \citep{sea09}, is used to further
characterize the stellar content of the LMC (\S4), and compare it with
existing classifications.

Eventually, our classification will be extended to cover all archival
IRS observations of point sources within the \emph{SAGE-LMC} footprint
(Woods et al., in prep., Paper II). The classification of each of
these additional $\sim$750 sources will be part of the data delivery
of the \SSp\ legacy project to the \emph{Spitzer Science Center} and
the
community\footnotemark[1]\footnotetext[1]{\url{http://ssc.spitzer.caltech.edu/legacy/sagespechistory.html}}.
The classification will be used to benchmark a colour-classification
scheme that will be applied to all $\sim$6.5 million point sources in
the \emph{SAGE-LMC} survey (Marengo et al., in prep).

\begin{figure}
\includegraphics[width=84mm,clip]{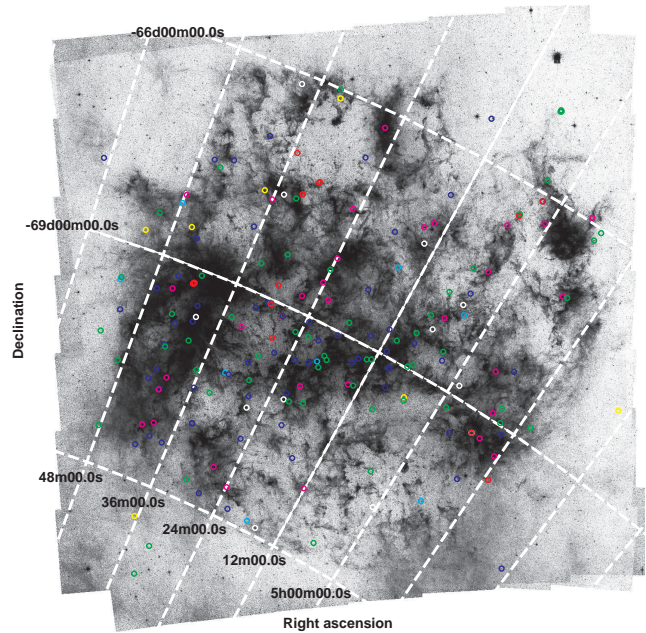}
\caption{The \SSp\ sources distributed on the sky, overlaid upon a
  \emph{SAGE-LMC} 8\,$\mu$m map. The colour of the points represent our
  classifications: YSO \& H{\sc ii} (magenta), STAR (cyan), O-AGB,
  O-PAGB \& O-PN (blue), C-AGB, C-PAGB, C-PN (green), RSG (red),
  GAL (yellow), OTHER \& UNK (white). See text for class definitions.}
\label{fig:sage8um}
\end{figure}

\section{The classification method}
\label{sec:classmethods}

A full description of the \SSp\ project and the techniques used in the
reduction of the IRS data utilised in this work can be found in
\citet{kem10} and the documentation accompanying the \SSp\ database
deliveries to the \emph{Spitzer} Science Center. The 197 objects in
the \SSp\ sample \citep[observed with 196 IRS staring mode pointings;
  see][]{kem10} are classified using the \emph{Spitzer} IRS spectrum
for each object; the \emph{U}, \emph{B}, \emph{V}, \emph{I}, \emph{J},
\emph{H}, \emph{K}, IRAC and MIPS photometry; a calculation of
bolometric luminosity; variability information; cluster membership and
other information found in the literature. The \emph{UBVI} photometry
comes from the Magellanic Clouds Photometric Survey
\citep[MCPS;][]{zar04}; \emph{JHK$_s$/K$^\prime$} photometry comes
from both the Two Micron All Sky Survey catalog
\citep[2MASS;][]{skr06} and the Infrared Survey Facility (IRSF) survey
\citep{kat07}; IRAC and MIPS photometric data were taken from the
\emph{SAGE-LMC} database \citep{mei06}. The \SSp\ sample was matched
to Optical Gravitational Lensing Experiment (OGLE-III) catalogs of
variable stars \citep{sos08,sos09a,sos09b} and to the Massive Compact
Halo Objects (MACHO) database \citep*{alc98,fra05,fra08} to obtain
periods. Bolometric luminosities were calculated over the range of
available photometric points (typically U-band to IRAC 8\,$\mu$m or
MIPS 24\,$\mu$m) and compared with published values
\citep[e.g.,][]{sri09}. A literature search was also performed for
each object to retrieve other information useful in the purposes of
classification, including (but not limited to) determination of the
stellar type, luminosity, the age of nascent cluster (if the star was
found to be a member of a cluster of stars), H$\alpha$\ detections,
etc. Appendix~\ref{litsurv} provides a brief summary of this survey
for each object. Many of the objects in the sample were
newly-discovered in the \emph{SAGE-LMC} survey, and hence have not
been well-studied in the literature. We also matched our source list
to recent lists of YSO candidates, viz. \citet{whi08}, \citet{gru09},
\citet{sea09}.

\begin{table}
\caption{Classification groups \label{tab:groups}}
~\\
\begin{tabular}{cl}
\hline
Code & Object type\\
\hline
{\sc YSO-1} to {\sc YSO-4} & Young stellar object \\
{\sc STAR}  & Stellar photosphere \\
{\sc C-AGB}  & C-rich AGB star \\
{\sc O-AGB}  & O-rich AGB star \\
{\sc RSG}   & Red supergiant \\
{\sc C-PAGB} & C-rich post-AGB star \\
{\sc O-PAGB} & O-rich post-AGB star \\
{\sc C-PN}   & C-rich planetary nebula \\
{\sc O-PN}   & O-rich planetary nebula \\
{\sc HII}   & H{\sc ii} region \\
{\sc GAL}   & Galaxy \\
{\sc OTHER} & Object of known type \\
{\sc UNK}   & Object of unknown type\\
\hline
\end{tabular}
\end{table}

We adopt the following categories for the source classification. Low
and intermediate mass ($M<8$\,M$_\odot$) post-Main Sequence stars are
classified by chemistry (O- or C-rich) and by evolutionary stage
(Asymptotic Giant Branch; post-Asymptotic Giant Branch; and Planetary
Nebula), hence our groupings {\tt O-AGB, O-PAGB, O-PN, C-AGB, C-PAGB,
  C-PN}. More massive red supergiants have a class of their own, {\tt
  RSG}. Young stellar objects can be identified ({\tt YSO}). Stars
showing a stellar photosphere, but no additional dust or gas features
are classified as {\tt STAR}. We also distinguish galaxies ({\tt
  GAL}), and H{\sc ii} regions ({\tt HII}). We have the classification
{\tt OTHER} for objects of \emph{known type} which do not fit into
another category (e.g., R Coronae Borealis stars). These objects are
usually identified by searching the astronomical literature for
pertinent information and similar spectra. A literature search was
performed for all \SSp\ objects, and is summarised in Appendix
\ref{litsurv}. Finally, we use {\tt UNK} for objects which cannot be
classified (unknown objects) due to low signal-to-noise data, or
  unidentifiable spectral features.  Table~\ref{tab:groups}
summarises the classification groups. The following sections discuss
the description of the individual categories, along with the
classification criteria.

\subsection{Young stellar objects}
\label{YSOspec}
A robust sub-classification of young stellar objects by evolutionary
stage or mass involves complex SED-fitting of multi-band photometry so
that some distinction between classes can be made
\citep[e.g.,][]{whi08}. Given that such an in-depth treatment would be
out of place in this work, we classify YSO spectra phenomenologically
into four groups (see below and our further comments in
\S\ref{sec:sourceclass}).

The spectra of YSOs are characterised by oxygen-rich dust features
superimposed onto a cold dust continuum, and often exhibit strong
silicate features at 10\,$\mu$m and in the 18--20\,$\mu$m region,
either in emission or absorption \citep[e.g.,][]{fur06,fur08}. In
Galactic sources silicate absorption superimposed on a very red
continuum is indicative of embedded protostellar objects
\citep[e.g.,][]{fur08}. These objects are traditionally classified as
Class I sources \citep[based on their IR spectral index;][]{lad87} and
more recently as Stage I sources \citep[based on their modelled
  mass-accretion rates;][]{rob06}.The 10-$\mu$m feature can also be
self-absorbed at critical optical depths.
%Self-absorption can also be seen in the 10-$\mu$m feature, due to
%silicate emission from the warmer regions of a somewhat more evolved
%YSO being absorbed by cooler material further away from the forming
%star.

Ice absorption features are another common feature in spectra of
embedded YSOs. The IRS spectra can show prominent ice features at
5$-$7\,$\mu$m and 15.2\,$\mu$m that are attributed to a mixture of
H$_2$O, NH$_3$, CH$_3$OH, HCOOH, H$_2$CO and CO$_2$\ ices respectively
\citep[e.g.,][]{oli09}. At shorter wavelengths, ice features of water
and CO are found in the 3--5-$\mu$m range
\citep[e.g.,][]{shi10,oli09,shi08}.  Also common in the spectra of
many YSOs are emission features attributed to polycyclic aromatic
hydrocarbons (PAHs). The infrared YSO spectra show a superposition of
ice, dust and PAH features that can be difficult to disentangle.

At later stages, the envelopes of the YSOs become less dense and
hotter, their continua are bluer, and eventually emission from the
circumstellar disc dominates the SED and silicate emission becomes
conspicuous. Such objects are usually classified as Class II
\citep{lad87} or Stage II objects \citep{rob06}. Amongst such objects
are Herbig Ae/Be (HAeBe) stars. These intermediate-mass
(2--8\,M$_\odot$) YSOs have hot \citep[$>$7\,000\,K;][]{cox00} central
stars which are able to illuminate PAH molecules in their environs,
and thus often show a mixture of PAH emission and silicate dust
emission in their mid-infrared spectra
\citep{kel08,ack10}. Spectroscopically it can often be challenging to
distinguish these more evolved YSOs with dusty discs from post-AGB
stars in the IR. One could resolve this degeneracy with complementary
data, perhaps looking for signs of accretion (pre-Main Sequence) or
chemical enrichment (post-AGB) in optical observations, or correlating
positions with known star-forming clusters or molecular clouds.
Several groups have spectroscopically identified YSOs in the LMC
\citep[e.g.,][]{vlo05c,oli06,shi08,oli09,sea09,shi10}.

\subsection{Stellar photospheres}

Most Main Sequence and sub-giant stars show no significant emission in
excess over that from the stellar photosphere alone and present
largely featureless IRS spectra (\textquotedblleft naked
stars\textquotedblright). Stars of spectral class K or earlier present
infrared spectra similar to 10\,000-K blackbodies, because the hotter
spectra are on the Rayleigh-Jeans tail, and the H$^-$\ ion dominates
the opacity in the cooler spectra \citep[e.g.,][]{eng92}.  For A-type
and earlier-type stars hydrogen absorption lines are usually present,
and are noticeable in spectra with good signal-to-noise, even at the
low resolution of the spectra presented here.  When the spectrum is
plotted in Rayleigh-Jeans units ($\lambda^2F_\nu$ vs. $\lambda$), a
true Rayleigh-Jeans tail will appear as a horizontal line
\citep*{coh92}.

Stars with only a small infrared excess with no obvious dust features
may also be classified as naked stars.  These correspond to the
\textquotedblleft type F\textquotedblright\ sources of \citet{vol89}
in the Infrared Astronomical Satellite (IRAS) Low Resolution
Spectrometer (LRS) spectra. Stars with molecular absorption features
but no overt dust features have been classified as AGB stars.

Naked stars in the \SSp\ survey may include foreground Main Sequence
stars, Cepheid variables and other luminous stars in the LMC.  These
different types of stars can be distinguished on the basis of their
short wavelength colours, but often cannot be readily distinguished
based on the IRS spectrum alone. Interested readers are referred to
\citet{her02} for a classification of stellar photospheres.

\subsection{Stars on the Asymptotic Giant Branch}
\label{OAGBdesc}

AGB stars are characterised by an infrared excess at wavelengths
longer than a few microns due to circumstellar dust or a combination
of molecular absorption features and photometric variability. The dust
and molecular features reveal the particular chemistry of the star.

Carbon-rich AGB stars have undergone a series of thermal pulses which
dredge carbon produced by the triple alpha sequence from the interior
to the photosphere \citep{ibe83}.  When carbon atoms outnumber oxygen
atoms, the formation of CO ties up all of the oxygen, resulting in
carbon-rich molecules and dust grains.  CO, C$_3$, C$_2$H$_2$, and HCN
can produce strong absorption features \citep*[$\sim$4--8.5 and
  13.5--14\,$\mu$m, e.g.,][]{jor00,mat06}.  Dust is usually present,
and amorphous carbon and graphite dominate the composition
\citep[e.g.,][]{mar87}.  The emission from amorphous carbon is
featureless, but trace elements like SiC produce features at
$\sim$11.3\,$\mu$m, either in emission or absorption
\citep[e.g.,][]{tre74,gru08,spe09}.  MgS dust produces a broad
emission feature at $\sim$30\,$\mu$m
\citep*[22--38~$\mu$m;][]{goe85,hon02,zij06,slo06,lag07}.

As long as the C/O ratio remains below unity, silicates are the
dominant dust component in the spectra of AGB stars, with features at
10 and 18\,$\mu$m, either in emission or absorption
\citep*[e.g.,][]{gil68, woo69, mer76}. Silicate self-absorption at
10\,$\mu$m indicates an extreme or optically thick O-rich AGB star
\citep[e.g.,][]{syl99,tra99}. A molecular absorption at 8\,$\mu$m due
to the fundamental vibrational mode of SiO is an indication of
oxygen-rich chemistry, and is often manifested as a slight inflection
in the spectrum. Water absorption or emission can cause a broad
feature in the region 6.4--7.0\,$\mu$m. Alumina or spinel has a
feature at 13--14\,$\mu$m which can be seen in emission or
absorption. A continuous, featureless mid-IR excess can be caused by
metallic iron dust \citep{mcd10}.

AGB stars of both chemistries can be variable, and often have regular,
well-defined periods from 100 days to over 1\,000 days for extreme
carbon and OH/IR stars \citep[e.g.,][]{woo92,whi03}. The
{\textquotedblleft classical\textquotedblright} luminosity limit for
AGB stars based on the core-mass--luminosity relationship is given as
$M_{\mathrm{bol}}$ = $-$7.1\,mag \citep{woo83,smi95}, although
evolutionary calculations including hot-bottom burning allow for AGB
stars as bright as $M_{\mathrm{bol}}$ = $-$8.0\,mag
\citep{wag98,her05,poe08}.

\subsection{Red supergiants}

Red supergiants (RSGs) are in general more luminous than AGB stars,
although there is some overlap in the range $-$7.1$\sol
M_{\mathrm{bol}} \sol-$8.0\,mag \citep{woo83,woo92,vlo05a,gro09}.
We consider a suitable O-rich star to be an RSG if its bolometric
magnitude is greater than the classical AGB luminosity limit of
$M_{\mathrm{bol}}=-7.1$, or if it resides in a cluster too young for a
low-mass star to have reached the AGB. They exhibit a similar dust
chemistry to O-rich AGB stars, although to our knowledge no supergiant
has been associated with silicate absorption.

Red supergiants which reside in clusters can then be distinguished
from O-rich AGB stars providing the age of the cluster can be
determined. Ages of red supergiants vary from $\sim$3--30 million
years \citep{sch92} since they are more massive than AGB stars
($>$8\,M$_\odot$) and thus evolve more rapidly. An 8\,M$_\odot$ star
will spend $\approx$55\,Myr on the Main Sequence.

Red supergiants show no large-amplitude brightness variations, and
have previously been distinguished from O-rich AGB stars by means of
period-magnitude diagrams or amplitude of light-curve
\citep{woo92}. They are generally classified as irregular or
semi-regular pulsating variables.

\subsection{Post-AGB stars}

As the AGB mass-loss phase ceases, the circumstellar dust shell
continues to move outwards, thus gradually exposing the central
star. This generally results in a double-peaked SED with one peak due
to stellar emission and the other due to circumstellar dust
\citep[e.g.,][]{vwi03}. This double-peaked shape means that post-AGB
stars can be readily distinguished from AGB stars by means of
colour. Some post-AGB stars have SEDs with strong near-infrared
emission, pointing to the presence of hot dust in the system.  Some of
these particular stars reside in a binary system, which has led to the
formation of a stable dusty circumbinary disc
\citep{wat92,der06,gie08,vwi07,vwi09}.

Carbon-rich chemistry produces a dust-dominated continuum with a
variety of possible emission features, from PAHs, or MgS at
30\,$\mu$m, or the unidentified \textquotedblleft 21-$\mu$m
feature\textquotedblright\ \citep*{kwo89,hri09}. Oxygen-rich post-AGB
objects typically exhibit silicate emission features, but with a
strong contribution from crystalline grains, which produce narrower
features at 11, 16, 20, 23, 28, and 33\,$\mu$m \citep{gie08,gie09}.
This high degree of crystallinity and the presence of large grains
indicates a circumbinary dusty disc \citep[e.g.,
  {AFGL4106};][]{vlo99,mol99}. Some sources show mixed
chemistry with both carbon-rich and oxygen-rich molecules and dust,
such as the Red Rectangle \citep[HD~44179;][]{wat98} and IRAS
16279-4757 \citep{mat04}.
%PAHs may be seen in both cases, since the rapidly heating
%post-AGB star not only excites PAH molecules to emission, but may also
%cause their production by radiation-induced destruction of
%carbonaceous grains.

\subsubsection{RV Tauri-type stars}

RV Tauri stars are a particular class of post-AGB stars \citep{jur86}
which show a distinct variability pattern -- alternating deep and
shallow minima -- due to pulsations, since they cross the Population
II Cepheid instability strip. Typically they have (minimum-to-minimum)
periods of 20--68 days \citep{sos08}. They exhibit photospheric
depletion in refractory elements which is consistent with the presence
of a dusty circumstellar disc. This depletion phenomenon is commonly
observed in binary post-AGB stars with discs, which led \citet{vwi99}
and \citet{gie09} to suggest that these depleted RV Tauri stars must
also be binary stars surrounded by such discs. Presently it is not
possible to distinguish between a regular oxygen-rich post-AGB object
and a pulsating one (RV~Tau) via mid-infrared spectra; one must in
addition obtain a lightcurve from several years' observations.
Various catalogues exist which make this determination
straightforward, e.g., MACHO \citep{alc98} and OGLE-III
\citep{sos08}. Time-variable \UBcol\ or \BVcol\ colours could point to
pulsating behaviour, although this could also be due to other factors
(e.g., variable reddening due to the circumstellar disc).

\subsection{Planetary nebulae}

The material around the hot central stars of planetary nebulae is
readily ionized, and thus their spectra show evidence of forbidden
line emission from excited atomic species. Commonly-observed lines in
the \emph{Spitzer} IRS bandwidth are [ArII] (6.99\,$\mu$m), [ArIII]
(8.99\,$\mu$m), [SIII] (18.71 and 33.48\,$\mu$m) , [SIV]
(10.51\,$\mu$m), [SiII] (34.81\,$\mu$m), [NeII] (12.81\,$\mu$m),
[NeIII] (15.56\,$\mu$m), [NeV] (14.32 and 24.30\,$\mu$m), [NeVI]
(7.64\,$\mu$m) and [OIV] (25.91\,$\mu$m), where the latter three
\textquotedblleft high excitation\textquotedblright\ lines are
indicative of highly-ionizing photon fields. The radiation field in
PNe also readily excites PAH emission. \citet{sta07} showed that the
presence of carbon-rich dust (i.e., SiC, MgS or PAHs) in the
\emph{Spitzer} IRS spectrum correlated with carbon-rich planetary
nebulae, and so we can make a distinction between {\tt C-PN} and {\tt
  O-PN}. In PNe the dust continuum first rises towards longer
wavelengths, but then turns over ($\lambda$$\approx$30\,$\mu$m for
young PNe) due to a lack of large amounts of cold dust in the regions
around PNe \citep{ber08,ber09}.

\subsection{H{\sc ii} regions}\label{hiiregs}

H{\sc ii} regions form around young hot stars and appear very similar
to planetary nebulae in their mid-infrared spectra. They show the same
range of forbidden line emission, with the exception of the
higher-excitation lines [NeV] and [OIV], due to the less
strongly-exciting radiation field in H{\sc ii} regions. Another
differentiating feature is that the dust continuum at longer
wavelengths ($\lambda\sog$30\,$\mu$m) continues to rise and falls at
wavelengths far outside of the \emph{Spitzer} IRS range since H{\sc
  ii} regions are typically embedded in Giant Molecular
Clouds. \citet{buc06} find that H{\sc ii} regions in their sample are
more luminous than PNe, having an infrared (1--100\,$\mu$m) luminosity
of 0.2--5$\times$10$^5$\,L$_\odot$.

It should be noted that since \emph{Spitzer} IRS observations in the
LMC are likely only sensitive to more massive YSOs, there is a
continuum of properties from embedded YSO to ultra-compact H{\sc ii}
region to classical H{\sc ii} region, and at the distance of the LMC
this may also become confused by spatial resolution issues. The IRS
has an angular resolution of the order of arcseconds, which at the
distance of the LMC would mean a resolution of 0.5\,pc at
best. Therefore (ultra)compact H{\sc ii} regions would not be resolved
\citep[cf.][]{chu02}.  Thus there may be some ambiguity between
different classes here, especially when, for instance, weak silicate
absorption is veiled by PAH emission. For this reason, we classify
(ultra)compact H{\sc ii} regions as YSOs (see discussion in
\S\ref{sec:sourceclass}) and classify featureless spectra with
low-excitation forbidden line emission as H{\sc ii} regions.

\subsection{Galaxies}

Quiescent galaxies are usually elliptical and have very low rates of
star formation. Thus their infrared spectra appear to be similar to
the spectrum of a K and M star combined, since these star classes are
what dominate the stellar population in these galaxies. Active
galaxies are presently star-forming, and thus show emission lines from
[NeII] and frequently [NeIII]. Those active galaxies with active
galactic nuclei (AGN) will show [NeV] emission. Active galaxies also
show PAH emission. These narrow features are red-shifted in spectra of
galaxies, allowing a good determination of distance. Broader features
(e.g., silicate emission at 10\,$\mu$m) are also red-shifted, but a
precise determination of redshift is more difficult. Mid-infrared
spectra of galaxies are discussed in more detail in \citet{hao07}, for
example. The galaxies in our sample are unresolved, and treated as
point sources.

\subsection{Other types of object}

Less-common objects were not given their own classes, but examples are
described below and plotted under the catch-all of \textquotedblleft
{\tt OTHER}\textquotedblright\ in Figs.~\ref{fig:cmdbasic} and
\ref{fig:blumf34}. Their classifications are given common astronomical
abbreviations in Table~\ref{tab:psclass}.

\subsubsection{R Coronae Borealis stars}

R Coronae Borealis stars (R CrB) are hydrogen-deficient supergiants
which are irregularly variable. Thought to be formed by the merger of
two white dwarf stars or by a final helium-shell flash, they go
through episodic dimming due to the formation of what is proposed to
be carbon dust in their atmospheres \citep{cla96,cla02}. In the
mid-infrared they appear featureless, and they exhibit a red SED akin
to an evolved star \citep[e.g.,][]{lam01,kra05}.

\subsubsection{B supergiants}

B supergiants (BSGs) are early-type dusty stars which usually present
a flat spectrum with weak silicate features \citep{buc09}. They also
show a deeply double-peaked SED with a blue stellar peak and a red
dust peak thought to be due to a dusty circumstellar disc or torus,
\citep[e.g.,][]{bol07,bon09,bon10}.

%%Cirrus hotspots.

\subsubsection{Wolf-Rayet stars}

Wolf-Rayet (WR) stars are evolved massive stars surrounded by a thick,
dusty envelope. Characterised by strong emission lines such as HeII,
[NeII], [NeIII], [SIV] and [OIV], they also show a red SED due to warm
circumstellar dust. WR stars show dust emission from three sources: 1)
self-produced carbon dust, only detected in late-type WC stars as a
7.7\,$\mu$m-type feature, attributable to a C--C stretch mode of
carbonaceous-type grains and consistent with a lack of C--H stretch or
bend modes given that their stellar winds are H-deficient
\citep*{chi02,che00}; 2) wind collisions between a WR wind and a
colliding O/B star wind \citep*{tut99} and 3) swept-up interstellar
dust. With the low spectral resolution of \emph{Spitzer}, it will be
hard to identify one of these three mechanisms.

%%WN stars?

%%Dust produced in interacting winds in massive binaries (e.g., Pinwheel
%%Nebula WR104)? \citep{tut99}

\subsection{Unknown objects}

Objects that we have not been able to classify with a high confidence
are classified as \textquotedblleft Unknown\textquotedblright\ ({\tt
  UNK}). Generally the cause of this is low signal-to-noise data,
where spectra have been too noisy to show distinct features, or where
a mispointing of the telescope has meant that our intended target was
missed and no target was located in the slit.

\subsection{Usage of the classification tree}

\begin{figure*}
%\epsscale{.80}
\includegraphics[angle=90,width=16cm]{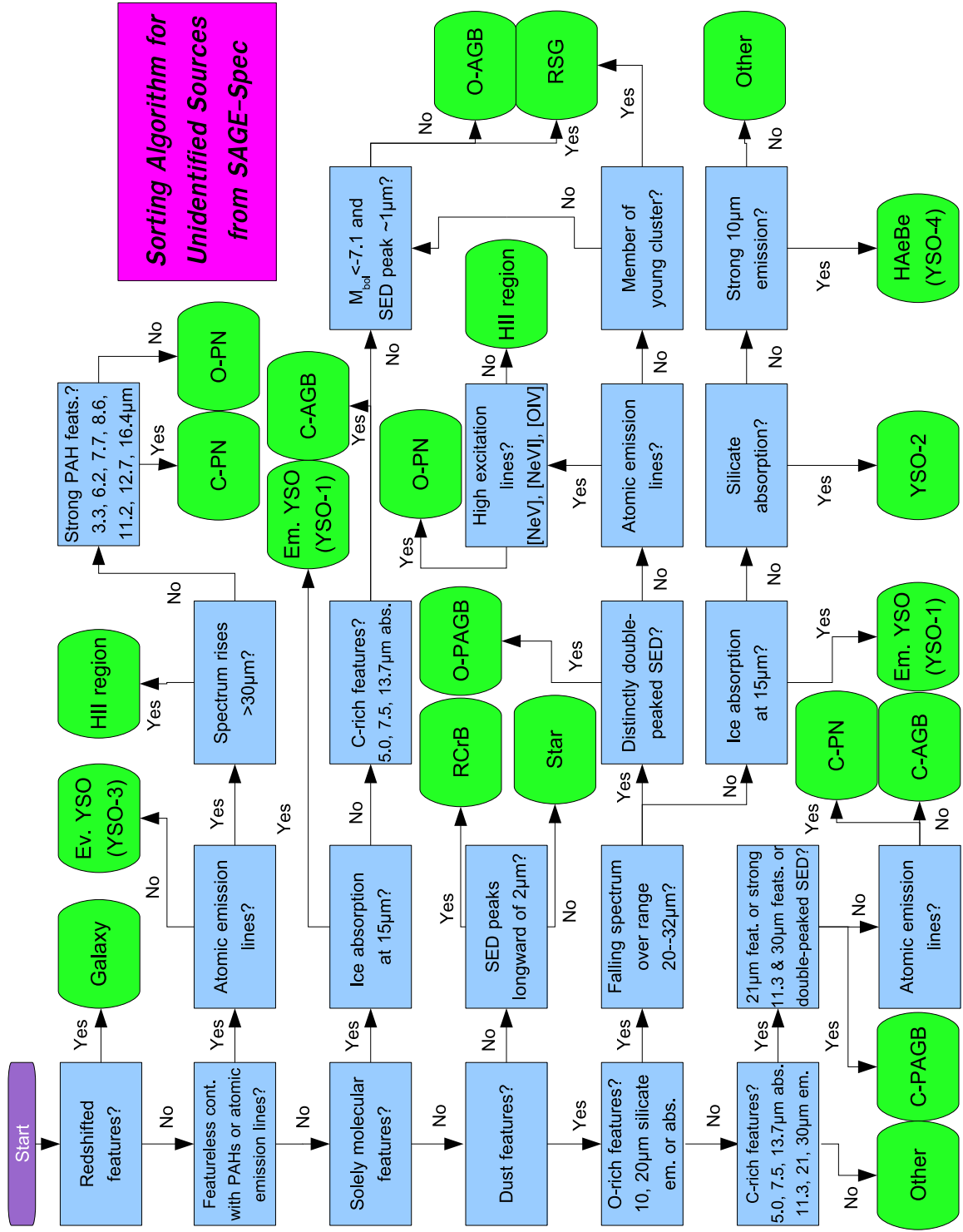}
\caption{The decision tree for the classification of the
  \SSp\ sample. For the purposes of the classification tree, the word
  \textquotedblleft spectrum\textquotedblright\ is used to indicate a
  plot of flux intensity (F$_\nu$) against wavelength ($\lambda$)
  showing the higher dispersion data from the spectrograph, and this
  is different to \textquotedblleft SED\textquotedblright, which is
  used to mean a plot of log($\lambda$F$_\lambda$) (or $\nu$F$_\nu$)
  against log($\lambda$) which includes all available photometry, as
  well as the spectral data. Abbreviations used in this figure:
  \textquoteleft cont.\textquoteright\ for continuum, \textquoteleft
  ev.\textquoteright\ for evolved, \textquoteleft
  em.\textquoteright\ for embedded, and \textquoteleft
  feat(s).\textquoteright\ for features. See \S\ref{sec:treeops} for
  more details.
\label{fig:psctree}}
\end{figure*}

\subsubsection{Operation}
\label{sec:treeops}

Our classification methodology is shown in the form of a decision tree
(Fig.~\ref{fig:psctree}). The tree aims to classify objects into
different categories via means of Yes/No selections. Not all of the
decisions rely solely on the appearance of the infrared spectrum in
question, but where possible this has been the main discriminant. Such
a rigorous method of classification lends itself to the classification
of large samples of (spectroscopic) infrared data. Thus this tree
could readily be applied to data from the \emph{Infrared Astronomical
  Satellite} (IRAS) and \emph{Infrared Space Observatory} (ISO)
satellites, and with a few adaptations could also be applied to AKARI,
\emph{Herschel Space Observatory} (\emph{Herschel}) and \emph{James
  Webb Space Telescope} (JWST) data, for example. It also lends itself
to automation, given that the requisite information on variability and
cluster membership, for example, can be gathered.  The classification
tree is designed to be simple to use; however, some of the decision
boxes may require slightly more explanation than that provided in the
caption to Fig.~\ref{fig:psctree}.

\textquotedblleft Featureless cont. with PAHs or atomic emission
lines?\textquotedblright\ partially filters out the spectra without
dust or molecular features, given the aforementioned (\S\ref{YSOspec})
issues with diffuse PAH and atomic features in the spectrum. Spectra
captured by this filter will generally comprise of a rising continuum
at longer wavelengths with several atomic emission lines or PAH
features, and this indicates that the emitting object is a planetary
nebula or H{\sc ii} region. Other featureless spectra are collected by
the \textquotedblleft Dust features?\textquotedblright\ box.

\textquotedblleft Solely molecular features?\textquotedblright\ is
intended to filter spectra with dust and molecular features from
spectra with just molecular features. Ideally, this filter captures
embedded YSOs with evident ice absorption features but no (or weak)
silicate features, and also evolved stars (C-AGB/O-AGB/RSG) which do
not appear significantly dusty.

\textquotedblleft Falling spectrum over range
$\sim$20--32\,$\mu$m?\textquotedblright\ is intended to distinguish
between YSOs and evolved oxygen-rich objects. Generally, if a spectrum
falls after the 20\,$\mu$m silicate feature the emitting object is an
evolved star, with only moderate amounts of cold dust
\citep{vlo10}. However, if a spectrum rises or remains level it is
more often than not a young object. The particular range of
$\sim$20--32\,$\mu$m was chosen to exclude $\lambda>32$\,$\mu$m to
avoid any far-IR rise due to contaminating sources in the IRS Long-Low
slit. If spectral data in the region 20--32\,$\mu$m is missing, one
should use the 24\,$\mu$m photometry as a replacement.

\begin{figure}
%\epsscale{num}
\includegraphics[width=84mm]{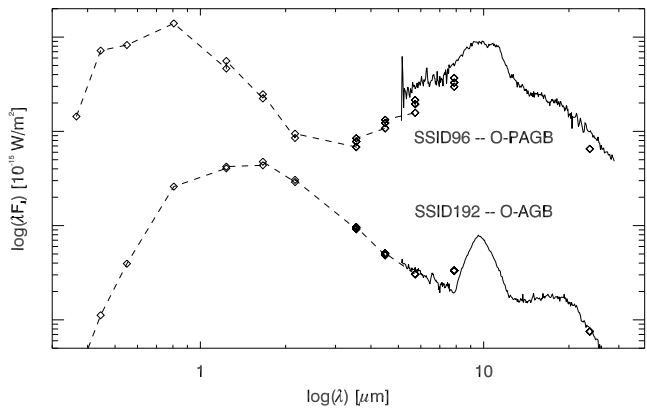}
\caption{An illustration of a double-peaked SED (post-AGB) compared to
  a stellar peak with a large dust excess (AGB). See text.}
\label{fig:doublepeaks}
\end{figure}

\textquotedblleft Double-peaked SED\textquotedblright\ (in two
locations) is used to separate post-AGB objects. Here the crucial
distinction to make is between an SED with a stellar component which
has a large infrared excess, and an SED with a stellar and a dust peak
with a minimum between them. An illustration is given in
Fig.~\ref{fig:doublepeaks}.

\textquotedblleft Other\textquotedblright\ (in two locations) is a
catch-all for unusual or uncommon objects. The classification of the
objects which fall into this category is then performed by referring
to the astronomical literature (e.g., with the help of
SIMBAD\footnotemark[2]\footnotetext[2]{\url{http://simbad.u-strasbg.fr/simbad/}})
or by comparing spectra of similar, known objects. If an object cannot
reliably be classified in this way, or if the suggested classification
is not in accord with the \emph{Spitzer} IRS spectrum, it is deemed
\textquoteleft unknown\textquoteright\ and classified as {\tt UNK}.

\subsubsection{Caveats and data issues}

The decision tree in Fig.~\ref{fig:psctree} is a useful formalisation
of a difficult task. As a simplification of a complex problem, it is,
however, open to error, and this error can be minimised by being aware
of the following:

\emph{Spectral identification error}. The largest source of error is
incorrect identification of spectral features. The correct
identification of features is crucial, and the error reduces only with
experience in looking at spectra or very careful inspection of the
spectra. One particular class of object which may prove difficult to
spot is low-redshift galaxies ($z\sol0.02$). The potential shift
in wavelength of distinguishing dust features such as PAH emission and
silicate emission features due to different physical effects alone
(e.g., ionization of PAHs, differing dust grain shapes) may mask the
redshift of the emitting object.

\emph{Low signal-to-noise data/mispointings}. Noise can often
masquerade as real features in low-contrast spectra. For instance, the
spectrum for SSID42 could potentially show PAH features redshifted to
z$\sim$1.57. However, given the low levels of flux and the lack of an
obvious point source in 8\,$\mu$m images, it seems that this spectrum
is solely noise. Spectra can also be contaminated by nearby objects
thereby altering the continuum shape and spectral
features. \emph{Spitzer}'s IRS instrument has two low-resolution
modules named Short-Low (SL) and Long-Low (LL), each of which has two
orders, SL1, SL2 and LL1, LL2. Each slit of SL has an aperture of
3\farcs6$\times$57\arcsec\ (with a 22\arcsec\ separation). The two LL
slits are 10\farcs6$\times$168\arcsec\ apertures. With such a large
difference in apertures between SL and LL it is possible that SL and
LL may observe different objects. This was determined to be the case
for one observation, where supergiant {GV 60} (SSID16), the intended
target, was observed with SL, whereas LL picked up emission from a
nearby Wolf-Rayet star, {LH$\alpha$120-N~82} (alternatively named
{Brey~3a} or {IRAS~04537-6922}) (SSID17). Thus not only is it possible
to add serendipitous detections to the sample, but it is also possible
that contamination may affect the observed spectrum, especially in
LL. A more full discussion of these two objects can be found in
\citet[][object \#4]{vlo10}.  Furthermore, issues in data reduction
can introduce spurious features. For \emph{Spitzer} IRS data an
obvious place for this to occur is at the joins between modules. For
instance, the join between SL1 and LL2 falls at $\sim$14\,$\mu$m,
which can complicate the continuum determination to the blue of the
ice feature due to CO$_2$ ice. Similarly, the join between SL1 and SL2
at $\sim$8\,$\mu$m could affect the recognition of an SiO feature in a
stellar spectrum.

\emph{Foreground objects}. We take no account of foreground objects in
the decision tree (Fig.~\ref{fig:psctree}). Since the only place where
we take distance into account in the tree is indirectly through the
luminosity discriminant between oxygen-rich AGB stars and red
supergiants, source classification is not particularly affected by
distance. It may be that the {\tt RSG} category is contaminated by
foreground oxygen-rich red giant branch and asymptotic giant branch
stars, however these objects should be distinguishable by means of
colour, variability or radial velocity measurements.

\emph{Limited data}. In the case where spectral coverage is limited
(e.g., Long-Low data is not available) or other information is
lacking, the decision tree may fail to end in a firm
classification. In this case the classification must be limited to the
classes remaining along the branch of the tree where the failure
occurred.

%Once the 197 \SSp\ spectra had been classified via the
%\textquotedblleft comprehensive\textquotedblright\ method, they were
%classified using the decision tree (Fig.~\ref{fig:psctree}). This
%resulted in a success rate of ???\%, where success is determined in
%terms of a match to the \textquotedblleft
%comprehensive\textquotedblright\ classification. (Success rate was
%$\sim$75\% for v0.3, $\sim$80\% for v.0.4. v0.5 is untested, as yet)

%To assess the effectiveness of this scheme, once the \SSp\ sample had
%been classified by the \textquotedblleft
%comprehensive\textquotedblright\ scheme (see
%Sect.~\ref{sec:sourceclass}), we then processed all the spectra with
%the classification tree (Sect.~\ref{sec:treeclass}) and compared the
%results.

\subsubsection{Quality control}

IRAC 8\,$\mu$m images were examined for all 196 pointings in order to
flag the presence of nearby objects and the \emph{Spitzer} Basic
Calibrated Data (BCDs) were thoroughly examined for multiple sources,
cosmic-ray hits and diffuse emission not localised to the point
source. A full discussion of the errors associated with the
observations and the steps taken to correct them can be found in the
\SSp\ data delivery document \citep{woo10}. PAH emission is seen
frequently in the spectra, but arises in many cases due to the
presence of diffuse ISM along the line of sight to our intended
target. Similarly, spectra can include background emission lines at
15.6, 18.7, 33.5, 34.8 and 36.0\,$\mu$m, arising from diffuse gas in
the LMC. These are due to [Ne{\sc iii}], [S{\sc iii}], [S{\sc iii}],
[Si{\sc ii}] and [Ne{\sc iii}] lines respectively, and are not always
subtracted effectively due to non-uniform variation across the
slit. As such, the design of the classification tree
(Fig.~\ref{fig:psctree}) is devised so as not to use the presence of
PAH emission as an indication of a carbon-rich nature, except in
  the case of carbon-rich PNe \citep{sta07}.

\section{\SSp\ point source classification}
\label{sec:sourceclass}

\subsection{Calculation of $M_\mathrm{bol}$}

Bolometric magnitudes were calculated for all \SSp\ sources using the
method described by \citet{slo08}. We adopt a distance to the LMC
  in our calculations of 50\,kpc \citep[e.g.,][]{sch08,kel06,fea99},
  and thus a distance modulus of 18.5\,mag \citep[e.g.,][]{alv04}.
The \citet{slo08} technique involves an integration of the IRS
spectrum and photometry points, fitting a 3\,600\,K Planck Function to
the optical photometry, and adding a Rayleigh-Jeans tail to the
long-wavelength data. For redder sources with large amounts of cold
dust, this will mean that the bolometric magnitude we derive will be a
lower limit. For stellar photospheres, low dust-excess
oxygen-rich AGB stars and red supergiants, we also calculate the
bolometric magnitude by fitting a MARCS stellar atmosphere model
\citep{gus75,gus08} to the SED. This method provided a better fit to
the optical and near-IR photometry which cover the peak of the energy
distribution of these objects.  Such a method has been previously used
by \citet{mcd09}, \citet{boy10} and McDonald et al. (2010, submitted),
and we refer the reader to those papers for details of the MARCS
  models, and for an evaluation of errors. In general, the agreement
between the \citet{slo08} method and \citet{mcd09} method was good to
within 0.1\,mag. We prefer the \citet{mcd09} method values for the
above-mentioned three categories. As a further check, we compared the
luminosities derived for AGB stars with those calculated by
\citet{sri09} for their sample of LMC AGB stars. Again, agreement was
reasonably good, although the values of \citet{sri09} were generally
0.15\,mag brighter across the board. This is likely due to the use of
a different zero-point flux. Results of our calculations are shown in
Table~\ref{tab:lums}.

\begin{figure*}
%\epsscale{2.15}
\includegraphics{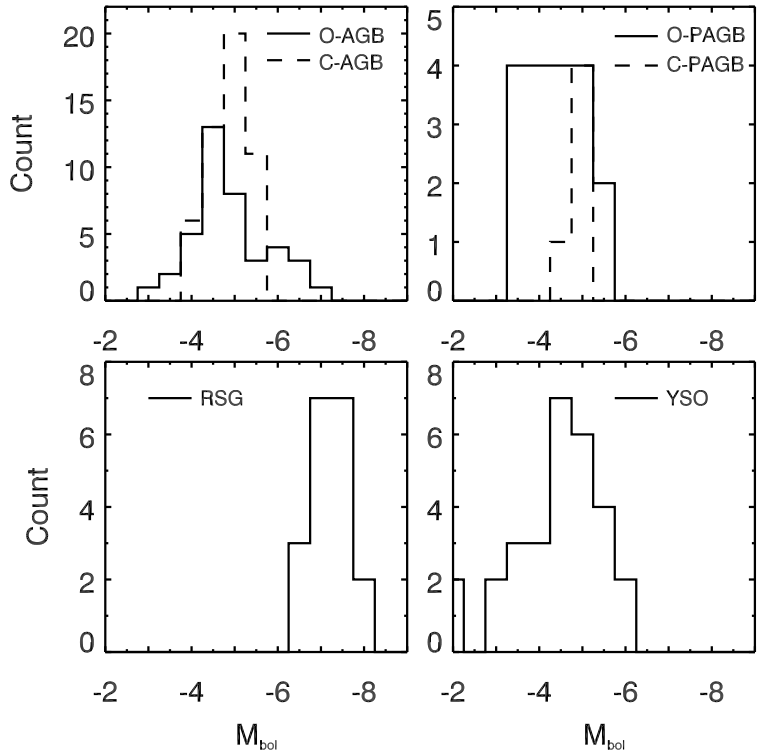}
\caption{Luminosity functions for AGB stars, post-AGB stars, RSGs and YSOs.}
\label{fig:lumfuncs}
\end{figure*}

Fig.~\ref{fig:lumfuncs} shows histograms of bolometric magnitudes for
AGB stars, post-AGB stars, RSGs and YSOs. Most C-AGB stars have a
bolometric magnitude of $\sim$-5\,mag, whilst O-AGB stars have two
peaks, the one at higher magnitude possibly due to O-AGBs
  currently undergoing Hot Bottom Burning \citep{boo92}. C-PAGB stars
show a peak in bolometric magnitude which coincides with that for
C-AGB stars.

%\begin{figure*}
%\epsscale{2.15}
%\plotone{lumplot.eps}
%\caption{Bolometric magnitudes for the \SSp\ sample, calculated using five different methods, P, L, S, I \& G (see text for details).}
%\label{fig:lums}
%\end{figure*}

\begin{table*}
\begin{minipage}{16cm}
\caption{Bolometric magnitudes for \SSp\ sources. Effective
  temperatures derived from fitting the SED of small-dust-excess oxygen-rich
  stars are given in parentheses. Typical errors are $\pm$100\,K. \label{tab:lums}} \small
\begin{tabular}{l@{~}l|l@{~}l|l@{~}l|l@{~}l|l@{~}l|l@{~}l}
\hline
SSID & $M_\mathrm{bol}$ & SSID ($T_\mathrm{eff}$) & $M_\mathrm{bol}$ &SSID ($T_\mathrm{eff}$) & $M_\mathrm{bol}$ &SSID ($T_\mathrm{eff}$) & $M_\mathrm{bol}$ &SSID ($T_\mathrm{eff}$) & $M_\mathrm{bol}$ &SSID & $M_\mathrm{bol}$\\
\hline
\cline{1-2}
C-AGB	&		&	156	&	-4.67	&	197 (3\,106\,K)	&	-4.78	&	168	&	-4.66	&	37 (3\,985\,K)	&	-6.66	&	42	&	-1.82	\\\cline{1-2}
51	&	-5.70	&	41	&	-4.66	&	148 (3\,010\,K)	&	-4.63	&	187	&	-4.50	&	35 (3\,986\,K)	&	-6.29	&	184	&	-1.54	\\\cline{9-10}
31	&	-5.70	&	132	&	-4.56	&	13 (3\,269\,K)	&	-4.53	&	11	&	-4.36	&	O-PAGB	&		&	138	&	-1.04	\\\cline{9-10}\cline{11-12}
87	&	-5.61	&	103	&	-4.54	&	63 (2\,980\,K)	&	-4.49	&	90	&	-4.36	&	162	&	-5.60	&	OTHER	&		\\\cline{11-12}
30	&	-5.54	&	3	&	-4.51	&	67 (3\,143\,K)	&	-4.47	&	109	&	-4.34	&	177	&	-5.31	&	17 (W-R)	&	-5.44	\\
167	&	-5.52	&	49	&	-4.49	&	96	&	-4.47	&	164	&	-4.33	&	56	&	-5.02	&	50 (BSG)	&	-5.35	\\
33	&	-5.52	&	141	&	-4.46	&	77 (3\,122\,K)	&	-4.44	&	5	&	-4.26	&	115	&	-4.93	&	69 (BSG)	&	-5.04	\\
18	&	-5.47	&	47	&	-4.42	&	110 (2\,886\,K)	&	-4.43	&	21	&	-4.23	&	118	&	-4.86	&	94 (RCrB)	&	-4.60	\\
48	&	-5.45	&	60	&	-4.39	&	1 (3\,524\,K)	&	-4.43	&	106	&	-4.21	&	131	&	-4.85	&	26 (RCrB)	&	-3.55	\\\cline{11-12}
145	&	-5.41	&	80	&	-4.37	&	152 (2\,925\,K)	&	-4.38	&	97	&	-3.83	&	73	&	-4.75	&	C-PAGB	&		\\\cline{11-12}
190	&	-5.35	&	179	&	-4.23	&	59 (5\,437\,K)	&	-4.38	&	114	&	-3.63	&	161	&	-4.74	&	150	&	-5.05	\\
105	&	-5.30	&	46	&	-4.17	&	72 (3\,085\,K)	&	-4.36	&	163	&	-3.32	&	192	&	-4.28	&	196	&	-5.00	\\
86	&	-5.25	&	57	&	-4.05	&	91	&	-4.71	&	137	&	-3.30	&	111	&	-4.26	&	52	&	-4.89	\\
119	&	-5.25	&	53	&	-4.04	&	89 (3\,542\,K)	&	-4.33	&	158	&	-2.98	&	85	&	-4.18	&	84	&	-4.88	\\
12	&	-5.22	&	191	&	-3.91	&	166 (3\,660\,K)	&	-4.20	&	44	&	-2.91	&	29	&	-4.03	&	64	&	-4.50	\\\cline{11-12}
136	&	-5.11	&	43	&	-3.83	&	99 (3\,538\,K)	&	-4.16	&	149	&	-2.10	&	157	&	-3.95	&	O-PN	&		\\\cline{3-4}\cline{11-12}
120	&	-5.10	&	O-AGB	&		&	124 (2\,561\,K)	&	-4.02	&	25	&	-1.77	&	107	&	-3.94	&	186	&	-4.10	\\\cline{3-4}\cline{7-8}
189	&	-5.09	&	121	&	-7.04	&	54	&	-3.98	&	RSG	&		&	75	&	-3.72	&	100	&	-2.29	\\\cline{7-8}
15	&	-5.06	&	165 (2\,561\,K)	&	-6.68	&	182	&	-3.85	&	27 (4\,140\,K)	&	-8.19	&	95	&	-3.55	&	174	&	-2.22	\\
125	&	-5.06	&	82 (3\,183\,K)	&	-6.43	&	8 (2\,561\,K)	&	-3.74	&	117 (3\,977\,K)	&	-7.91	&	28	&	-3.43	&	153	&	-1.33	\\\cline{11-12}
139	&	-5.04	&	180	&	-6.50	&	79	&	-3.68	&	135 (3\,638\,K)	&	-7.68	&	113	&	-3.31	&	C-PN	&		\\\cline{9-10}\cline{11-12}
45	&	-4.94	&	142 (3\,686\,K)	&	-6.01	&	185	&	-3.08	&	16 (3\,159\,K)	&	-7.57	&	STAR	&		&	92	&	-4.85	\\\cline{5-6}\cline{9-10}
9	&	-4.93	&	61 (2\,987\,K)	&	-5.97	&	YSO	&		&	116 (3\,781\,K)	&	-7.52	&	19 (5\,250\,K)	&	-7.56	&	144	&	-4.66	\\\cline{5-6}
7	&	-4.89	&	38	&	-6.17	&	101	&	-6.13	&	171 (4\,487\,K)	&	-7.47	&	195 (4\,892\,K)	&	-6.97	&	175	&	-4.17	\\\cline{11-12}
126	&	-4.88	&	173 (3\,139\,K)	&	-5.91	&	20	&	-5.85	&	147 (3\,652\,K)	&	-7.43	&	81 (3\,514\,K)	&	-5.01	&	HII	&		\\\cline{11-12}
55	&	-4.86	&	6	&	-5.54	&	108	&	-5.63	&	122 (3\,871\,K)	&	-7.42	&	133 (3\,187\,K)	&	-4.98	&	104	&	-6.23	\\
83	&	-4.85	&	178 (3\,512\,K)	&	-5.68	&	34	&	-5.51	&	134 (4\,188\,K)	&	-7.32	&	32 (3\,471\,K)	&	-4.71	&	71	&	-1.58	\\\cline{11-12}
140	&	-4.82	&	68 (3\,500\,K)	&	-5.34	&	102	&	-5.42	&	169 (4\,008\,K)	&	-7.17	&	76 (3\,439\,K)	&	-4.67	&	UNK	&		\\\cline{11-12}
23	&	-4.80	&	130 (2\,561\,K)	&	-5.25	&	183	&	-5.33	&	170 (3\,851\,K)	&	-7.10	&	88 (3\,575\,K)	&	-4.31	&	78	&	-5.04	\\
65	&	-4.78	&	143 (3\,284\,K)	&	-5.22	&	62	&	-5.03	&	129 (3\,985\,K)	&	-7.09	&	188 (3\,588\,K)	&	-4.27	&	155	&	-4.18	\\\cline{9-10}
66	&	-4.78	&	58 (3\,552\,K)	&	-5.13	&	40	&	-5.01	&	123 (3\,844\,K)	&	-7.03	&	GAL	&		&	146	&	-3.73	\\\cline{9-10}
194	&	-4.78	&	93	&	-5.10	&	10	&	-4.86	&	128 (4\,041\,K)	&	-6.99	&	154	&	-3.14	&	39	&	-2.09	\\
181	&	-4.73	&	176 (2\,587\,K)	&	-4.99	&	74	&	-4.86	&	172 (3\,524\,K)	&	-6.84	&	193	&	-2.64	&	112	&	-1.14	\\
98	&	-4.71	&	22 (3\,965\,K)	&	-4.97	&	70	&	-4.82	&	127 (4\,078\,K)	&	-6.84	&	151	&	-2.58	&	24	&	---	\\
36	&	-4.67	&	159 (3\,536\,K)	&	-4.79	&	14	&	-4.77	&	4 (3\,693\,K)	&	-6.68	&	2	&	-2.14	&	160	&	---	\\
\hline
\end{tabular}
\end{minipage}
\end{table*}

\subsection{Classification results}

The results of the classification process are summarised in
Table~\ref{tab:psclass}, along with \SSp\ identifier (SSID),
co-ordinates and alternate designations for the sample objects, and
class counts are tallied in Table~\ref{tab:classtally}.

\begin{table*}
\vbox to220mm{\vfil Landscape table to go here.
\caption{}
\vfil}
\label{tab:psclass}
\end{table*}

\begin{table}
\caption{Classification groups and tally \label{tab:classtally}}
\begin{tabular}{lc|lc}
\hline
Type & Count & Type & Count\\
\hline
C-AGB    & 50    & GAL       & 7 \\
O-AGB    & 40    & OTHER     & 5 \\
YSO      & 29    & C-PAGB    & 5 \\
RSG      & 19    & O-PN      & 4 \\
O-PAGB*  & 18    & C-PN      & 3 \\
(*inc. RV~Tau)& (9) & H{\sc ii} & 2 \\
STAR     & 8     & UNK       & 7\\
\hline
\end{tabular}
\end{table}

\subsubsection{YSOs}

\begin{figure*}
%\epsscale{num}
\includegraphics[width=8.0cm]{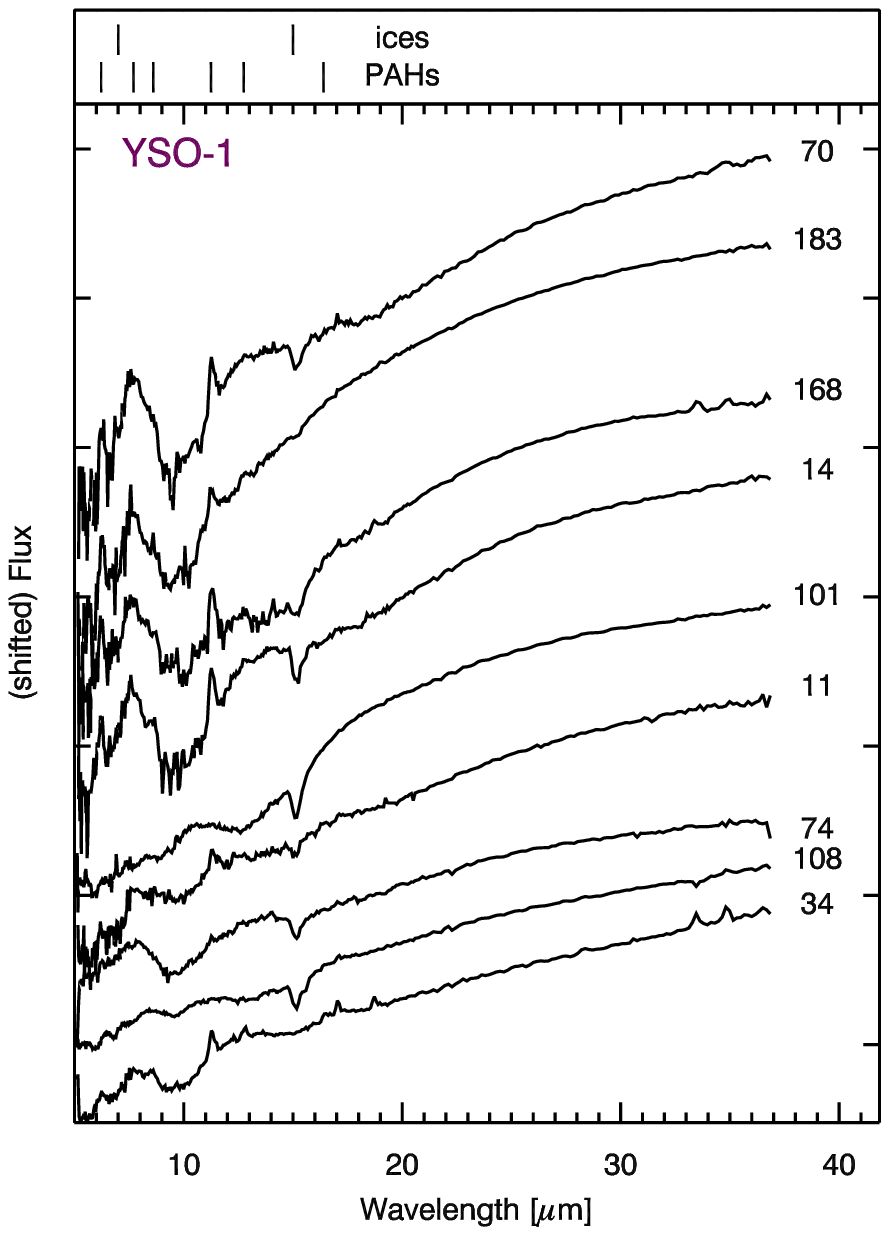}
\includegraphics[width=8.0cm]{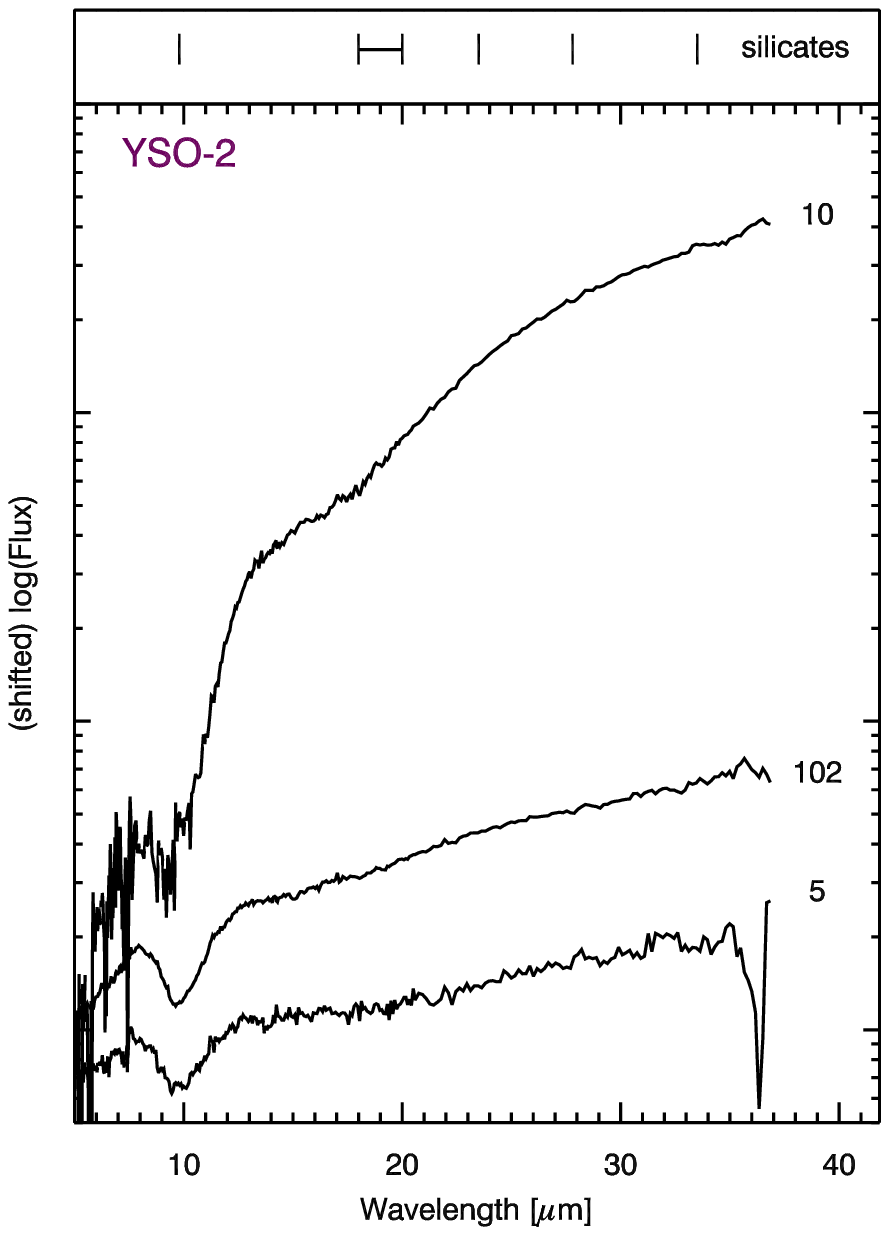}
%\plottwo{YSO-1.eps}{YSO-2.eps}
%\end{figure*} 
%\begin{figure*}
%\epsscale{num}
\includegraphics[width=8.0cm]{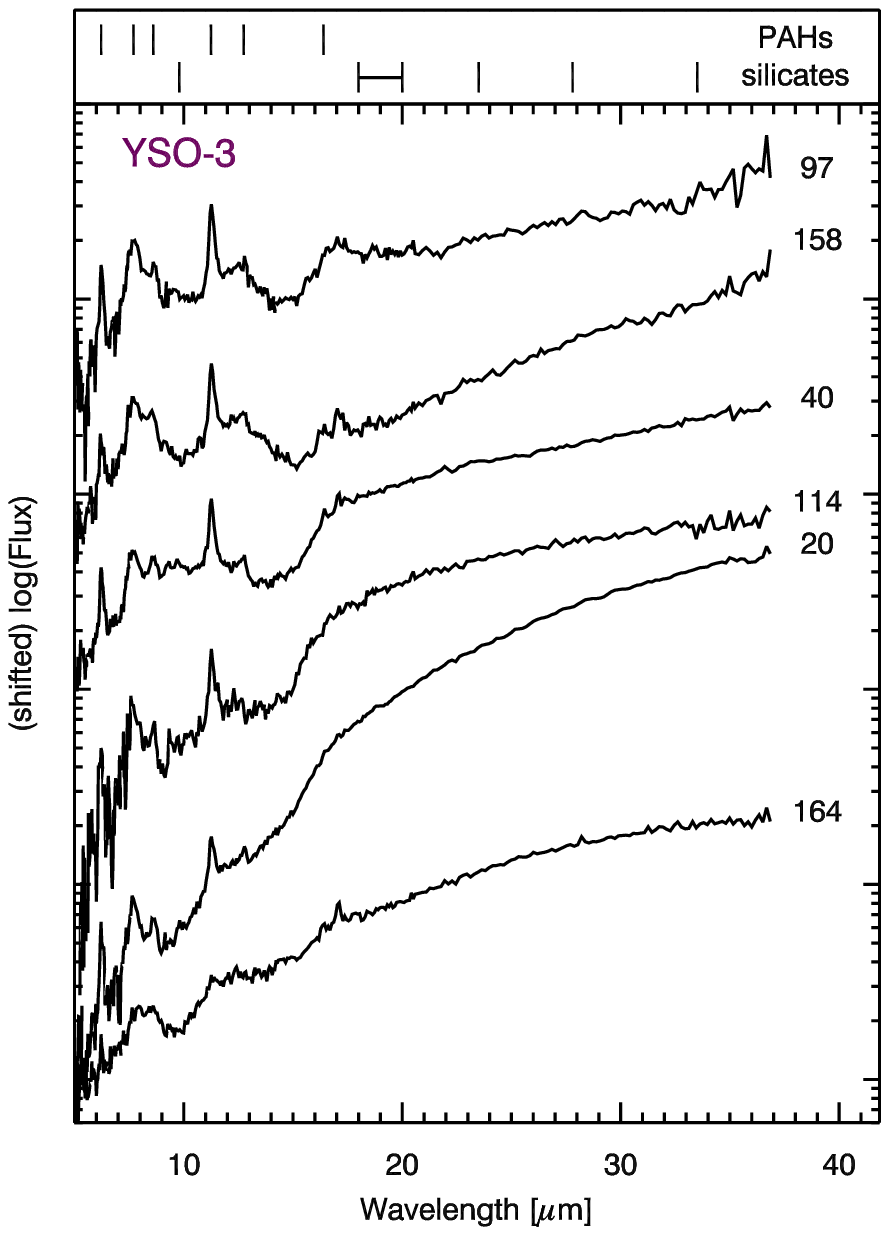}
\includegraphics[width=8.0cm]{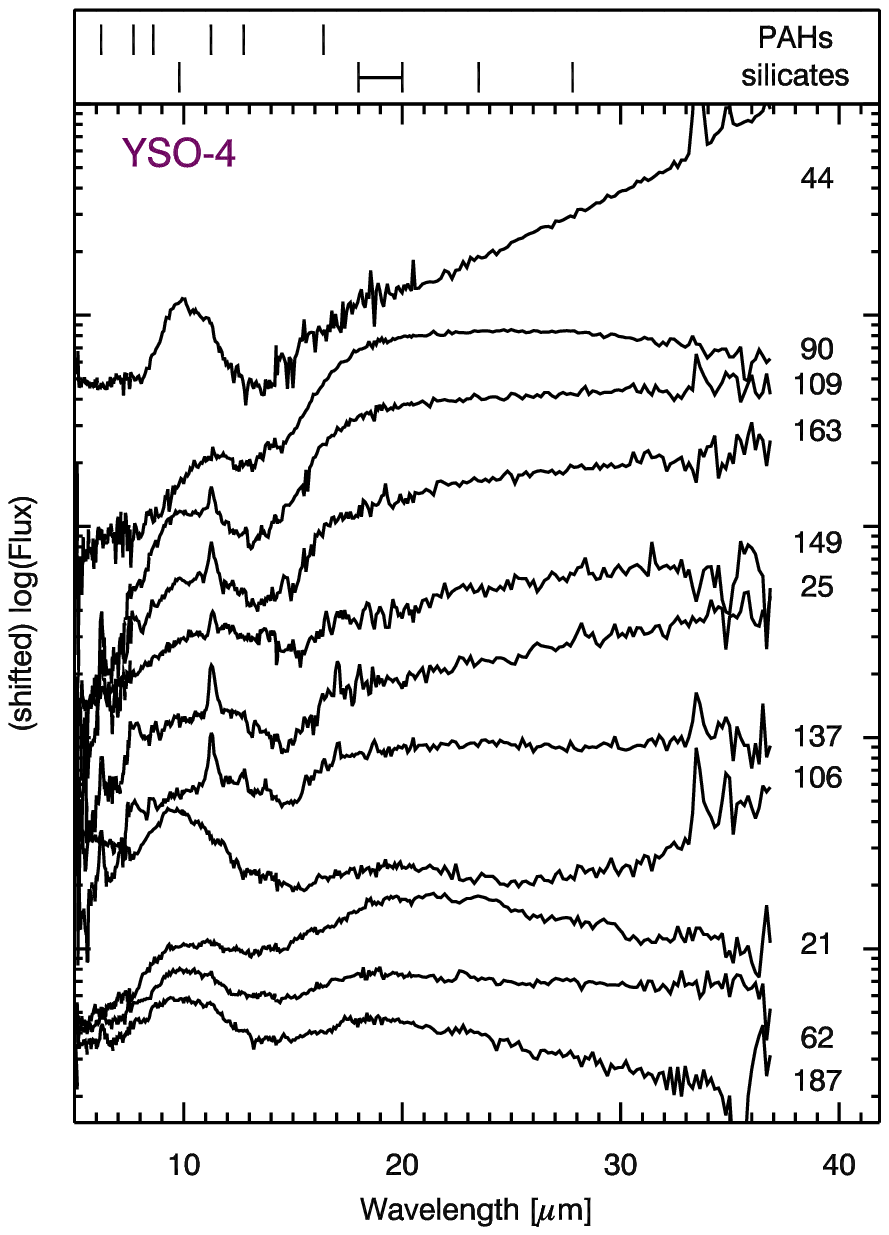}
%\plottwo{YSO-3.eps}{YSO-4.eps}
\caption{\SSp\ YSO spectra, sorted into four groups. These (except for
  those in group {\tt YSO-1}) and subsequent spectra in
  Figs.~\ref{fig:speceros}--\ref{fig:specunk} are presented as
  log($F_\nu$) versus $\lambda$\ for display purposes.  Group 1
  contains YSOs with ices, Group 2 has YSOs with silicate absorption
  (but no ices), Group 3 has YSOs with PAH emission (but no ices or
  overt silicate absorption) and Group 4 contains YSOs with silicate
  emission. Numeric labels refer to SSID (see
  Table~\ref{tab:psclass}).}
\label{fig:specyso}
\end{figure*}

The 30 YSO spectra in the \SSp\ sample broadly fall into two
categories: those with a rising F$_\nu$ spectrum toward longer
wavelengths, and those with a flat or declining spectrum
(Fig.\ref{fig:specyso}). The first group is significantly larger, with
23 YSOs falling into this group. Many of these objects are still
enveloped in dust, and the varying slopes of their continua are
evidence of dust at a range of temperatures. The remaining seven
objects represent more evolved YSOs with silicate features in emission
superimposed on a hotter dust continuum. Additional care needs to be
taken when identifying objects belonging to the latter group, as some
oxygen-rich evolved stars also exhibit silicate emission (see
\S\ref{OAGBdesc}).

We use mid-IR spectral features to classify the 30 YSOs in distinct
groups:
\begin{enumerate}
\item YSOs with ices (Fig.~\ref{fig:specyso}, {\tt YSO-1})
\item YSOs with silicate absorption (Fig.~\ref{fig:specyso}, {\tt YSO-2})
\item YSOs with PAH emission (Fig.~\ref{fig:specyso}, {\tt YSO-3})
\item YSOs with silicate emission (Fig.~\ref{fig:specyso}, {\tt
  YSO-4}).
\end{enumerate}
The classification of the spectra into the first three of these groups
is hierarchical, in the sense that a YSO is classified in Group 1 if
its spectrum has ice absorption features, irrespective of the presence
of silicate and/or PAH features; most (but not all) YSOs in this group
exhibit silicate absorption and PAH emission. The spectra of objects
in Group 2 show no evidence for ice features but exhibit silicate
features in absorption, while the spectra of YSOs in group 3 have PAH
emission without ice or overt dust features.  This phenomenological
classification approach is similar to that employed by
\citet{sea09}. The icy objects in Group 1 are the YSOs described in
\citet{oli09} while the other YSOs are newly identified in this work:
three in Group 2, six in Group 3 and eleven in Group 4.

Groups 1--3 roughly represent an evolutionary sequence for massive
YSOs: from sources deeply embedded in the cold molecular material that
exhibit strong ice and dust absorption features in their mid-IR
spectra \citep[e.g.,][]{boo08}, to (ultra-)compact H{\sc ii} sources,
whose spectra are dominated by PAH and fine structure line emission,
where the central object becomes hotter and the dusty envelope becomes
progressively more tenuous \citep[e.g.,][]{chu02}. As discussed
previously and extensively in \citet{sea09} and \citet{oli09}, the
issue of spatial resolution at the distance of the LMC needs to be
considered carefully when trying to diagnose the evolutionary state of
a YSO. The IRS slits are wide enough that emission originating from
regions external to the YSO envelope can contaminate the observed
spectrum (i.e., as is likely the case for objects that sit near to
H{\sc ii} regions). This is likely to be the explanation for the
difference in appearance between our Group 1 YSOs, for example, and
the Galactic Class I YSOs discussed by \citet{fur08}. Many of their
icy YSOs do not show PAH features, whereas our sample do, and this is
likely due to the coverage of the IRS slits, which cover a larger
region in the LMC than they do in the Galaxy. It is also possible that
the observed spectrum originates from a small compact cluster not from
a single source.

Group 4 represents a more evolved state of lower mass YSOs, likely
HAeBe stars.  This stage is reached once the accretion rate from the
cold envelope is depressed, as most of the dusty material is now found
in a circumstellar disc.  The shape of the SED for these objects is
also sensitive to other factors like disc inclination angle. The
silicate emission features in such objects can exhibit signatures of
dust processing (e.g., crystallization or modification of the grain
size distribution), as dust grains are subjected to higher
temperatures. PAH emission is also common in Galactic HAeBe stars
\citep[e.g.,][] {kel08}. From the sample of 11 objects in Group 4,
four objects show silicate emission superimposed on a rising
continuum. \citet{fur08} observed objects with similar spectral
properties in the Galaxy and suggested that such objects are likely
examples of a transition between Class I and Class II YSOs, already
with a significant contribution from a dusty disc but still embedded
in a relatively low-density envelope. The remaining 7 LMC YSOs in this
group have a flat or downward sloping continuum and thus are likely
the LMC analogs of Galactic Class II sources \citep{fur06}.

\subsubsection{Stellar photospheres}

Two of the eight of the sample classified as {\tt STAR} have an SED
peak between 1--2\,$\mu$m and calculation of \emph{J}-\emph{H} and
\emph{H}-\emph{K} shows that they are likely foreground K
giants. These are SSID19 and 195. The remaining six have significantly
cooler effective temperatures of $\approx$3\,400--3\,600\,K. This,
combined with their \emph{J}-\emph{H} and \emph{H}-\emph{K} values,
would indicate that SSID32, 76, 81, 88, 133 and 188 are M4 or M5
giants. SSID81 and 188 are both classified by \citet{kon01} as carbon
stars, but we see no indication of carbon-rich molecular or dust
features in the IRS spectra. SSID88 and 133 show PAH features from
foreground emission; SSID88 is not visible in \emph{SAGE-LMC}
8\,$\mu$m images, and SSID133 is a faint point-source.

\subsubsection{Carbon-rich AGB stars}

\begin{figure}
%\epsscale{num}
\includegraphics[width=8.4cm]{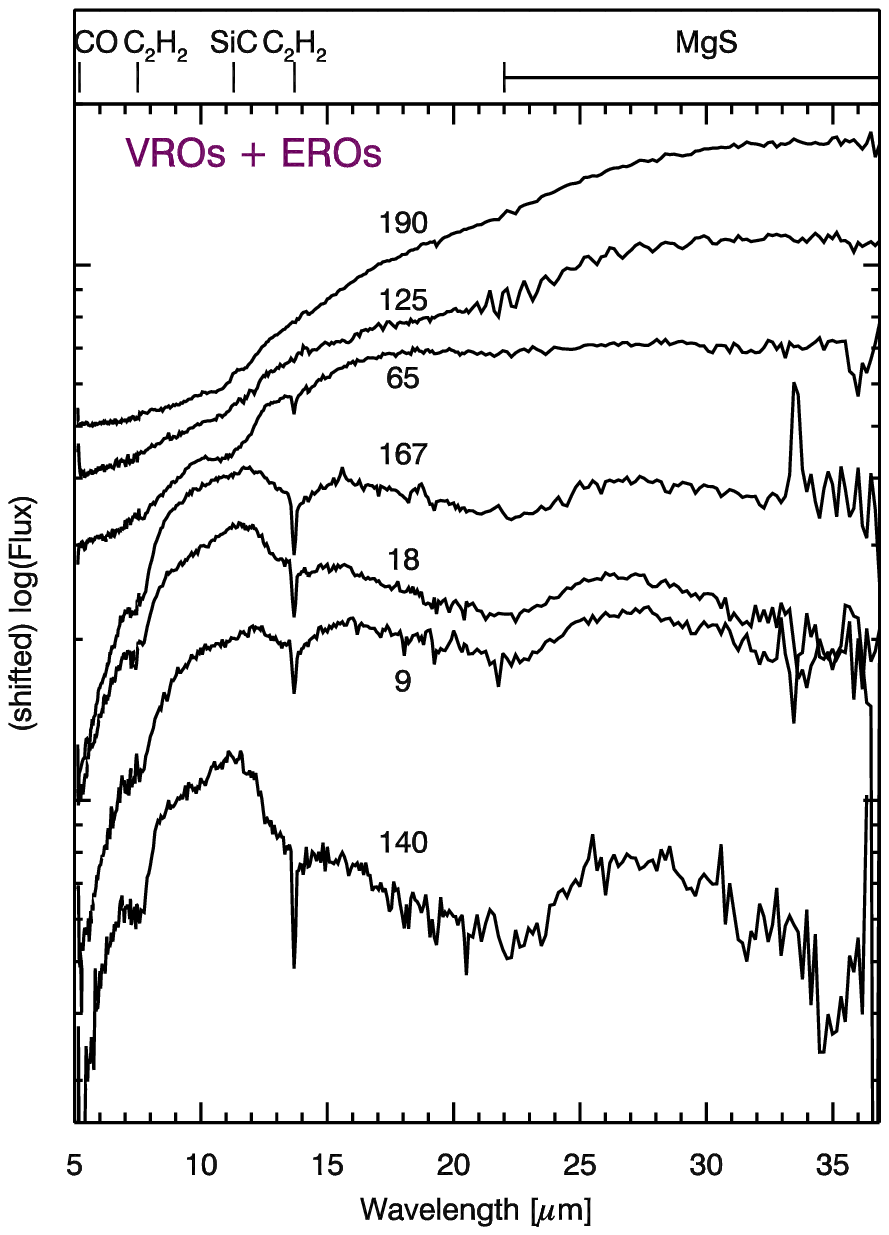}
\caption{Spectra of the VRO and ERO carbon-rich AGB stars.}
\label{fig:speceros}
\end{figure}

The sample of carbon star spectra ranges from almost-photospheric with
mild acetylene absorption to extremely red objects \citep[EROs;
  according to the definition of][]{gru08}. The colour criteria for
EROs, that they have extremely red mid-IR colours
([4.5]$-$[8.0]$>$4.0), and that they all fall in a narrow range of
brightness (7.0$<$[8.0]$<$8.5), are met by three of the sample,
SSID65, 125 and 190. All three have a strong MgS features at
$\sim$30\,$\mu$m (Fig.~\ref{fig:speceros}).  These criteria are also
met by two carbon-rich post-AGB objects, SSID84 and 196. We identify
four objects which we dub very red objects (VROs) which are not quite
as extreme in colours as EROs. They meet the criteria,
2.0$<$[4.5]$-$[8.0]$<$4.0 and 6.0$\leq$[8.0]$\leq$7.0, and have very
pronounced SiC features. These objects are SSID9, 18, 140 and
167. Figure~\ref{fig:speceros} shows the seven VRO and ERO
spectra. All seven VROs and EROs would be classed as \textquotedblleft
extreme AGB\textquotedblright\ stars by \citet{blu06}, \citet{sri09}
and others, who use the criteria J$-$[3.6]$\geq$3.1, [3.6]$\leq$10.5.

In general, we note that the C$_2$H$_2$\ bands appear stronger in the
dusty LMC C-rich AGB stars than in Galactic analogs, such as
{AFGL 3068} and {IRC+10216} \citep{jus98,jus00}. This
has been noticed and discussed in much depth previously
\citep[e.g.,][]{vlo99,mat02,mat05,vlo06,slo06,spe06,zij06,vlo08}. This
tendency also appears to be present for the less dusty objects in our
sample, although in those cases it is more difficult to distinguish
between the photospheric and circumstellar molecular bands. In these
less-dusty objects, the 11.3\,$\mu$m feature appears weaker than in
typical C-rich Galactic AGB stars \citep[e.g,][]{zij06}. The EROs show
clear instances of SiC absorption, and these features are stronger
than in any known Galactic object. As near-infrared colours are similar
in both samples, this does not appear to be solely a total optical
depth effect. The 30\,$\mu$m feature strength varies widely from
object to object, in some cases being very weak even when the SED
indicates a large total dust optical depth
\citep*[cf.,][]{lei08,lag07}. The cause of this wide variation is not
known; in the Galactic objects one does not see such a wide range of
30\,$\mu$m feature strengths in the dustier objects \citep{hon02}.

\subsubsection{Oxygen-rich AGB stars}

\begin{figure}
%\epsscale{num}
\includegraphics[width=8.4cm]{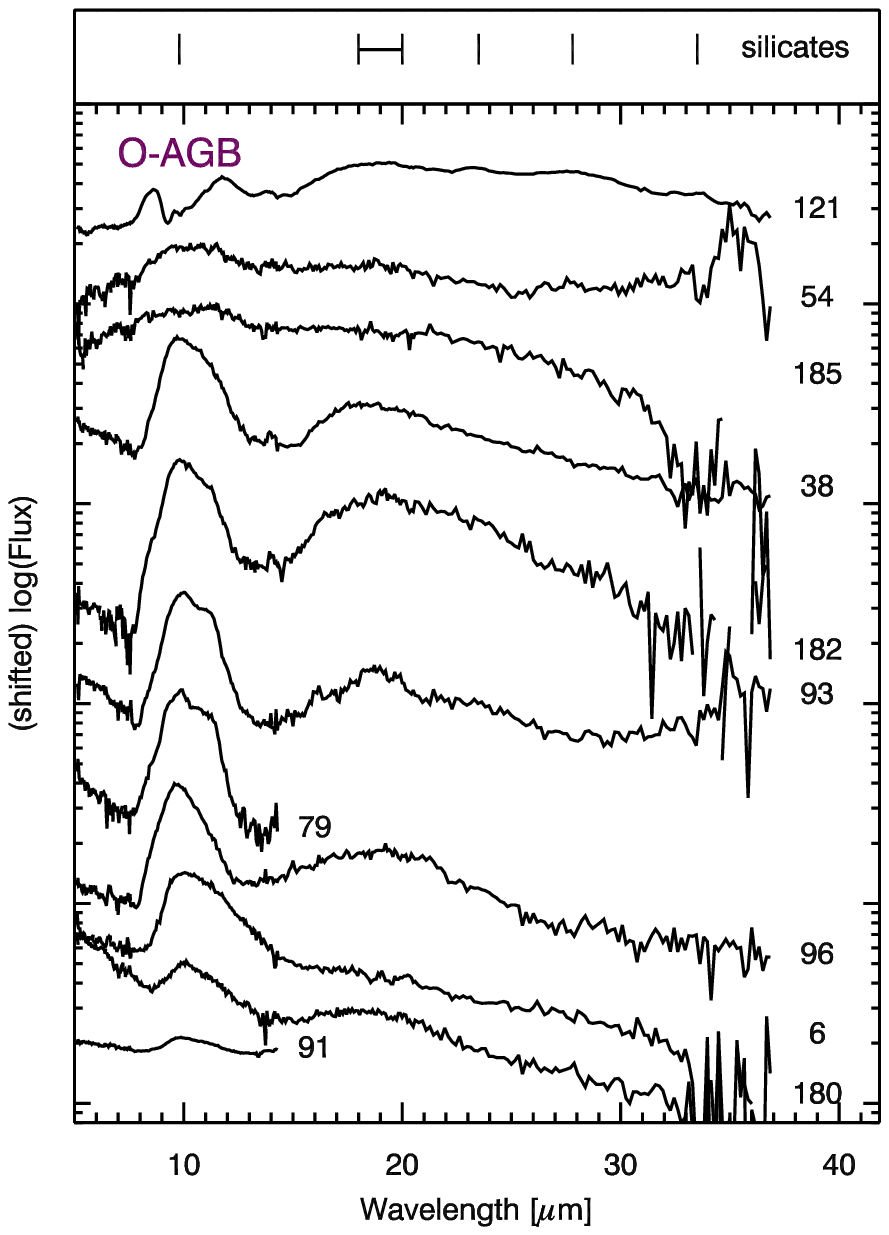}
\caption{Spectra of the heavily dust-enshrouded oxygen-rich AGB
  stars.}
\label{fig:specoagbs}
\end{figure}

Again, a single segregation can be made in this group which results in
two populous subgroups. Several {\tt O-AGB}s (SSID13, 58, 68, 124,
142, 173, 178) appear as stellar photospheres with an inflection at
8\,$\mu$m due to SiO and have little if any IR excess. The remaining
objects in this group exhibit dust emission, most notably 10\,$\mu$m
silicate features. In some cases this is weak (SSID1, 59, 89, 143,
176) but in other cases (see Fig.~\ref{fig:specoagbs}) both 10 and
20\,$\mu$m features are pronounced or self-absorbed. Curiously, SSID6
shows a strong and broad 10\,$\mu$m feature but no discernible sign of
a 20\,$\mu$m feature.

SSID91, 180 and 96 show silicate features which peak short of
10\,$\mu$m; the shape of feature suggests that the silicates are
largely amorphous in form. In these three objects there is relatively
little IR excess. SSID96, along with SSID82, is considered in more
detail in \citet{sar10}, and modelling shows the presence of a
CO$_2$\ gas feature at 15\,$\mu$m. SSID79, 93, 182 and 38 show more
developed 10\,$\mu$m features, with an increasing prominence of the
11\,$\mu$m crystalline silicate feature and the 20\,$\mu$m feature,
indicating an enhanced dust production and evolutionary stage. SSID93
and 182 also show weak 16 and 19\,$\mu$m features, probably due to
forsterite. SSID54 appears to be a highly evolved star, showing
evidence of a molecular sphere with broad line emission, in
particular, H$_2$O at 6.7\,$\mu$m, and a significant amount of dust,
indicated by the flattening of the silicate peaks due to the onset of
silicate self-absorption. Finally, SSID121 ({IRAS05298$-$6957})
is a very highly-evolved star with an initial mass of 4\,M$_\odot$, a
large dust excess, silicate in absorption, a high mass-loss rate, and
1612\,MHz OH maser emission \citep{woo92,tra99,vlo01,vlo10}. Its
spectrum appears very similar to those of very embedded oxygen-rich
dust sources in the LMC shown in Fig.~10 of \citet{slo08}. Narrow
absorption features superimposed on the broad amorphous silicate
absorption would indicate the presence of enstatite in the dust. This
sequence of increasing evolution and dust production for O-AGBs has
been discussed previously in the context of Galactic stars by
\citet{slo95} and \citet{spe00}, among others. O-AGBs with the lowest
mass-loss rates produce Al-bearing dust, whilst those with higher
mass-loss rates produce Mg- and Fe-bearing amorphous silicates
\citep{slo03}, which produce the features that we see here. Also,
O-AGBs with a lesser dust excess exhibit stronger amorphous silicate
features compared to crystalline \citep{syl99}, and as mass-loss rates
increase, crystalline features become more dominant due to the
optically thicker dust shell \citep{kem01}, also seen in our
sample. There is no clear example in these spectra of the narrow
13\,$\mu$m feature often seen in Galactic O-AGBs \citep*{slo96}.

\subsubsection{Red supergiants}

\begin{figure}
%\epsscale{num}
\includegraphics[width=8.4cm]{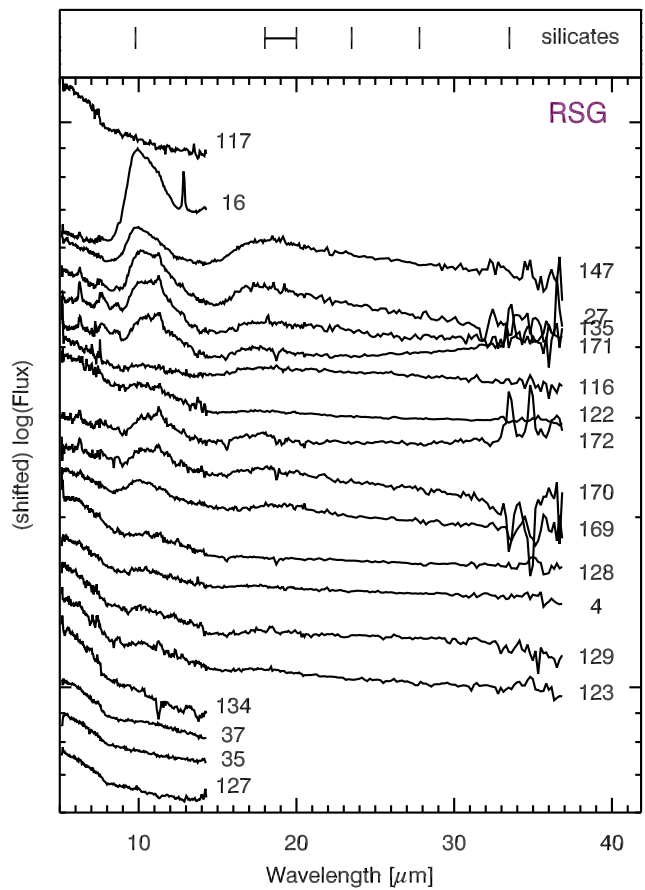}
\caption{Spectra of the RSGs contained in the \SSp\ sample.}
\label{fig:specrsgs}
\end{figure}

The RSG spectra show a range of oxygen-rich dust features
(Fig.~\ref{fig:specrsgs}).  Five of the sources (SSID35, 37, 117, 127,
and 134) appear to be nearly dust-free, although the spectral coverage
stops at 14\,$\mu$m. Others have particularly weak or absent
20\,$\mu$m features, a phenomenon that has not been noted in Galactic
samples. The sources include examples of classic silicate features at
10\,$\mu$m (SSID16, 27, 147), amorphous alumina (SSID4, 128), as well
as both (SSID116, 123); in several cases, the signal-to-noise ratio
was insufficient to pin down the type of dust present, other than its
oxygen-rich nature. At least six objects have PAH emission features,
although these could potentially be interstellar in the case of SSID16
and 171, where \emph{SAGE-LMC} maps show some faint and diffuse
8\,$\mu$m emission. Only one source (SSID16) shows emission from a
fine-structure line ([NeII] at 12.81\,$\mu$m), which is consistent
with the small number of RSGs with ionized lines seen previously in
the LMC \citep{buc06,buc09,vlo10}. The RSGs in those samples tend to
have much stronger 10\,$\mu$m features, probably due to one of their
selection criteria of an MSX 8\,$\mu$m detection \citep{kas08}.

\subsubsection{Oxygen-rich post-AGB stars}

Half of the O-PAGB sample of 18 objects is composed of RV~Tau type
stars (SSID29, 73, 85, 95, 107, 131, 157, 161, 192), identified by
obtaining periods from the literature, which appear almost identical
to the other O-PAGB spectra (SSID28, 56, 75, 111, 113, 115, 118, 162,
177). However, the lightcurves \citep{sos08} of these objects
distinguish the pulsating RV~Tau stars from other oxygen-rich post-AGB
stars. SSID73, 131 and 161 are investigated in detail in
\citet{gie09}. Of the spectra of the nine non-RV~Tau objects, three
show very square, blocky features indicative of a high degree of
crystalline silicates (SSID56, 75, 177), whilst all nine show some
degree of crystallinity. It should be noted that the far-infrared rise
in the spectrum of SSID95 is likely to be due to an encroaching red
source in the LL slit, and that the spectrum of SSID28 is likely
contaminated by emission from nearby {MSX LMC 1271}. SSID161 is an
unusual object in that it does not show 10\,$\mu$m silicate
emission. It does show PAH emission, which arises along the
line-of-sight, and this may mask a weak 10\,$\mu$m feature. It has
been classified as an RV~Tau by \citet{sos09a} and by us. In this case
LL was unfortunately not observed.

\subsubsection{Carbon-rich post-AGB stars}

\begin{figure}
%\epsscale{num}
\includegraphics[width=8.4cm]{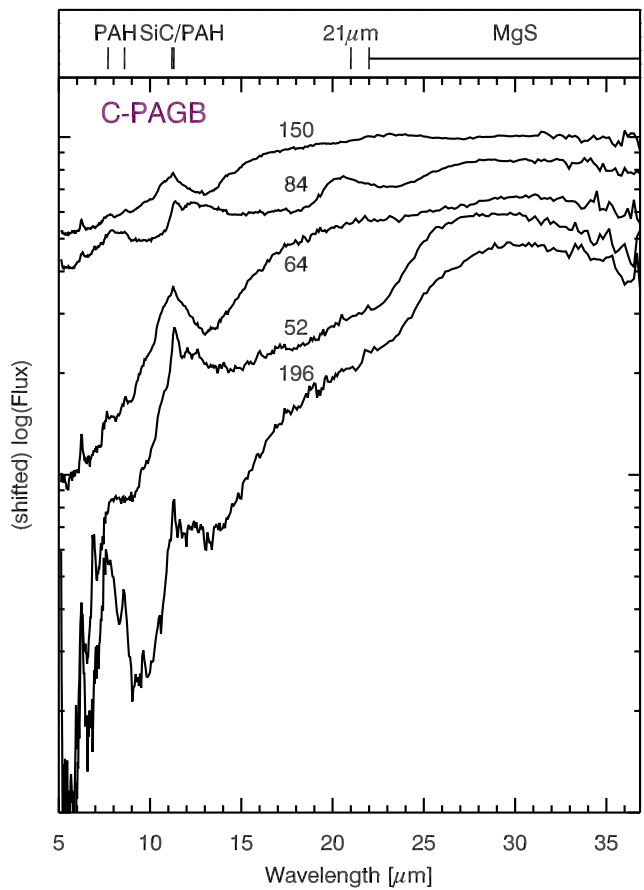}
\caption{Spectra of two C-PAGB objects with very triangular SiC
  features, as well as SSID84, which has a pronounced 21\,$\mu$m
  feature.}
\label{fig:speccpagb}
\end{figure}

This small group of five objects proves to be very interesting
(Fig.~\ref{fig:speccpagb}). All five spectra show a prominent MgS
feature at 30\,$\mu$m, whilst SSID84 additionally shows a 21\,$\mu$m
feature, only found in post-AGB carbon-rich objects
\citep{hri09}. SSID64 and 150 appear very similar to each other, with
a very sharp SiC feature at 11.3\,$\mu$m that appears triangular and
could be somewhat self-absorbed. These triangular features are also
seen in PNe \citep[e.g.,][]{ber09}, where they are attributed to a
superposed PAH band. The 11.2\,$\mu$m PAH band is dominated by neutral
PAHs, whilst the other mid-IR PAH bands arise from ionized PAHs.

\subsubsection{Oxygen-rich planetary nebulae}

The four oxygen-rich PNe (SSID100, 153, 174 and 186) show either very
weak or absent PAH emission (Fig.~\ref{fig:specpne}). Weak PAH
emission in these cases was determined to come from diffuse emission
regions along the line-of-sight, through analysis of the
\emph{SAGE-LMC} 8\,$\mu$m images. Two of the sample (SSID174 and 186)
show strong [NeV] lines at 14.3 and 24.5\,$\mu$m and [OIV] lines at
25.9\,$\mu$m. The spectrum of SSID174 contains a high-excitation
[NeVI] line. SSID153 shows only a weak [ArII] line since Long-Low data
are not available.

\subsubsection{Carbon-rich planetary nebulae}

\begin{figure}
%\epsscale{num}
\includegraphics[width=8.4cm]{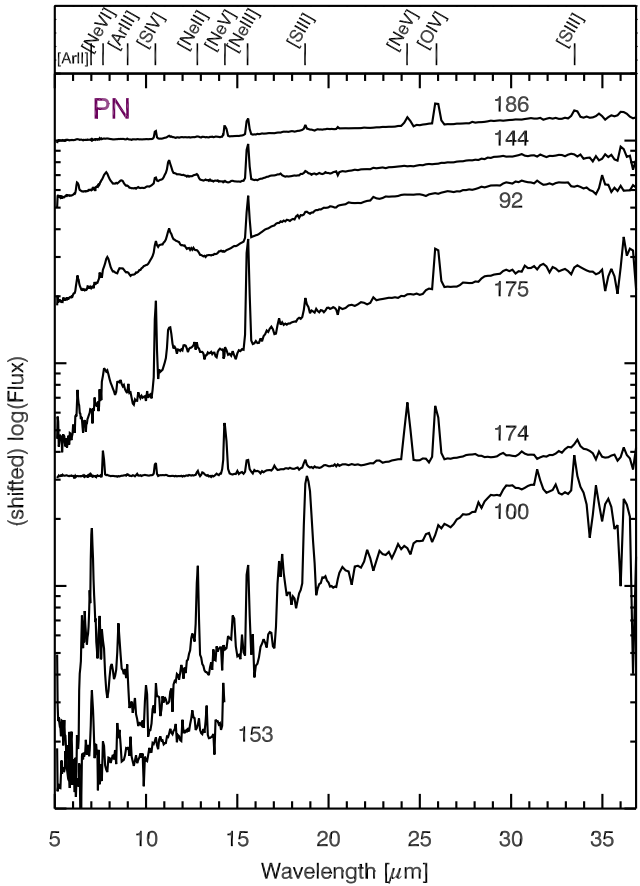}
\caption{Spectra of the PNe contained in the \SSp\ sample.}
\label{fig:specpne}
\end{figure}

The three carbon-rich PNe in the \SSp\ sample (SSID92, 144 and 175)
show reasonably strong PAH emission, along with [NeIII], [SIII] and
[SIV] lines.  SSID175 also shows a strong [OIV] line
(Fig.~\ref{fig:specpne}).

\subsubsection{H{\sc ii} regions}

Only two objects are classified as H{\sc ii} regions, SSID71 and 104,
although due to spatial resolution issues (\S\ref{hiiregs}) some H{\sc
  ii} regions may be confused with evolved YSOs ({\tt YSO-3}s). They
both show a [SIII] line and a continuum which rises with
wavelength. The weak PAH features appear to be due to nebulous
material along the line of sight to the objects, apparent in
\emph{SAGE-LMC} 8\,um images of the regions concerned.

\subsubsection{Galaxies}

\begin{figure}
%\epsscale{num}
\includegraphics[width=8.4cm]{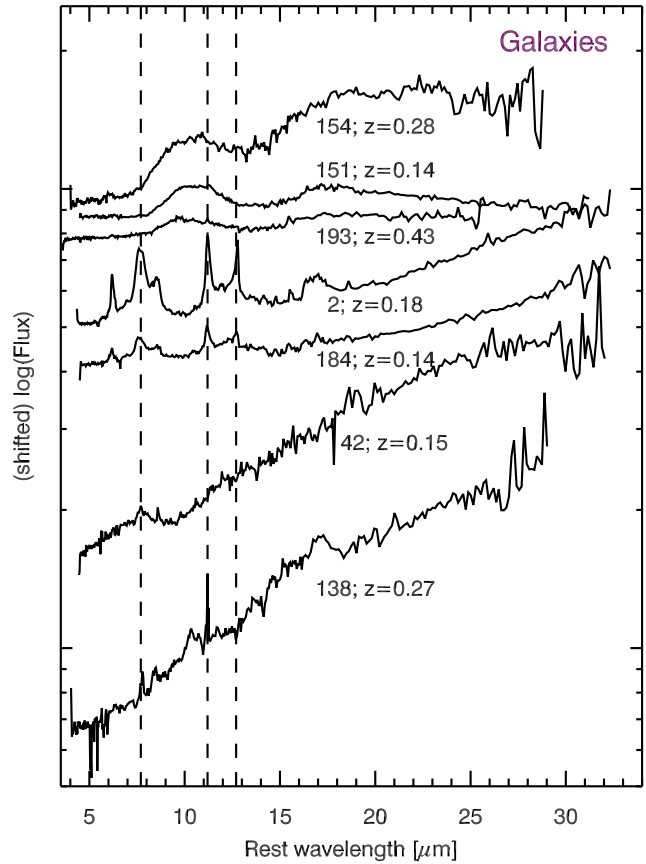}
\caption{Spectra of the galaxies contained in the
  \SSp\ sample. Spectroscopically-determined redshifts are indicated
  on the figure. Dashed vertical lines indicate the rest-frame
  positions of narrow PAH features, used to align some of the
  spectra. Redshift errors for the upper three spectra are on the
  order of $\pm$0.05; the lower four spectra have narrow features,
  reducing the error in redshift determination to $\pm$0.01.}
\label{fig:specgals}
\end{figure}

All seven galaxy spectra are shown in rest-frame in
Fig.~\ref{fig:specgals}. Three galaxies show strong silicate features
-- SSID151, 154 and 193, which are a hallmark of type 1 active
galactic nuclei \citep[AGN; cf.,][]{hao05,sie05,shi06}. Two galaxies
show strong PAH features (SSID2, 184) and SSID2 clearly shows the
[NeV] line at 14.32\,$\mu$m, which can be indicative of an accreting
black hole, since star-forming galaxies rarely show lines with such
high ionization potentials. Four of the galaxies (the lower four in
Fig.~\ref{fig:specgals}) exhibit rising SEDs towards longer
wavelengths, again suggesting that these are AGN, similar to those
described by \citet{buc06}. It should be mentioned that \citet{vlo10}
speculate that SSID97 is also a galaxy with z=0.27 or 0.54, based on
the presence of a potential oxygen line in a MIPS-SED
spectrum. However, the IRS spectrum shows a number of PAH features at
the correct rest wavelengths, which leads us to classify SSID97 as
{\tt YSO}.

\subsubsection{Others}

\begin{figure}
%\epsscale{num}
\includegraphics[width=8.4cm]{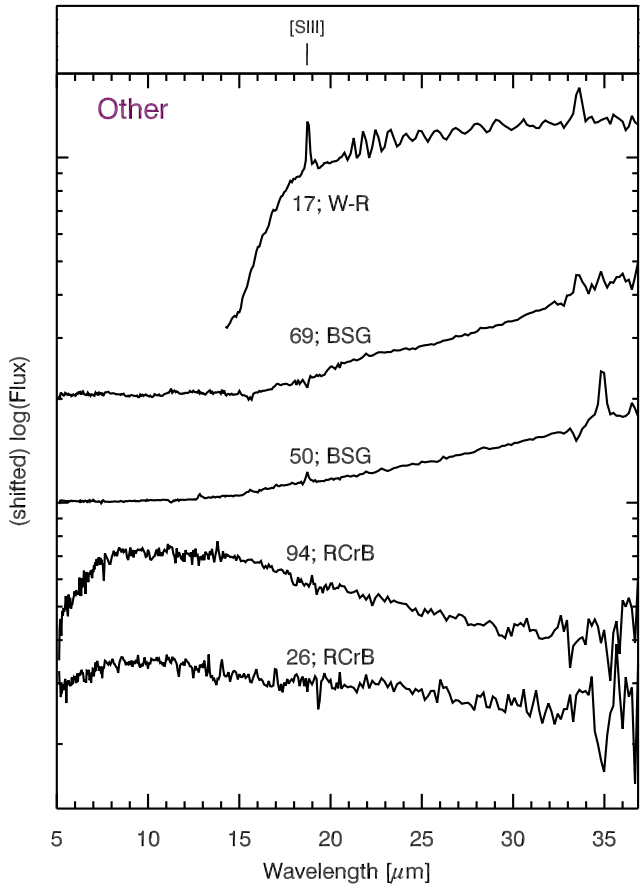}
\caption{Spectra of the objects in the \textquotedblleft
  Other\textquotedblright\ group, which contains a Wolf-Rayet star,
  two B supergiants and two candidate RCrB stars.}
\label{fig:specothers}
\end{figure}

One potential and one subsequently-confirmed \citep{sos09a} RCrB stars
are found in the sample, SSID26 and 94. Both show red, featureless
spectra (Fig.~\ref{fig:specothers}), and SSID26 should be further
investigated through optical spectra or monitoring of its light curves
to confirm RCrB status. The two B supergiants in the sample, SSID50
and 69, show SEDs that plunge to a minimum at $\sim$8--9\,$\mu$m. They
also show spectra that rise steeply long-ward of 15\,$\mu$m. SSID17 is
a known Wolf-Rayet star, and exhibits an SED which rises steeply at
$\sim$15\,$\mu$m, levels off, and then decreases. There is an emission
line due to [SIII] at 18.7\,$\mu$m, however this may be sky emission;
features in the LL2-wavelength portion of the spectrum are affected by
data issues. All these spectra are shown in Fig.~\ref{fig:specothers}.

\subsubsection{Unidentified objects}

The observation of SSID24 was pointed to within 0\farcs33 of planetary
nebula {RP 1805}, which has a diameter in the H$\alpha$\ image of
5\arcsec\ \citep{rei06}, however no object is seen in \emph{SAGE-LMC}
8\,$\mu$m images and no convincing features are seen in the IRS
spectrum. We report this as a non-detection. Similarly SSID160 is not
visible in \emph{SAGE-LMC} 8\,$\mu$m images and no features are seen
above the noise in the IRS spectrum.

\begin{figure}
%\epsscale{num}
\includegraphics[width=8.4cm]{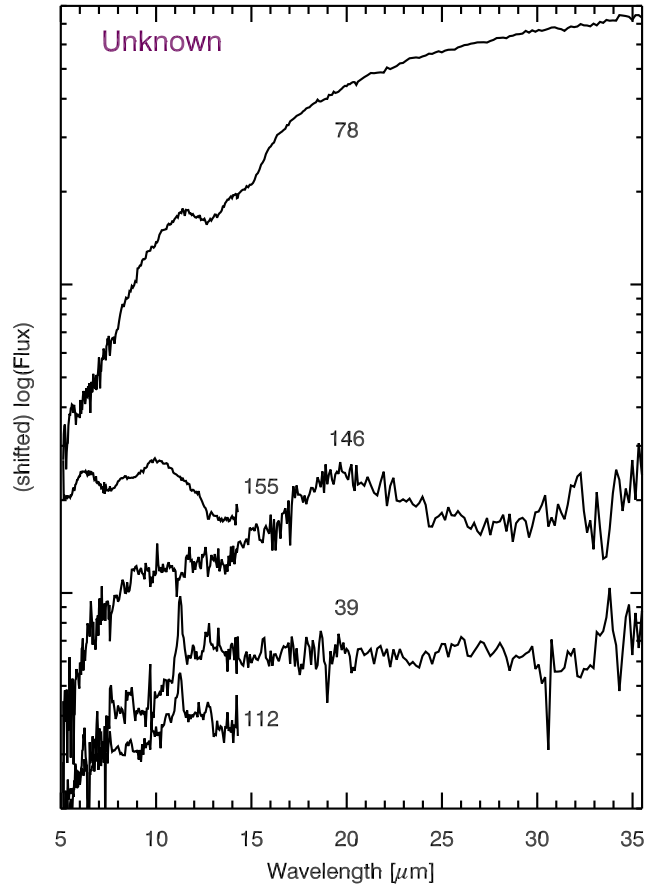}
\caption{Spectra of unidentified objects.}
\label{fig:specunk}
\end{figure}

SSID78 is an unusual object in that it presents a rising spectrum
toward longer wavelengths which is featureless apart from a dip in the
continuum from 11.5--16.0\,$\mu$m
(Fig.~\ref{fig:specunk}). \citet*{deg87} consider it to be a candidate
AGN, and when plotted on Fig.\ref{fig:cmdgals} it would certainly lie
within the \textquotedblleft AGN wedge\textquotedblright. SSID146 is
also an unusual object, in that it shows very broad 20\,$\mu$m
emission and very broad but weak 10\,$\mu$m emission. \citet*{sas00}
identifies it as an X-ray source, {SHP LMC 256}, of unknown
physical nature. Its SED is double-peaked, which could be indicative
of a post-AGB object, however, it is rather blue. There may be
interesting mineralogy, containing iron or magnesium oxides.

SSID39 and SSID112 both present very weak spectra, with no
recognisable features apart from a 11.2--12.7\,$\mu$m PAH complex,
which may be due to foreground emission. Both observations were
pointed towards PNe, {RP 1878} and {RP 589} respectively, but no clear
emission lines were detected.

SSID155 is optically identified as a carbon star \citep{kon01}, but
shows unusual features, potentially including an absorption at
7.5\,$\mu$m and either emission at 10\,$\mu$m or absorption at
9\,$\mu$m. This could indicate a mixed-chemistry object.

%SSID88 appears to be an AGB star of some kind, but its chemistry is
%obscured by strong PAH features which arise along the line of sight to
%the star. \citet{sri09} tentatively classify this object as O-rich,
%and in colour-magnitude diagrams SSID88 lies at the base of the AGB,
%making a determination of its chemistry difficult.

\subsubsection{Comparision of results with source selection}

At this point it is useful to make a comparison between our original
selection of candidates \citep[][and recapped here]{kem10} and our
final classification to test the validity of our source selection
criteria. The best-selected category were the post-AGB objects (86\%),
with one C-AGB and one O-AGB creeping into the sample. These were
selected from a list of candidates \citep{woo01} and the MACHO catalog
\citep{alc98}. Also well-selected were non-clustered C-AGBs (78\%) and
O-AGBs (72\%), which were taken from the sample of \citet{sri09}. Five
{\tt STAR}s crept into these two categories, which may in fact be
relatively dust-free AGB stars. Categories which yielded poor
selections were clustered AGB stars, which only successfully produced
17 AGB stars from 34 (old, intermediate and young AGB stars were
chosen from three metallicity bins) but also 12 RSGs; planetary
nebulae \citep[5 out of 12, picked from ][]{rei06,lei97}, three of
which we classed as {\tt UNK}; and YSO candidates, which only resulted
in 26 YSOs from a sample of 88 \citep[selected from][based on
  detectability at $\lambda>$14\,$\mu$m]{whi08}. The biggest
contaminants in the YSO category were 20 C-AGB, 16 O-AGB stars and
seven RSGs. This shows the difficulty in selecting good YSO
candidates. Somewhat surprisingly, many of the failed YSO candidates
were not galaxies, as often expected, but evolved and post-AGB stars.
The 13 Colour--Magnitude Diagram (CMD) \textquotedblleft
fillers\textquotedblright\ produced 4 C-AGBs, 3 O-AGBs, 3 galaxies, 2
YSOs and one unknown object, possibly an H{\sc ii} region.

\section{Stellar populations in the Large Magellanic Cloud}
\label{sec:stellarpop}

\begin{figure*}
%\epsscale{num}
\includegraphics[width=16.27cm]{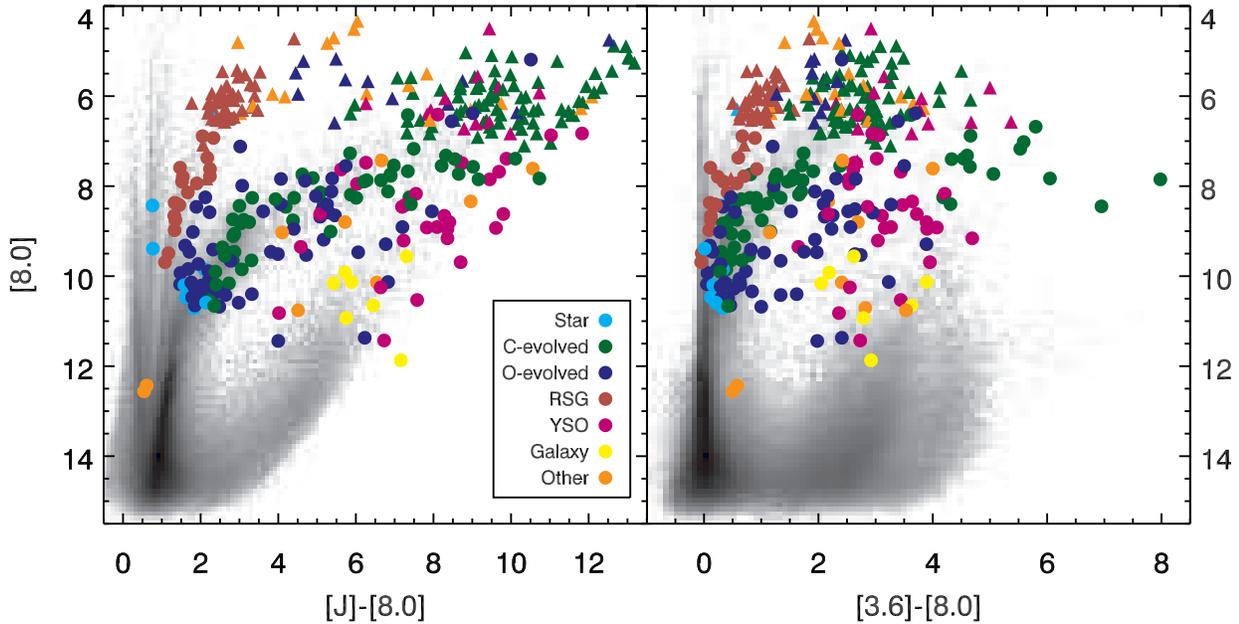}
\caption{Two plots showing the distribution of all sources in
  colour--magnitude space. The \SSp\ sample is shown as filled circles,
  whilst the brighter \citet{kas08} sample \citep[incorporating the
    updates of][]{buc09} are filled triangles. Here we merge the {\tt
    O-AGB}, {\tt O-PAGB}, {\tt O-PN}, {\tt C-AGB}, {\tt C-PAGB}, and
  {\tt C-PN} classifications into broad groups.}
\label{fig:cmdbasic}
\end{figure*}

\begin{figure*}
%\epsscale{num}
\includegraphics[width=16.27cm]{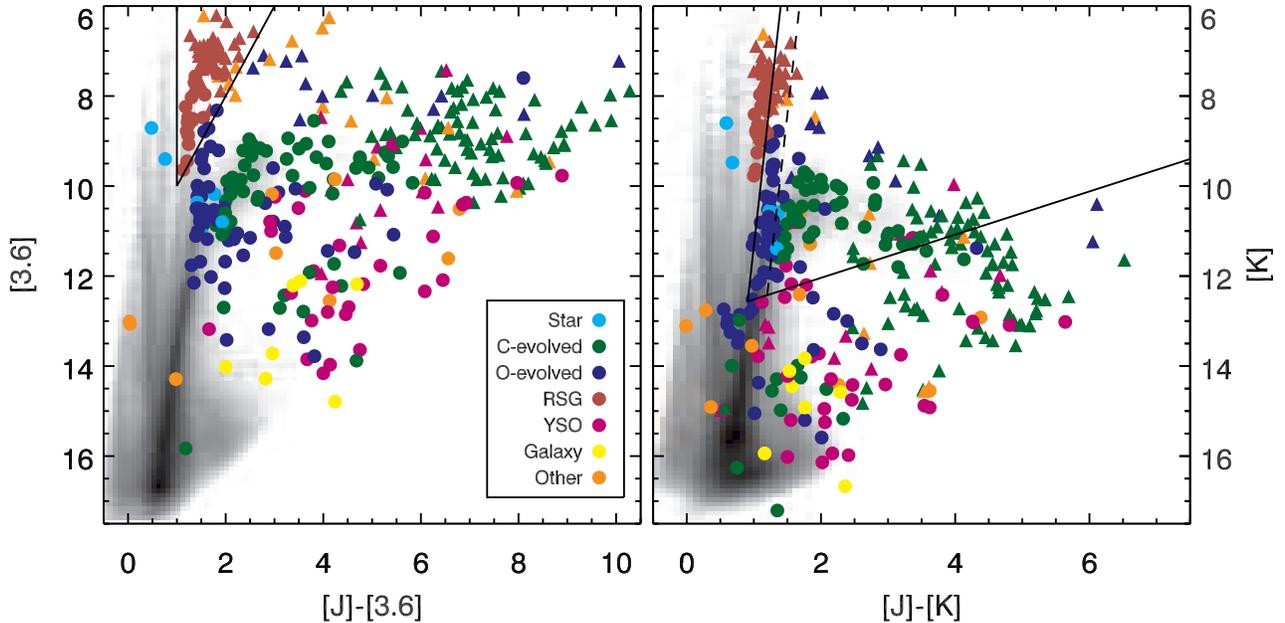}
\caption{Two further CMDs, with cuts for selecting RSGs and
  O-AGBs. Symbols are as for Fig.~\ref{fig:cmdbasic}. {\tt RSG}s can
  clearly be distinguished from other oxygen-rich evolved stars in the
  [3.6] vs. [J]$-$[3.6] diagram (left panel). In the right panel we
  use the cuts used by \citet{cio06} to select AGB stars (solid lines)
  and to distinguish O-AGB from C-AGB (dashed line).}
\label{fig:blumf34}
\end{figure*}

\begin{figure*}
%\epsscale{num}
\includegraphics[width=16.27cm]{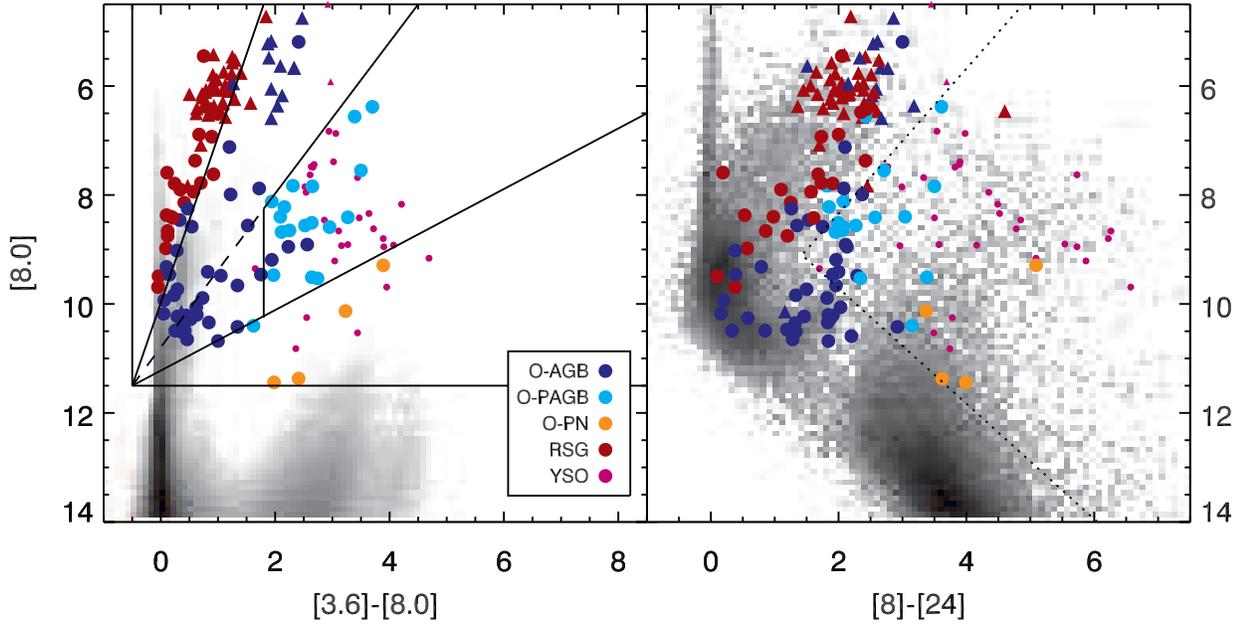}
\caption{A colour--magnitude diagram showing the classes of O-rich
  evolved objects. {\tt YSO}s are shown as small magenta points, since
  they overlap in colour--magnitude space with the more evolved O-rich
  objects. Other symbols are as for Fig.~\ref{fig:cmdbasic}. The
  dotted line in the right panel is a cut used by \citet{whi08} to
  select YSOs. The background shows a Hess diagram of the
  \emph{SAGE-LMC} sample.}
\label{fig:cmdorich}
\end{figure*}

\begin{figure*}
%\epsscale{num}
\includegraphics[width=16.27cm]{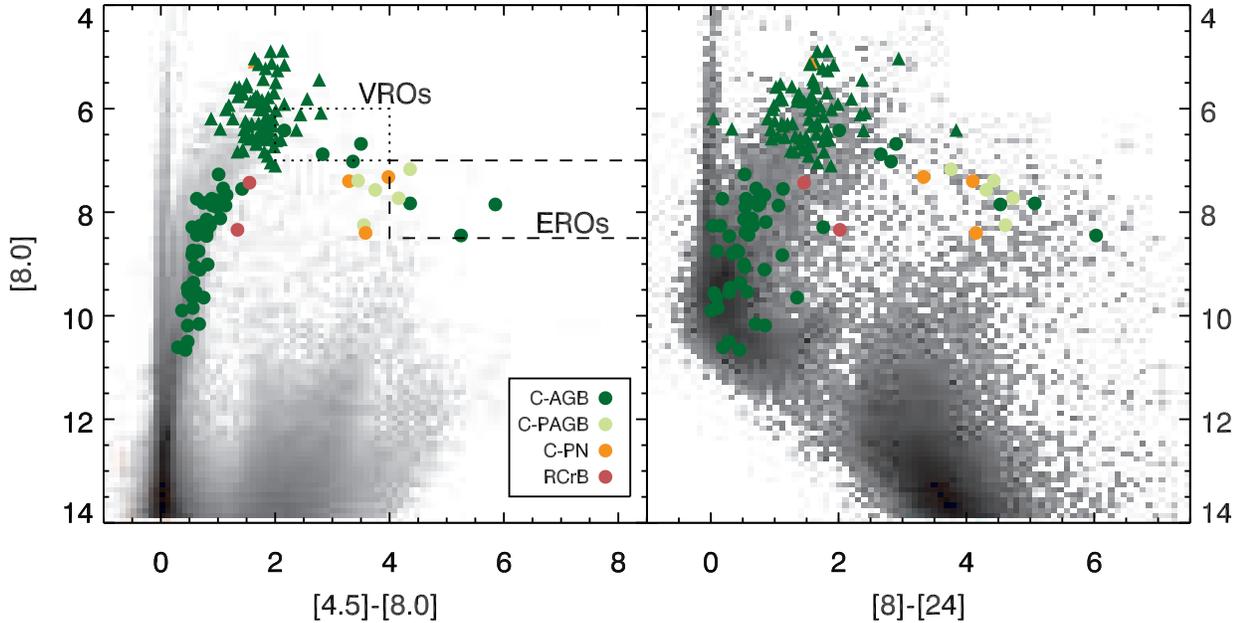}
\caption{A colour--magnitude diagram showing the classes of C-rich
  evolved objects and RCrB stars. Symbols are as for
  Fig.~\ref{fig:cmdbasic}. Boxes show the selection of VROs and EROs
  in the left-hand panel. The background shows a Hess diagram of the
  \emph{SAGE-LMC} sample.}
\label{fig:cmdcrich}
\end{figure*}

\begin{figure*}
%\epsscale{num}
\includegraphics[width=16.27cm]{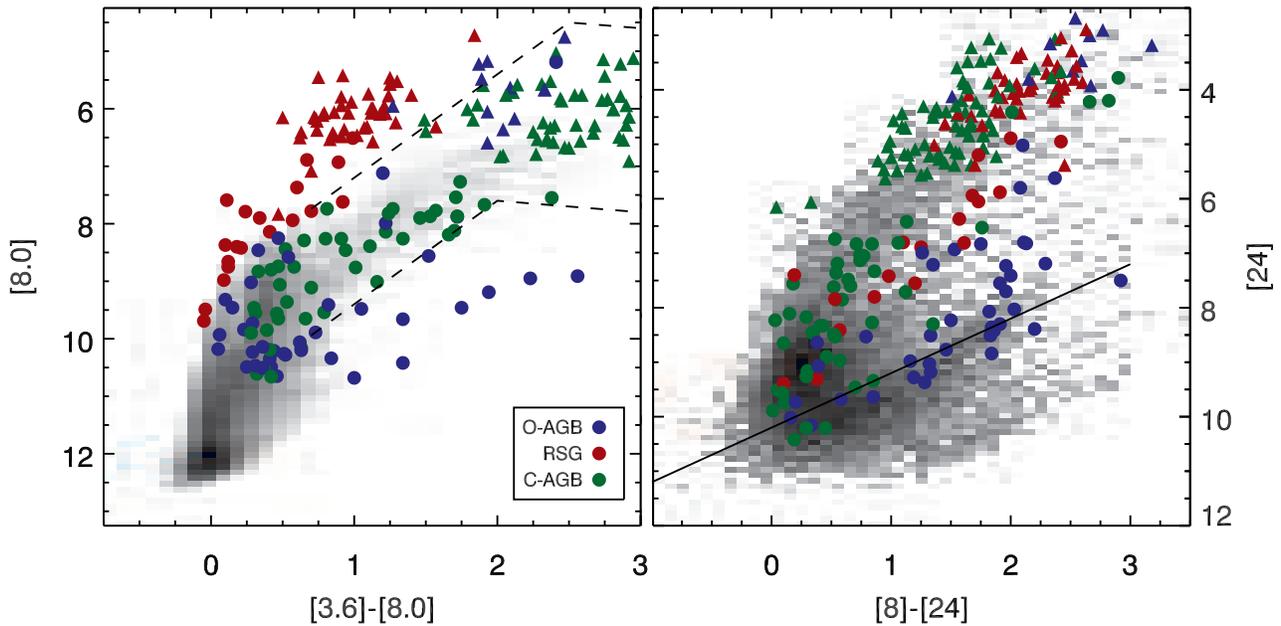}
\caption{A colour--magnitude diagram showing the distribution of the
  {\tt O-AGB}, {\tt RSG} and {\tt C-AGB} classes plotted over the
  evolved star sample of \citet{sri09}. Symbols are as for
  Fig.~\ref{fig:cmdbasic}. The dashed cuts separate C-AGB and O-AGB
  stars according to \citet{mat09}. In the right panel the cut
  differentiates the faint O-AGB stars \citep[below the
    line;][]{blu06} and the bright O-AGB stars.}
\label{fig:cmdagbs}
\end{figure*}

\begin{figure*}
%\epsscale{num}
%\plotone{colmagysogal.eps}
\includegraphics[width=16.27cm]{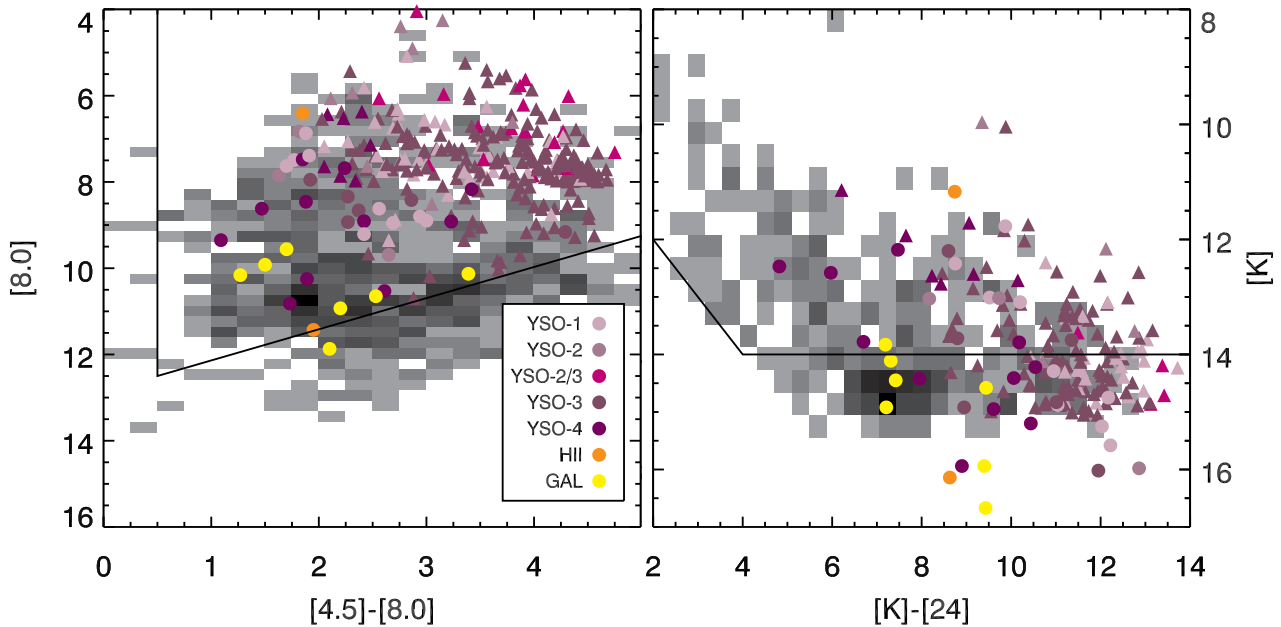}
\caption{A colour--magnitude diagram showing the distribution of the
  {\tt YSO}, {\tt HII} and {\tt GAL} classes. In this case, the filled
  triangles show the YSO sample of \citet{sea09}, and the YSO sample
  of \citet{whi08} is plotted as a Hess diagram.}
\label{fig:cmdysogal}
\end{figure*}

\begin{figure}
%\epsscale{num}
\includegraphics[width=8.4cm,clip]{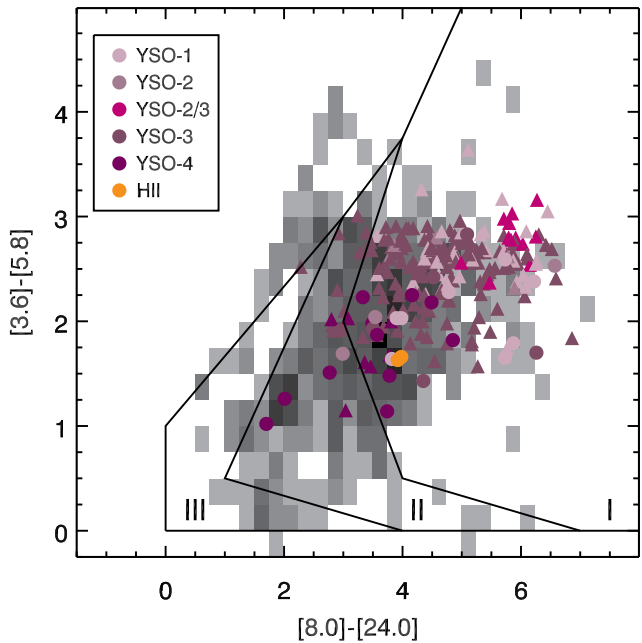}
\caption{A colour--colour diagram showing the finer grades of
  classification of YSOs (circles), in comparison to the sample of
  \citet{sea09} (triangles) and the modelling of \citet{rob06} (solid
  lines). The areas marked by Roman numerals refer to the different
  YSO Stages which dominate those regions, as described in the latter work.}
\label{fig:cmdysos}
\end{figure}

A spectral survey such as \SSp\ can in some way act as a check on the
many different classifications based on photometry that have come out
of the \emph{SAGE} collaboration
\citep[e.g.,][]{blu06,whi08,sri09,bon09} and other groups who have
studied the LMC \citep*{ega01,kas08,buc09,mat09}. Whereas photometric
surveys can be performed more quickly (in terms of telescope-hours)
and spectral surveys in general take more time, classification by
photometry can be fraught with inaccuracy, often due to overlapping or
co-located classes in colour--magnitude diagrams (e.g., YSOs and
galaxies). In Figs.~\ref{fig:cmdbasic}--\ref{fig:cmdysos} we present
our versions of the colour--magnitude and colour--colour diagrams
presented in the various studies of the LMC performed recently, and
discuss the accuracy of their photometric classifications.

The two panels in Fig.~\ref{fig:cmdbasic} show [8.0] versus
[J]$-$[8.0] and [3.6]$-$[8.0], respectively. The objects from the
\SSp\ sample are over-plotted on a Hess
diagram\footnotemark[3]\footnotetext[3]{Hess diagrams are 2D
  histograms where the number density is represented by the brightness
  of each pixel. Darker pixels are more dense.} of the entire
\emph{SAGE-LMC} sample with the relevant 2MASS/IRSF and IRAC
magnitudes. In Figs.~\ref{fig:cmdbasic}--\ref{fig:cmdagbs} we also
include classifications from the samples of \citet{buc06},
\citet{kas08} and \citet{buc09}, who targeted bright sources with
8\,$\mu$m MSX detections. These objects were observed in Spitzer
Cycles 1--3, and were not part of the \SSp\ observational sample. Note
the large population of objects in the \emph{SAGE-LMC} sample with
[8.0]$\sog$11 that are not probed by the \SSp\ sample; these are
mainly background galaxies and YSOs not selected by \SSp\ due to
signal-to-noise constraints. Two objects which stand out in this plot
are the two B supergiants, at [8.0]$\approx$12.5\,mag. Also we see
that of our sample of stellar photospheres the two K giants are
distinct from the other M-stars, which cluster at the base of the
AGB. The right panel shows reasonably good separation between
different groups of objects in the \SSp\ sample, particularly between
C-AGBs ([3.6]$-$[8.0]$\sol$3) and more evolved carbon-rich and extreme
objects. This gap is filled by the brighter Kastner sample, and the
conjunction of the two shows the full extent of C-rich (post-)AGB
evolution. The distinction between O-AGBs and RSGs can also be seen in
this figure, but is clearer in the left panel of
Fig.~\ref{fig:blumf34}, which is a representation of Fig.~3 from
\citet{blu06}. In this arrangement RSGs lie in a sequence slightly
bluer (in J-[3.6]) than the O-AGB stars, and so the distinction is not
entirely luminosity-based. This colour criterion, then, gives us a
very useful tool in making the difficult distinction between RSGs and
O-AGBs. In Fig.~\ref{fig:blumf34} we can make a cut along [J]=1 and
[3.6]=12$-$2([J]$-$[3.6]) to select the RSGs from the \SSp\ sample and
the \citet{kas08} sample. This cut also corroborates with the
optically-selected RSG sample of \citet{bon09}. This selection is
extremely clean -- only four non-RSGs are contained within: two
potential WR stars, one bright B[e] star \citep{buc09} that was
formerly classified as a C/O-AGB \citep{kas08} and one unknown object
\citep{kas08}. No contaminants from the \SSp\ sample are contained in
this cut.. The right-hand panel of Fig.~\ref{fig:blumf34} shows the
colour-cut used by \citet{cio06} to select candidate AGB stars from
those detected in the DENIS survey of the LMC \citep{cio00a}. The cut
is successful over the range it was intended (9.5$<$[K]$<$12.6,
0.9$<$[J]$-$[K]$<$2.0), although it may include some red giant branch
(RGB) stars since the tip of the RGB is at [K]=12.3 \citep{nik00}.

Also in Figs.~\ref{fig:cmdbasic} and \ref{fig:blumf34} we see the
large overlap of YSOs with galaxies and extreme and evolved carbon
stars. This leads to a great degeneracy in photometric classifications
in the region of colour space covered by YSOs, and this is evident
particularly in Appendix~\ref{litsurv} with the overlapping and
conflicting classifications of \citet{whi08} (YSOs) and \citet{sri09}
(AGBs), among others, and also in the discrepancy between our original
source selection and final classification discussed earlier. It is
clear that although our selection was good in that it selected YSOs,
it also does not discriminate enough to be uncontaminated.

The next two figures focus in more detail on the O-rich and C-rich
evolved objects. Figure~\ref{fig:cmdorich} shows a clockwise
progression in colour space from
RSG$\rightarrow$O-AGB$\rightarrow$O-PAGB$\rightarrow$O-PN. These
O-rich objects become redder and fainter in [8.0] with age. SSID186,
the orange point at [8]$\sim$9.3\,mag, may be misclassified since it
falls within the colour-space of carbon-rich PNe. Interestingly, the
YSOs in the \SSp\ sample are redder than [3.6]$-$[8.0]$>$1.5, and thus
only the O-PAGB and O-PN groups are contaminated. The dotted lines in
the right-hand panel show the selection criterion for YSOs used by
\citet{whi08} in this CMD. The selection of YSOs is very good,
although one must be wary of contamination of YSO samples with O-rich
post-AGB objects. Figure~\ref{fig:cmdcrich} shows similar colour
spaces but for carbon-rich evolved objects. This diagram enables us to
separate the VROs and EROs from the remainder of the carbon-rich AGB
stars. These seven extreme stars fall red-ward of 2$<$[4.5]$-$[8.0].

Figure~\ref{fig:cmdagbs} shows the AGB and RSG stars in [8.0] vs.
[3.6]$-$[8.0] colour--magnitude space (left panel). Such a diagram has
been used by various authors \citep{blu06,sri09,mat09} to investigate
the separation of oxygen-rich and carbon-rich AGB stars. Plotted in
the left panel are cuts made by \citet{mat09} to separate carbon-rich
stars out from oxygen-rich (dotted lines). In general, these cuts work
well for our sample, with only seven oxygen-rich stars contaminating
the carbon-rich sample for [3.6]$-$[8.0]$>$0.75. Thus contamination is
small, but by rare, extremely interesting objects that would be missed
by conservative cuts.  Comparison with Fig.~\ref{fig:cmdbasic} shows
that this region of the CMD may also contain YSOs, and O-PAGB objects.
The right-hand panel of Fig.~\ref{fig:cmdagbs} shows that the fainter
evolved stars are missing from the \SSp\ and Kastner et
al. samples. Coverage of this region of colour-space should improve
with the addition of data from the IRS archive (Paper II).

To show the distribution of YSOs in colour--magnitude and
colour--colour spaces we have constructed Fig.~\ref{fig:cmdysogal} and
\ref{fig:cmdysos}. In Fig.~\ref{fig:cmdysogal} we show the \SSp\ and
\citet{sea09} YSO samples plotted on top of the \citet{whi08}
sample. We have classified the \citet{sea09} YSOs into our groupings,
although in some cases it was hard to distinguish between YSO-2 and
YSO-3. Surprisingly, there is no clear distinction between groups of
YSOs, and no evident gradient is seen either, although on average {\tt
  YSO-4}s are bluer and fainter than the other classes. It is not
clear why this is the case. A similar result is found in
Fig.~\ref{fig:cmdysos}, where almost all of the sampled YSOs are
classed as Stage I on the colour--colour diagram of \citet{rob06}. The
solid lines in Fig.~\ref{fig:cmdysogal} are cuts used by \citet{kir09}
to distinguish between YSOs and background galaxies. These cuts are
not particularly successful for the LMC, with only one of the
\SSp\ galaxies being isolated from the YSOs in the left-hand panel,
and the YSO sample being bisected by the cut in the right-hand panel.

\begin{figure}
%\epsscale{num}
\includegraphics[width=8.4cm]{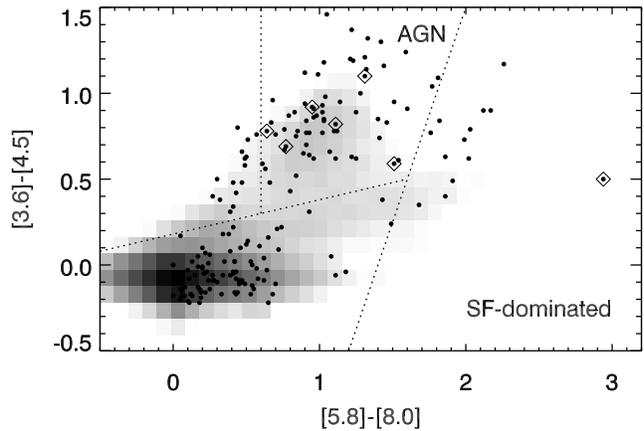}
\caption{Further classification of galaxies (points surrounded by
  diamonds), using a colour--colour diagram from \citet{ste05} and
  \citet{gor08}. Six of the seven background galaxies in the
  \SSp\ sample are found in the AGN wedge. The seventh, SSID2, is
  found in the region dominated by star formation. Overplotted on a
  Hess diagram of the \emph{SAGE-LMC} catalog, with other
  \SSp\ objects plotted as black points.}
\label{fig:cmdgals}
\end{figure}

In Fig.~\ref{fig:cmdgals}, we plot the seven galaxies in the
\SSp\ sample in colour--colour space. \citet{ste05} were able to show
that an empirically-derived wedge-shaped region in this colour space
was dominated by luminous AGN. Six of the seven \SSp\ galaxies reside
in this region, while the seventh, SSID2, lies in the region dominated
by star-forming galaxies \citep{gor08}. This is supported by the
strong PAH features in the spectrum of SSID2. However, SSID2 seems
also to show weak [NeV] emission, indicating that it may have both an
AGN and a starburst.

%Discuss missing objects - MSX detection limit, faint O-AGBs, low mass YSOs

%Revise figures

%Discuss young stars, c.f., Whitney

%Discuss evolved stars, c.f., Blum \& Srinivasan

%Discuss PN, c.f., Hora

%Discuss massive stars, c.f., Bonanos

%Plot up Buchanan's RSG/OAGB/CAGB boxes.

\section{Application to other surveys}

The classification system shown in Fig.~\ref{fig:psctree} is not
limited in its application to \emph{Spitzer} data; it can also be
applied to JWST, ISO and even IRAS LRS data, allowing for the lower
resolution and signal-to-noise ratio of that instrument. One must be
careful to allow for differing resolutions, since, for example, at a
resolution of 5\arcsec\ a compact H{\sc ii} region ($\approx$1\,pc)
may look like a YSO. The scheme can be modified slightly to
additionally make use of data from \emph{Herschel} and the AKARI
satellite, which will be especially useful in discriminating between
classical H{\sc ii} regions and planetary nebulae using the
long-wavelength shape of the SED. Similarly, AKARI Infrared Camera
data at shorter wavelengths (2.5--5\,$\mu$m) can be used to provide
additional confirmation of YSO status when 3.05\,$\mu$m H$_2$O and
4.27\,$\mu$m CO$_2$\ ice are detected \citep[e.g.,][]{shi08}, and also
encompasses gaseous CO absorption near 5\,$\mu$m and C$_2$H$_2$/HCN
bands near 3\,$\mu$m, which are useful in identifying C-rich AGB
stars. The CMDs presented here and Fig.\ref{fig:cmdysos} will be of
use when classifying point sources from the Wide-Field Infrared Survey
Explorer (WISE) all-sky survey at 3.3, 4.7, 12 and 23\,$\mu$m.

The decision tree can also be applied to other external galaxies, or
Galactic point sources. The only overt distance factor in the scheme
is the cut in bolometric luminosity between O-AGBs and RSGs. This
differentiation is difficult to make, and in systems where distances
are highly uncertain (i.e., the Galaxy) the luminosity-based selection
criterion could be replaced by the [J]=1 and [3.6]=12$-$2([J]$-$[3.6])
CMD cut discussed in \S\ref{sec:stellarpop}, for instance.

\section{Summary}

We have classified 197 objects observed as part of the
\SSp\ \emph{Spitzer} Legacy Program according to their object type
using a decision-tree method which discriminates according to spectral
features and other ancillary data. Classification using spectra is
more robust than that using photometric colours and can be used to
resolve degeneracies in colour--magnitude space, e.g., between YSOs and
galaxies. Our classification is in agreement with, and acts as an
extrapolation of, brighter (at 8\,$\mu$m) sources in the LMC
\citep{kas08,buc09}, and the combination of both has highlighted the
extent of carbon-rich (post-)AGB evolution, for example, in
colour--magnitude space. Several colour cuts have been established or
confirmed to distinguish between source classes. The \SSp\ sample will
be expanded to include all observations of point sources in the LMC in
the \emph{Spitzer} archive, which number approximately 750
\citep[including those of][]{kas08,buc09}. Such a large sample will
allow us to:
\begin{description}
\item select samples of similar objects for future studies,
\item obtain a global picture of the mass budget of the LMC through
different stellar populations, and their respective contributions to
the ISM,
\item improve existing colour classifications, and adapt results
for use in other galaxies, e.g., the Small Magellanic Cloud, and
\item use our point-source classifications (a biased sample) as seeds
  to statistically classify the \emph{SAGE-LMC} sample (an unbiased
  sample) of $\sim$6.5 million point sources (Marengo et al., in
  prep.).
\end{description}

\section*{Acknowledgments}

R.~Sz. acknowledges support from grant N203 511838 (MNiSW). This paper
utilizes public domain data obtained by the MACHO Project, jointly
funded by the US Department of Energy through the University of
California, Lawrence Livermore National Laboratory under contract
No.~W-7405-Eng-48, by the National Science Foundation through the
Center for Particle Astrophysics of the University of California under
cooperative agreement AST-8809616, and by the Mount Stromlo and Siding
Spring Observatory, part of the Australian National University. This
publication makes use of data products from the Two Micron All Sky
Survey, which is a joint project of the University of Massachusetts
and the Infrared Processing and Analysis Center/California Institute
of Technology, funded by the National Aeronautics and Space
Administration and the National Science Foundation. This publication
makes use of data products from the Optical Gravitational Lensing
Experiment OGLE-III online catalog of variable stars. This research
has made use of the {\sc VizieR} catalog access tool, CDS, Strasbourg,
France. This research has made use of the {\sc SIMBAD} database,
operated at CDS, Strasbourg, France. This research has made use of
NASA's Astrophysics Data System Bibliographic Services.

\appendix
\section{Literature-based classification support for \SSp\ objects}
\label{litsurv}

\emph{{NGC 1651 SAGE IRS 1} (SSID1)}. {NGC 1651} is an
intermediate-aged globular cluster
\citep[2$\times$10$^9$\,yr;][]{mac03}, and SAGE IRS 1 is a particular
member of that cluster with a K-band magnitude of 11.2\,mag
\citep[2MASS;][]{cut03} and a period of 101 days
\citep{sos09b}. \citet{sri09} classify this object as an O-rich AGB
star. This, and the cluster age, supports our classification of {\tt
  O-AGB}.

\emph{{SSTISAGEMC J043727.61$-$675435.1} (SSID2)}. This object is
classified as a galaxy by \citet{gru09} and \citet{vlo10}, which
matches our classification of {\tt GAL}.

\emph{{SSTISAGEMC J044627.10$-$684747.0} (SSID3)}. This object
has a period given by MACHO and OGLE-III of $\sim$400 days
\citep{alc98,sos09b}. This fits with our classification of {\tt
  C-AGB}.

\emph{{SSTISAGEMC J044718.63$-$694220.6} (SSID4)}.  This object is
visibly bright in an uncrowded region, with \BVcol=1.8\,mag and
\VRcol=1.4\,mag \citep{reb83}. It is classified as a star of spectral
type M based on a low-resolution objective prism survey, and has a
radial velocity measurement concurrent with being in the LMC
\citep{pre85}. This supports our classification of {\tt RSG}.
%The only SIMBAD
%association near this position is RM 1-1 (E. Rebeirot et al., 1983,
%A\&AS, vol. 51, 277) which is located at 04:47:16 -69:42.4 and has
%V=13.4.  This is clearly the same star from the DSS image of the
%field, since no other star of such brightness is visible within about
%50 arc-seconds of the SAGE position.

\emph{{SSTISAGEMC J044837.76$-$692337.0} (SSID5)}.  \citet{sri09}
classify this object as an extreme AGB star, based on \emph{Spitzer}
colours. This object is modelled by \citet{whi08}, who classify it as
a high-probability Stage I YSO \citep[meaning that this YSO is
  embedded in an infalling envelope;][]{rob06}. This work supports our
classification of {\tt YSO}.

\emph{{SSTISAGEMC J044934.31$-$690549.3} (SSID6)}. This object has
recently been identified as LH$\alpha$\ 120-S 67 \citep{ski09} owing
to a mistake in the original \citet{hen56} co-ordinates. Both authors
and \citet{and64} list this object as a weak emission-line
object. \citet{whi08} categorise it as a YSO, but do not put it in
their \textquotedblleft high
probability\textquotedblright\ class. \citet{vij09} also categorise it
as a YSO candidate. However, we classify this object as {\tt O-AGB}.

\emph{{MSX LMC 1128} (SSID7)}. This Long Period Variable star (LPV)
has a period of 418--445\,days \citep{hug89,fra05,fra08}, and is
classified by \citet{ega01} as carbon-rich, based on \emph{JHK$_s$A}
colours from 2MASS and MSX. This supports our classification of {\tt
  C-AGB}.

\emph{{SSTISAGEMC J045128.58$-$695550.1} (SSID8)}. This variable star
has a lengthy period of 884--911\,d \citep{sos08,fra05,fra08}, which
is congruous with it being an evolved {\tt O-AGB}.

\emph{{IRAS 04518$-$6852} (SSID9)}. {IRAS 04518-6852}
has been identified by \citet{vlo97} as C-AGB based on IRAS colours,
and as an extreme AGB star by \citet{sri09} and \citet{gru09}, based
on \emph{Spitzer} colours. However, \citet{ega01} classify this object
as an H{\sc ii} region, based on 2MASS and MSX \emph{JHK$_s$A}
colours. This latter classification would seem to be spurious, since
the IRS spectrum shows no indication of atomic emission lines. Hence
our classification as {\tt C-AGB}.

\emph{{SSTISAGEMC J045200.36$-$691805.6} (SSID10)}. This is
classified as a YSO by \citet{gru09} and more specifically, a
high-probability Stage I YSO by \citet{whi08}, which supports our
classification of {\tt YSO}.

\emph{{SSTISAGEMC J045228.68$-$685451.3} (SSID11)}. This object is
classified as a YSO (high probability -- HP, Stage I) by \citet{whi08}
based on an SED fit to photometry points, and by \citet{oli09} based
on modelling of the spectrum. This supports our classification of {\tt
  YSO}.

\emph{{KDM 764} (SSID12)}. {KDM 764} is identified as a
carbon-rich AGB star from the detection of C$_2$\ in its UK Schmidt
Telescope (UKST) spectrum \citep{kon01}. \citet{sri09} also classify
this object as C-AGB. The OGLE-III period is given as 445 days
\citep{sos09b} whereas the MACHO period ranges between 250--873 days
\citep{fra05,fra08}. This supports our classification of {\tt C-AGB}.

\emph{{SSTISAGEMC J045309.39$-$681710.8} (SSID13)}. This LPV
has a period of 916--1,122 days \citep{sos09b,fra05,fra08} and is
classified by OGLE-III \citep{sos09b} and \citet{sri09} as an
oxygen-rich AGB star. Very faint visually, this star has
\VRcol\ typical of a mid-M Giant \citep{mas02}, in agreement with our
classification of {\tt O-AGB}.

\emph{{IRAS F04532$-$6709} (SSID14)}. This star is classified as a
YSO by \citet{oli09}, based on modelling of the spectrum, and also by
\citet{whi08} and \citet{gru09}, supporting our classification of {\tt
  YSO}.

\emph{{SSTISAGEMC J045328.70$-$660334.4} (SSID15)}. This object has
an extremely long period of 2710 days from MACHO measurements
\citep{alc98}.  \citet{sri09} classify this as {\tt C-AGB}, in
agreement with our classification.

\emph{{GV 60} (SSID16)}. This object is listed as an M3 star in
SIMBAD, which is supported by photometry from \citet{wes81}. It is
also included in the RSG catalog of \citet{mas03}, who give
$M_\mathrm{bol}$=$-$9.39\,mag.  This would support our classification of
{\tt RSG}.

\emph{{LH$\alpha$\ 120-N 82} (SSID17)}. Alternatively called
{Brey~3a} \citep*{bre99}, this object is a known WR star
\citep*{hey90}, although its exact spectral type remains controversial
\citep{mof91,hey92}. We classify it as {\tt OTHER}.

\emph{{SSTISAGEMC J045344.24$-$661146.0} (SSID18)}. This object is
classified as C-AGB by \citet{vlo97} and \citet{lou97}, using IRAS
colours and by \citet{gru09} using \emph{Spitzer} photometry. This
matches our classification of {\tt C-AGB}.

\emph{{SSTISAGEMC J045422.82$-$702657.0} (SSID19)}. This object is
a hot foreground star with \BVcol=1.033, and a stellar type of K5
\citep{ski09}. We classify it as {\tt STAR}.

\emph{{SSTISAGEMC J045526.69$-$682508.4} (SSID20)}. This
\emph{SAGE-LMC} source was classified by \citet{whi08} as a
high-probability Stage I YSO, and similarly by \citet{gru09}. It sits
within 20$\arcsec$ of the H{\sc ii} region, {IRAS 04555-6829}. We
similarly classify this object as {\tt YSO}.

\emph{{SSTISAGEMC J045534.07$-$655701.3} (SSID21)}. This
\emph{SAGE-LMC} source was classified as a YSO by \citet{whi08} with
high-probability (Stage I), and with lower confidence by \citet{gru09}
(who suggest this object could potentially be a PN). No other
information in the literature could be found. We agree with
\citet{whi08} and \citet{gru09}, in our classification of {\tt YSO}.

\emph{{SSTISAGEMC J045623.21$-$692749.0} (SSID22)}. This
\emph{SAGE-LMC} source was classified by \citet{whi08} as a
high-probability YSO. No other information in the literature could be
found. We classify it as {\tt O-AGB}.

\emph{{KDM 1238} (SSID23)}. {KDM 1238} is classified by
\citet{kon01} as C-AGB by a detection of C$_2$. It has periods which
range from 133 days \citep{sos09b} to 543 days \citep{alc98}.  This
information supports our classification of {\tt C-AGB}.

\emph{{RP 1805} (SSID24)}. Classified as a PN by \citet{rei06}
based on optical spectra, with the comments that it was faint,
circular and diffuse. Not detected in our observations, hence
classified as {\tt UNK}.

\emph{{SSTISAGEMC J050032.61$-$662113.0} (SSID25)}. Classified as a
YSO by \citet{gru09}, in agreement with our classification ({\tt
  YSO}).

\emph{{RP 1631} (SSID26)}. Classified as a PN by \citet{rei06}
based on optical spectra, with the comments that it was bright,
circular and small. However, we believe this object to be an RCrB
star, and thus classify it as {\tt OTHER}.

\emph{{MSX LMC 1271} (SSID27)}. {MSX LMC 1271} is likely
a member of the compact cluster {NGC~1805}, which has an age of
8--45\,Myr \citep{wol07,liu09}. WFPC2 images from the \emph{Hubble
  Space Telescope} show a bright star close to the observed
co-ordinates. Both these pieces of information are congruous with
{MSX LMC 1271} being a red supergiant, {\tt RSG}.

\emph{{SSTISAGEMC J050224.17$-$660637.4}
  (SSID28)}. \citet{whi08} class this object as one of their
high-probability YSOs, still in the embedded stages of evolution
(Stage I). However, its spectrum has the distinctly double-peaked
shape of an oxygen-rich post-AGB object. This object may potentially
belong to the young cluster {NGC~1805}, in which case it would
almost certainly be a YSO, but we choose to tentatively classify it as
{\tt O-PAGB}.

\emph{{HV 2281} (SSID29)}.  This variable object is classified as
an RV~Tau star, based on the MACHO light curve
\citep{alc98}. OGLE-III also classifies this object as an RV~Tau
\citep{sos08}. We reach an identical conclusion, classifying
{HV 2281} as {\tt O-PAGB}. \citet{gru09}, however, class this
as a stellar photosphere.

\emph{{KDM 1656} (SSID30)}. {KDM 1656} is classified by
\citet{kon01} as C-AGB by a detection of C$_2$, and also by
\citet{sri09}, based on infrared colours. This object has a long period
of 1035 days \citep{alc98}. This supports our classification of {\tt
  C-AGB}.

\emph{{KDM 1691} (SSID31)}. Classified by \citet{kon01} as a
carbon star, \citet{wes78} also detect CN bands in visible and
near-infrared photometry and spectra. \citet{sri09} group this object
with the C-rich AGB stars. MACHO and OGLE-III calculate a period of
$\sim$510 days for this object \citep{fra05,fra08,sos09b}. Thus
{KDM 1691} is classified as {\tt C-AGB}.

\emph{{LMC-BM 11-19} (SSID32)}. This object exhibits highly
divergent periods of 98 days \citep{sos09b} and 894--1082 days
\citep{fra05,fra08}, although the OGLE-III data on this object are not
particularly good . This object is identified as a carbon star by
\citet{bla90} and \citet{lou03} based on near-infrared CN bands. Due
to the lack of identifiable carbon-rich features in the IRS spectrum,
we classify this object as {\tt STAR}.

\emph{{LMC-BM 12-14} (SSID33)}. Again, this object has
differing periods of 220 days \citep{sos09b} and 405--894 days
\citep{fra05,fra08}, and is included in the carbon star catalog of
\citet{bla90} and \citet{sri09}. This supports our classification of
      {\tt C-AGB}.

\emph{{SSTISAGEMC J050354.55$-$671848.7} (SSID34)}. This object was
classified as a YSO by \citet[][Stage I]{whi08}, \citet{oli09} and
\citet{gru09}, fortifying our classification of {\tt YSO}.

\emph{{NGC 1818 WBT 5} (SSID35)}. \citet{wil95} measure the
V-band magnitude and calculate the \BVcol\ colour (1.745\,mag) for this
object, which are consistent with it being a late-type star of
spectral type $\sim$M0--M2. \citet{fra05} and \citet{fra08} measure
periods of 360--1707 days. \citet{wol07} give a cluster age for
{NGC 1818} of 14--40\,Myr. This supports our classification of
       {\tt RSG}.

\emph{{SSTISAGEMC J050407.72$-$662505.9}
  (SSID36)}. \citet{sri09} classify this object as a carbon-rich AGB
star based on $J$-$K_\mathrm{S}$ infrared colours, and we agree with
the classification, {\tt C-AGB}. However, this star is co-located on
the sky with globular cluster {NGC 1818}, a cluster of RSGs. It
has a period of $\sim$840 days \citep{fra05,fra08}.

\emph{{NGC 1818 WBT 3} (SSID37)}. Similarly to WBT~5,
\BVcol=1.684\,mag \citep{wil95} is typical of a late-type star with a
spectral type $\sim$M0. It has a period of 360 days
\citep{fra05,fra08}. This would support our classification of {\tt
  RSG}.

\emph{{MSX LMC 61} (SSID38)}. This object is classified by
\citet{ega01} as an O-AGB star, based on J$-$K vs. K$-$A colours, which
matches our classification ({\tt O-AGB}). It has a period of 580 days
\citep{fra05,fra08,sos09b}.

\emph{{RP 1878} (SSID39)}. \citet{rei06} list this as a bright,
round, small PN, with a diameter of $5\farcs3$\ in H$\alpha$\ and a
velocity of $v_{\rm helio} = 252.3$\,km s$^{-1}$. A weak IRS spectrum
with no distinctive features leads us to classify this object as {\tt
  UNK}.

\emph{{IRAS 05047$-$6642} (SSID40)}. This object has a short
period of 20 days according to \citet{fra05} and \citet{fra08}, and
was classified as a high-probability YSO by \citet{whi08}. However,
\citet{sri09} classify it as an extreme AGB star, and \citet{gru09} as
a stellar photosphere, or a PN. We agree with \citet{whi08},
classifying as {\tt YSO}.

\emph{{SSTISAGEMC J050503.21$-$692426.5} (SSID41)}. Little is
known about this object, apart from its period
\citep[217\,d;][]{sos09b}. We classify it as {\tt C-AGB}.

\emph{{SSTISAGEMC J050517.08$-$692157.0} (SSID42)}. Classified as a
potential galaxy or YSO by \citet{gru09}, we classify this object as
{\tt GAL}.

\emph{{LMC-BM 13-2} (SSID43)}. Measured to have a period of 206
days by \citet{sos09b}, this {\tt C-AGB} star has CN bands in its
near-infrared spectrum \citep{bla90}.

\emph{{SSTISAGEMC J050558.23$-$680923.6} (SSID44)}. This
\emph{SAGE-LMC} object was classified as a high-probability YSO by
\citet{whi08}. We also classify it as {\tt YSO}.

\emph{{SSTISAGEMC J050607.50$-$714148.4} (SSID45)}. Apart from a
period of 346 days \citep{fra05,fra08}, little is known about this
object. We classify it as {\tt C-AGB}.

\emph{{KDM 1961} (SSID46)}. Near-infrared C$_2$ bands were detected
in this star by \citet{kon01}, backing up our classification of {\tt
  C-AGB}. \citet{sri09} agree.

\emph{{KDM 1966} (SSID47)}. Similarly to the last object, the Swan
C$_2$ bands were detected in this star by \citet{kon01}, backing up
our classification of {\tt C-AGB}. \citet{sri09} agree once more.

\emph{{SSTISAGEMC J050620.12$-$645458.6} (SSID48)}. \citet{sri09}
classify this star as a carbon-rich AGB star, and no other information
was found about this object. Our classification of {\tt C-AGB} is in
agreement with Srinivasan et al.

\emph{{SSTISAGEMC J050629.61$-$685534.9} (SSID49)}. Apart from a
determination of period
\citep[154--295\,d;][]{sos09b,fra05,fra08,ita04}, little is known
about this object. \citet{sri09} class it as C-AGB, which agrees with
our classification, {\tt C-AGB}.

\emph{{SSTISAGEMC J050639.14$-$682209.3} (SSID50)}. \citet{san70}
lists this star as having an early-type spectrum, which is refined
further by \citet{rou78} to that of a B4 supergiant. B$-$V=-0.03\,mag
and U$-$B=-0.70\,mag \citep{iss79} are consistent with a B5--B8
supergiant, hence our classification of {\tt OTHER}.

\emph{{SHV 0507252$-$690238} (SSID51)}. This semi-regular variable
star has a number of defined periods: 1353\,d \citep[][I-band]{hug89},
254\,d, 431\,d, 563\,d \citep[][ OGLE, 2MASS, DENIS]{gro04}, 507\,d
\citep[][ OGLE-III]{sos09b}, and is classified as a C-rich Mira
variable by \citet{sos09b}. We classify it as {\tt C-AGB}.

\emph{{SSTISAGEMC J050713.90$-$674846.7} (SSID52)}. This object is
classified by \citet{whi08} as YSO-HP, and by \citet{gru09} as a
(post-)AGB object. We classify it as {\tt C-PAGB}.

\emph{{SSTISAGEMC J050752.93$-$681246.5} (SSID53)}. Both
\citet{vij09} and \citet{sri09} include this object in their variable
star and evolved star catalogs, respectively. Both classify it as an
extreme or obscured AGB star. It has periods of 302--335 days
\citep{sos09b,fra05,fra08}. We classify it as {\tt C-AGB}.

\emph{{SSTISAGEMC J050759.35$-$683925.8} (SSID54)}. This star
has a period of 150--362 days \citep{sos09b,fra05,fra08} and is
classified as a semi-regular variable O-AGB by OGLE-III, which
supports our classification of {\tt O-AGB}. It has also been classed
as a YSO by \citet{whi08}, but with a low confidence.

\emph{{SSTISAGEMC J050826.35$-$683115.1} (SSID55)}. This star is
classified as an extreme-AGB star by both \citet{vij09} and
\citet{sri09}. We agree, with our classification of {\tt C-AGB}.

\emph{{SSTISAGEMC J050830.51$-$692237.4} (SSID56)}. This is
classified as O-AGB by \citet{sri09}, but has a relatively short
period for an AGB star \citep[115\,d;][]{sos09b}. We classify it as
{\tt O-PAGB}.

\emph{{KDM 2187} (SSID57)}. This object is classified as a C-AGB in
the catalog of \citet{kon01}, which was selected by the detection of
C$_2$ bands. We agree with this classification, {\tt C-AGB}.

\emph{{BMB-BW 180} (SSID58)}. Classified as an AGB star by the
DENIS consortium \citep{cio00a} based on near-infrared colours, this
object was also found to be mildly variable, with a period of 70 days
\citep{gro04,ita04}, which may indicate that it lies on the early
AGB. It has a spectral type of M2 \citep{gro04}, in line with our
classification of {\tt O-AGB}.

\emph{{NGC 1856 SAGE IRS 1} (SSID59)}. {NGC 1856} is a
cluster which has an age of 120--151\,Myr \citep{wol07} and thus is
probably composed of relatively massive AGB stars. This object has a
positive (I$-$J) colour from DENIS data, and \citet{whi08} classify
this object as a YSO. We classify it as an {\tt O-AGB}.

\emph{{SSTISAGEMC J051028.27$-$684431.2} (SSID60)}. Both
\citet{vij09} and \citet{sri09} classify this object as an extreme AGB
star, based on 2MASS and \emph{Spitzer} colours. It has a period of 465
days \citep{sos09b}. We classify it as {\tt C-AGB}.

\emph{{SSTISAGEMC J051059.07$-$685613.7} (SSID61)}. This star is an
oxygen-rich AGB star according to \citet{sri09}, which agrees with our
classification {\tt O-AGB}.

\emph{{MSX LMC 209} (SSID62)}. {MSX LMC 209} has a 28-day
period according to \citet{sos09b}, and is classified by them as a
carbon-rich AGB star. \citet{ega01} classify it as a PN, based on
colours. It may be associated with the emission-line object
{LH$\alpha$ 120-S 160} \citep{hen56}. We class it as an
oxygen-rich protoplanetary nebula, {\tt O-PAGB}.

\emph{{SSTISAGEMC J051213.54$-$683922.8} (SSID63)}. This object is
categorised by \citet{sri09} as an oxygen-rich AGB star, which matches
our classification, {\tt O-AGB}.

\emph{{SSTISAGEMC J051228.19$-$690755.8} (SSID64)}. This object is
probably associated with LI-LMC 611 \citep{lou97} and
IRAS\,05127$-$6911, but unclassified by these authors. Both
\citet{whi08} and \citet{gru09} classify it as a YSO candidate, but
with low confidence. We classify it as {\tt C-PAGB}.

\emph{{IRAS 05133$-$6937} (SSID65)}. This object is one of the
thirteen EROs selected by \citet{gru08} based on extremely red mid-IR
colours. Follow-up observations show that this small sample is composed
of extreme carbon-rich AGB stars, confirming our classification ({\tt
  C-AGB}). This star is also classified by \citet{whi08} and
\citet{vij09} as a high-probability YSO, based on its infrared colours.

\emph{{OGLE 051306.52$-$690946.4} (SSID66)}. This star is a
long-period variable according to \citet{gro04}, with a period of 183
days. It has been modelled with a radiative transfer model by
\citet{sri10}, who pay particular attention to the acetylene features
in the IRS spectrum, thus supporting our classification of {\tt
  C-AGB}. \citet{vij09} and \citet{sri09} classify it as an extreme AGB
star.

\emph{{SSTISAGEMC J051339.94$-$663852.5} (SSID67)}. Classified as
O-rich AGB star by \citet{sri09}, which matches our classification,
{\tt O-AGB}.

\emph{{NGC 1866 Robb B136} (SSID68)}. Globular cluster {NGC
  1866} has an age of 100\,Myr \citep{bec83,bro89}, giving M5 star
\citep{aar85} {NGC 1866 Robb B136} the age typical of a massive
AGB star, and supporting our classification of {\tt
  O-AGB}. \citet{sri09} come to the same conclusion.

\emph{{BSDL 923} (SSID69)}. According to \citet{gou03} and
\citet{bic99}, this object is a member of the young stellar cluster
{LMC-N30}, suggesting that this is a massive star. We classify
it as a B supergiant, hence {\tt OTHER}.

\emph{{SSTISAGEMC J051347.72$-$693505.2} (SSID70)}. This object was
classified as a YSO by \citet[][high probability, Stage I]{whi08},
\citet{oli09} and \citet{gru09}, strengthening our classification of
      {\tt YSO}.

\emph{{SSTISAGEMC J051348.38$-$670527.0} (SSID71)}. \citet{whi08}
class this as a YSO-HP, Stage I, and \citet{gru09} think it to be a
probable YSO or galaxy. We classify it as {\tt HII} due to the
low-excitation emission lines in the IRS spectrum.

\emph{{SSTISAGEMC J051412.33$-$685058.0} (SSID72)}. Classified as
     {\tt O-AGB} in agreement with \citet{sri09}.

\emph{{HV 915} (SSID73)}.  This object is classified as an RV~Tau
star, based on the MACHO light curve \citep[MACHO
  79.5501.13;][]{alc98}, and \citet*{per03} confirm this (and
calculate a period of 48.5 days). \citet{vij09} also class this object
as O-rich. Hence our classification of {\tt O-PAGB} is upheld.

\emph{{SSTISAGEMC J051449.43$-$671221.4} (SSID74)}. This object was
classified as a YSO by \citet[][high probability, Stage I]{whi08} and
\citet{oli09}, based on the SED and IRS spectrum,
respectively. \citet{shi08} also classified it as a YSO based on the
detection of H$_2$O and CO$_2$\ ice with AKARI. This object is their
ST4, although it must be noted that the association of {IRAS
  05148-6715} with this object is likely erroneous. Our classification
of {\tt YSO} is upheld. \citet{sri09} classify this object as an
extreme AGB star, based on mid-infrared colours.

\emph{{SSTISAGEMC J051453.10$-$691723.5} (SSID75)}. This is a
possible YSO or naked star according to \citet{gru09}. We classify it
as {\tt O-PAGB}.

\emph{{SSTISAGEMC J051526.44$-$675126.9} (SSID76)}. This object has
a period of 107 days according to OGLE-III \citep{sos09b} and 441 days
according to MACHO \citep{fra05,fra08}. It is classified by the former
as an oxygen-rich semi-regular variable star. Due to a lack of evident
oxygen-rich features in the IRS spectrum, we classify it as {\tt
  STAR}.

\emph{{SSTISAGEMC J051612.42$-$704930.3} (SSID77)}. Classified as
     {\tt O-AGB} by us and by \citet{sri09}.

\emph{{IRAS 05170$-$7156} (SSID78)}. This object has an unusual IRS
spectrum and SED. It is considered to be an AGN candidate according to
\citet{deg87}, a YSO-HP, Stage I, according to \citet{whi08} and an
extreme AGB star by \citet{sri09} and \citet{gru09}. We classify it as
{\tt UNK}.

\emph{{SSTISAGEMC J051747.18$-$681842.6} (SSID79)}. This star is an
oxygen-rich AGB star according to \citet{sri09}, which is in accord
with our classification, {\tt O-AGB}.

\emph{{SSTISAGE1C J051803.28$-$684950.6} (SSID80)}. Considered an
extreme AGB by \citet{vij09} and \citet{sri09}, and a carbon-rich AGB
star with a period of 349 days by \citet{sos09b}. We classify it as
{\tt C-AGB}.

\emph{{KDM 3196} (SSID81)}. {KDM 3196} is included in the
carbon star catalogs of \citet{san77}, \citet{wes78} and
\citet{kon01}. This object may also be a CH star: a metal-poor,
carbon-rich giant, found in the halo of the Milky Way, and potentially
in the halo of the LMC \citep{har88,fea92}. We classify it as {\tt
  STAR} due to the lack of distinguishing features in the IRS
spectrum.

\emph{{HV 5715} (SSID82)}. This object is a long period variable,
with a period of 422 days \citep{wri71}. It is considered to be an
O-AGB by \citet{sri09}, and its IRS spectrum is modelled by
\citet{sar10}. This would support our classification of {\tt
  O-AGB}.

\emph{{SSTISAGEMC J051832.64$-$692525.5} (SSID83)}. \citet{ita04}
measure a period of 293 days for this object. \citet{sri09} consider
it to be a carbon-rich AGB star, as do we, {\tt C-AGB}. 

\emph{{IRAS F05192$-$7008} (SSID84)}. This object is probably
associated with MSX LMC 390 \citep{ega01} and IRAS 05193$-$7009. It
has been classified by \citet{ega01} as an H{\sc ii} region and as a
star by \citet{gru09}. We classify it as {\tt C-PAGB}.

\emph{{HV 2444} (SSID85)}.  This object was regarded as a Type II
Cepheid by \citet{wel87} and \citet{pay71}, and the OGLE-III catalog
further classifies this object as an RV~Tau stay with a period of 36
days \citep{sos08}. This supports our classification of {\tt O-PAGB}.

\emph{{SSTISAGEMC J051908.46$-$692314.3} (SSID86)}. Considered to
be a {\tt C-AGB} by both \citet{sri09} and us.

\emph{{2MASS J05191049$-$6933453} (SSID87)}. This object is a
variable star with a period of 452 days \citep{ita04}. We consider it
to be {\tt C-AGB}, in agreement with \citet{sri09}.

\emph{{2MASS J05194483$-$6929594} (SSID88)}. \citet{sri09}
tentatively classify this object as an oxygen-rich AGB star, but we
consider it to be a naked {\tt STAR}.

\emph{{SSTISAGEMC J052014.24$-$702931.0} (SSID89)}. This LPV,
with a 686-day period \citep{sos09b}, is oxygen-rich according to
\citet{sri09}, and we are concordant with that classification, {\tt
  O-AGB}.

\emph{{SSTISAGEMC J052023.97$-$695423.2} (SSID90)}. This object has
a very short period of 0.17\,d according to \citet{alc98} and is
classed as a high probability YSO by \citet{whi08}, in harmony with
us, {\tt YSO}.

\emph{{SSTISAGEMC J052051.83$-$693407.6} (SSID91)}. An LPV with a
lengthy period of $\approx$770 days \citep{sos09b}, this star is
considered to be {\tt O-AGB} by us and \citet{sri09}.

\emph{{LH$\alpha$\ 120$-$N 125} (SSID92)}. This object has been
recognized as a point-like emission nebula on objective prism
photographs of the LMC at H$\alpha$\ wavelengths by \citet{hen56}, and
tentatively classified as a PN by \citet{lin63}. The nature of this
object has been confirmed by \citet*{san78}, who classified it as a
medium-excitation PN using very deep blue- and red-sensitive objective
prism-plates. This PN is unresolved even on HST images
\citep{sha06}. Its chemical composition has been recently determined
by \citet{lei06}. \citet{gru09} also believe this to be a PN. We
agree, and designate it as a carbon-rich PN ({\tt C-PN}).

\emph{{SSTISAGEMC J052101.66$-$691417.5} (SSID93)}. This is a YSO
candidate according to \citet{whi08}, but an O-AGB according to
\citet{sri09}. We class this object as {\tt O-AGB}.

\emph{{HV 942} (SSID94)}. Considered to be an extreme AGB star by
\citet{sri09}, this object was confirmed to be an RCrB star by
\citet{sos09a}, in agreement with our classification, {\tt OTHER}.

\emph{{MACHO 78.6698.38} (SSID95)}. Identified as an RV~Tau star
with a period of 25 days by \citet{per03}, which agrees with our
classification of {\tt O-PAGB}.

\emph{{SSTISAGEMC J052206.92$-$715017.7} (SSID96)}. Our
classification of {\tt O-AGB} is concurrent with that of \citet{sri09}
for this object.

\emph{{SSTISAGEMC J052222.95$-$684101.2} (SSID97)}. This object is
considered to be a YSO-HP by \citet{whi08} and \citet{gru09}. We
concur, {\tt YSO}. However, \citet{vlo10} speculate that this is a
galaxy from the MIPS-SED spectrum.

\emph{{OGLE 052242.09$-$691526.2} (SSID98)}. This LPV has a period
of 128 days in the compilation of \citet{gro04}. \citet{vij09} and
\citet{sri09} consider it to be an extreme AGB star, whilst we
consider it to be {\tt C-AGB}.

\emph{{SHV 0523185$-$693932} (SSID99)}. {SHV
  0523185$-$693932} is a variable star with a period on the order of
200 days \citep{gro04}. This is typical of a star on the AGB, thus
supporting our classification of {\tt O-AGB}.

\emph{{LH$\alpha$\ 120$-$N 136} (SSID100)}. This object has been
recognized as a point-like emission nebula on objective prism
photographs of the LMC at H$\alpha$\ by \citet{hen56}, but classified
as an H$\alpha$\ emission-line object by \citet{lin63} since no other
nebular lines were detected on the analyzed plates. The object has
been classified as PN by \citet{san78} of very low excitation class
\citep{mor84}, which could be a member of SL434 LMC cluster
\citep{kon96}. This PN is resolved and has a round shape on HST
images, with radius of $\sim$0.55\,\arcsec\ \citep{sha01}. Its
chemical composition has been recently determined by \citet{lei06} and
shows an abundance pattern which may suggest low initial abundance of
its progenitor. Classified as PN by \citet{gru09} using \emph{Spitzer}
colours, in accord with our classification, {\tt O-PN}.

\emph{{IRAS 05240$-$6809} (SSID101)}. This object is ST7 in the
paper of \citet{shi08}, and they report an AKARI detection of
CO$_2$\ ice in this YSO. \citet{oli09} and \citet[][high probability
  YSO]{whi08} also count this as a YSO, affirming our classification
of {\tt YSO}. \citet{sri09} classify this as an O-AGB candidate.

\emph{{IRAS 05246$-$7137} (SSID102)}. Ambiguous colours have lead
\citet{ega01} and \citet*{vlo05b} to classify this object as an extreme
AGB star. However, \citet{whi08}, \citet{oli09} and \citet{vlo10}
classify this object as a YSO based on fits to the SED, the
\emph{Spitzer} IRS spectrum and the \emph{Spitzer} MIPS-SED spectrum,
respectively. We also classify this object as {\tt YSO}.

\emph{{SSTISAGEMC J052405.31$-$681802.5} (SSID103)}. Considered
an extreme AGB star by \citet{vij09} and \citet{sri09}, we consider it
to be {\tt C-AGB}.

\emph{{MSX LMC 464} (SSID104)}. \citet{ega01} classify {MSX LMC 464}
as an OH/IR star, based on colours, whilst \citet{kas08} classify it
as a potential H{\sc ii} region based also on colours. \citet{oli09}
consider it in their selection of Class I YSOs, but find no convincing
detection of ice in the \emph{Spitzer} IRS spectrum. This conflicts
with a previous classification based on photometry by \citet{whi08},
who report this object as a high probability YSO at Stage I of
evolution. Given the continuum shape and the [SIII] line in the IRS
spectrum, we classify this object as {\tt HII}, in agreement with
\citet{kas08}.

\emph{{OGLE 052445.53$-$691605.6} (SSID105)}. \citet{gro04} lists
this star as a long-secondary-period variable, with periods of 131
days and 399 days. \citet{sri09} classify it as an extreme AGB star,
whilst we classify it as {\tt C-AGB}. 

\emph{{LH$\alpha$\ 120$-$S 33} (SSID106)}. This H$\alpha$\ emitter
was detected by \citet{hen56}, \citet{and64} and \citet{boh74}. We
class it as {\tt YSO}.

\emph{{HV 5829} (SSID107)}.  This object was regarded as a Type II
Cepheid \citep{wel87}, and originally listed in \citet{pay71}. More
specifically, OGLE-III \citep{sos08} categorises this object as an
RV~Tau star, which backs up our classification of {\tt O-PAGB}.

\emph{{SSTISAGEMC J052546.51$-$661411.5} (SSID108)}. The SED and
ice features in the IRS spectrum of this object were modelled by
\citet{whi08} and \citet{oli09}. \citet{shi08} also were able to
measure H$_2$O and CO$_2$\ column densities in this YSO (ST3) using
AKARI. \citet{sri09} count this as an extreme AGB candidate, however
we concur with the classification of {\tt YSO}.

\emph{{SSTISAGEMC J052613.39$-$684715.0} (SSID109)}. We classify
this object as {\tt YSO}, in agreement with \citet{gru09}.

\emph{{OGLE 052620.25$-$693902.4} (SSID110)}. This is an LPV with a
period in excess of 800 days \citep[e.g.,][]{sos09b}. It is considered
to be O-rich by \citet{sri09}, which is in agreement with our
classification of {\tt O-AGB}.

\emph{{HV 2522} (SSID111)}. According to \citet{alc98},
\citet{pay71} classified this object as a Type II Cepheid, which would
imply that this apparent post-AGB object may be an RV~Tauri
object. However, without verification we class {HV 2522} as
{\tt O-PAGB}.
 
\emph{{RP 589} (SSID112)}. \citet{rei06} identify this object as a
circular, bright PN, with H$\alpha$\ to the east. They give a diameter
of $5\farcs3$\ in H$\alpha$\ and a velocity of $v_{\rm helio} =
261.1$\,km s$^{-1}$. The weak IRS spectrum presents no features to
suggest that this object is a PN, so we classify it as {\tt UNK}.

\emph{{SSTISAGEMC J052707.10$-$702001.9} (SSID113)}. Classified as
a YSO by \citet{whi08}, we classify this object as {\tt O-PAGB}.

\emph{{SSTISAGEMC J052723.14$-$712426.3} (SSID114)}. This object is
mentioned by \citet{vlo10} as a potential YSO based on the appearance
of the \emph{Spitzer} MIPS-SED spectrum. \citet{whi08} and
\citet{gru09} classify it similarly, with a high confidence, as do we:
      {\tt YSO}.

\emph{{LH$\alpha$\ 120$-$N 145} (SSID115)}. This object has been
recognized as a point-like emission nebula on objective prism
photographs of the LMC in H$\alpha$\ by \citet{hen56}, and tentatively
classified as PN with strong H$\alpha$\ emission by \citet{lin63}. The
nature of this object as PN has been confirmed by \citet{san78}. This
PN is not resolved on HST images \citep{sha06}. This is a peculiar
object with extremely high densities and temperature \citep{dop91} and
may be one of the youngest PN. Its chemical composition has been
determined several times with the most recent estimation by
\citet{lei06} showing a low N abundance. \citet{sri09} classify this
object as an extreme AGB, whilst \citet{gru09} choose PN. From the IRS
spectrum and SED, we classify this object as a very evolved {\tt
  O-PAGB}.

\emph{{HV 2551} (SSID116)}. This object is likely a red supergiant,
with spectral type K5-M0 \citep{oes99}. It also appears in the
catalogs of red supergiant candidates from \citet{wes81} and
\citet{san77}. We classify it as {\tt RSG}.

\emph{{W61 11$-$16} (SSID117)}. This star is given a spectral type
of M1 in the catalog of supergiants of \citet*{eli85} and K7 I in the
catalog of \cite{mas03}. The latter authors also calculate
$M_\mathrm{bol}$=$-$8.23\,mag, which supports our classification of
{\tt RSG}.

\emph{{SSTISAGEMC J052747.62$-$714852.8} (SSID118)}. \citet{whi08}
classify this star as a YSO, but with a low confidence. \citet{sri09}
consider it to be an extreme AGB star. \citet{hen56} lists this star
as an H$\alpha$\ emitter, which supports our classification of {\tt
  O-PAGB}.

\emph{{SHV 0528350$-$701014} (SSID119)}. This LPV has a period
on the order of 600 days according to \citet{gro04}, and is a
candidate for an obscured AGB star in that paper. \citet{sri09}, as
well, count this object amongst their extreme, or obscured, AGB
stars. We concur, with our classification of {\tt C-AGB}.

\emph{{OGLE 052825.96-694647.4} (SSID120)}. This star is
classified by \citet{kon01}, \citet{sri09} and \citet{gro04} as a
C-rich AGB star. The latter author notes that it has periods of 135
and 390 days. This supports our classification of {\tt C-AGB}.

\emph{{IRAS 05298$-$6957} (SSID121)}. This is a well-studied
oxygen-rich massive (4\,M$_\odot$) AGB star with a high mass-loss rate
\citep[][and references within]{vlo10}. \citet{vij09}, \citet{sri09}
and \citet{gru09} consider it to be an extreme AGB star. We agree with
this classification, {\tt O-AGB}.

\emph{{HV 5879} (SSID122)}. \citet{oes98} derive an effective
temperature of 3\,675\,K for this star from an unpublished optical
spectrum, which would indicate a spectral type of $\sim$M0Iab for a
supergiant. \citet{mas03} give a similar spectral type and effective
temperature. They also calculate $M_\mathrm{bol}$=$-$8.26\,mag.  These
classifications would suggest that our classification of {\tt RSG} is
correct.

\emph{{SP77 46-50} (SSID123)}. \citet{san77} identify this star as
spectral type M in their supergiant catalog, and photometry and colours
from \citet{mas02} and the 2MASS and DENIS databases would seem to
confirm our classification of {\tt RSG}. \citet{mas03} calculate
$M_\mathrm{bol}$=$-$7.44\,mag.

\emph{{SHV 0530472-690607} (SSID124)}. This object has a
well-defined period of 212 days \citep{hug89,sos09b}. We classify it
as {\tt O-AGB}.

\emph{{IRAS 05315$-$7145} (SSID125)}. This is an ERO according to
\citet{gru08} and \citet{gru09}, and was not detected by \citet{vlo97}
since it is very faint at K-band. We class this object as {\tt C-AGB}.

\emph{{KDM 4554} (SSID126)}. Classified as C-AGB by \citet{kon01}
based on the detection of C$_2$ bands, and an extreme AGB star
according to \citet{sri09}, this star is {\tt C-AGB} in our
classification scheme.

\emph{{NGC 2004 Robb B45} (SSID127)}. {NGC 2004} is a young
cluster with an age of 8--10\,Myr \citep{wol07,hod83}. This particular
star has a B$-$V colour which is consistent with an early M spectral
type. \citet{ben91} classify this object as a red supergiant based
upon the assumed V absolute magnitude. This information backs up our
classification of {\tt RSG}.

\emph{{NGC 2004 Wes 18-13} (SSID128)}. The spectral type given in
\citet{eli85} is K1Ib, with V=13.053 and B=14.465.  The B-V colour is
roughly consistent with the spectral type, or one a little later. This
affirms our classification of {\tt RSG}.

\emph{{NGC 2004 Wes 6-14} (SSID129)}. The spectral type given in
\citet{eli85} is K0Ib, with V=12.95 and B=14.60.  The B-V colour is a
bit red for the spectral type, nominally it would match a K5I star. A
bolometric magnitude of $-$8.1\,mag is calculated by \citet{mas03},
confirming our classification of {\tt RSG}.

\emph{{SSTISAGEMC J053128.44$-$701027.1} (SSID130)}. This is a YSO
candidate according to \citet{whi08}, but its $\approx$400 day period
\citep{fra05,fra08,sos09b} would indicate that it is an evolved
star. \citet{sri09} classify it as C-AGB, but we consider it to be
{\tt O-AGB}.

\emph{{MACHO 82.8405.15} (SSID131)}.  \citet{alc98} identified this
object as an RV~Tau star based on its MACHO light curve, and this
was confirmed by \citet{per03} and given a period of 47
days. \citet{rey07} found a depletion pattern which is typically found
in RV~Tau stars with a binary disc. This fortifies our classification
of {\tt O-PAGB}. \citet{gru09}, however, classify this object as a
stellar photosphere.

\emph{{KDM 4665} (SSID132)}. Categorised as a carbon-rich AGB star
by \citet{kon01} and \citet{sri09}, with which we agree, {\tt C-AGB}.

\emph{{SSTISAGEMC J053206.70$-$701024.8} (SSID133)}. This star is
assigned a period of $\sim$120 days in the MACHO and OGLE-III
catalogs, and classified as an O-rich AGB star in the latter, as well
as in \citet{sri09}. We classify it as {\tt STAR}, given its lack of
mid-infrared dust features.

\emph{{SSTISAGEMC J053218.64$-$673145.9} (SSID134)}. This star may be
associated with the nearby young cluster NGC~2011 (see SSID135). No
other information could be found on this object, which we classify as
{\tt RSG}.

\emph{{NGC 2011 SAGE IRS 1} (SSID135)}. {NGC 2011} is a young,
resolved open cluster for which photometry is given in the catalog of
\citet*{kum08} (SAGE IRS 1 is star \#4). The age of this cluster is
given by \citet{wol07}, \citet{hod83} and \citet{san95} as
5--6\,Myr. There is a star with V=12.90, B-V=1.97, and V-R=0.56 very
close to the \SSp\ position.  The colours are not consistent, the
\VRcol\ colour should be much larger for the measured
\BVcol\ (nominally $>$2 magnitudes). An integrated spectrum of {NGC
  2011} in the K-band shows a typical M-type supergiant CO band, and
from the 2MASS image of the cluster it is clear that this would be
dominated by SAGE IRS 1.  The only other object of comparable
brightness nearby is about 25\arcsec\ to the south which is too far
away to contribute to the near infrared spectrum reported by
\citet{oli98}. \citet{mas03} calculate
$M_\mathrm{bol}$=$-$7.74\,mag. These facts corroborate with our
classification of {\tt RSG}.

\emph{{KDM 4718} (SSID136)}. This star was classified as a
carbon-rich AGB star by \citet{kon01} based on the detection of C$_2$
bands and also by \citet{sri09} using Spitzer colours. We too classify
this object as {\tt C-AGB}.

\emph{{RP 774} (SSID137)}. {RP 774} is thought by
\citet{rei06} to be a PN based on H$\alpha$\ maps of the LMC. However,
it is classified as a YSO by \citet{gru09} which agrees with our
classification, {\tt YSO}.

\emph{{SSTISAGEMC J053253.36$-$660727.8} (SSID138)}. This is a
high-probability YSO according to \citet{whi08} and a background
galaxy according to \citet{gru09}. We agree with \citet{gru09},
classing this object as {\tt GAL}.

\emph{{KDM 4774} (SSID139)}. Categorized as a carbon-rich star,
according to \citet{kon01} and \citet{sri09}, with which we agree,
{\tt C-AGB}.

\emph{{MSX LMC 736} (SSID140)}. This object is classified by
\citet{ega01} as an H{\sc ii} region, based on infrared
colours. However, \citet{vij09}, \citet{gru09} and \citet{sri09}
categorise this star as an (extreme) AGB star. We classify it as {\tt
  C-AGB}.

\emph{{SSTISAGEMC J053318.58$-$660040.2} (SSID141)}. Classified as
an extreme AGB by \citet{sri09}, we specify that it is a {\tt C-AGB}.

\emph{{SSTISAGEMC J053343.27$-$705921.1} (SSID142)}. We classify
this star as {\tt O-AGB}, in agreement with \citet{sri09}.

\emph{{SSTISAGEMC J053343.98$-$705901.9} (SSID143)}.  We classify
this star as {\tt O-AGB}, in agreement with \citet{sri09}.

\emph{{LH$\alpha$\ 120$-$N 151} (SSID144)}. This object has been
recognized as a point-like emission nebula on objective prism
photographs of the LMC at H$\alpha$\ by \citet{hen56}, and tentatively
classified as a PN with strong H$\alpha$\ emission by
\citet{lin63}. The object has been recognized as a PN by \citet{wes64}
and classified as PN of medium excitation class by \citet{san78}. This
PN could be a member of the LMC cluster 1086 SL580 \citep{kon96}. It
is resolved and has a round shape on HST images with radius of
$\sim$0\farcs33 \citep{sha06}. Its chemical composition has been
determined several times with the most recent estimation by
\citet{lei06}. \citet{gru09} also classify this object as PN. We, too,
classify this object as a PN, designating it carbon-rich, {\tt C-PN}.

\emph{{SSTISAGEMC J053441.40$-$692630.6} (SSID145)}. This object
lies very close to planetary nebula {RP~793}, with which it is
confused in the SIMBAD database. However, it is clearly a {\tt C-AGB},
and \citet{sos09b} agree.

\emph{{SHP LMC 256} (SSID146)}. \citet{san95} identify a ROSAT
X-ray source near this object (6\arcsec\ away), which may be
associated, since it lies within the ROSAT pointing accuracy. This has
also been detected by the X-ray Multi-Mirror Mission (XMM/2XMMi), with
a spectral energy distribution peaking around 1 keV. The
\emph{Spitzer} IRS spectrum is extremely unusual, and hence we
classify this object as {\tt UNK}.

\emph{{HV 2700} (SSID147)}. SIMBAD gives the spectral type as M2Iab,
based on optical spectra in the red \citep*{woo83} and blue
\citep{hum79} parts of the spectrum. This would validate our
classification of {\tt RSG}.

\emph{{SSTISAGEMC J053548.07$-$703146.6} (SSID148)}. This LPV has a
period of nearly 1\,000 days \citep{fra05,fra08,sos09b}. \citet{sri09}
class it as an oxygen-rich AGB star, with which we agree, {\tt O-AGB}.

\emph{{SSTISAGEMC J053602.36$-$674517.3} (SSID149)}. This object is
listed as a possible PN by \citet{rei06}, but it exhibits no clear
emission lines in the IRS spectrum. We classify this object as {\tt
  YSO}.

\emph{{IRAS 05370$-$7019} (SSID150)}.  This object is
associated with IRAS 05370$-$7019 and LI-LMC 1424 and regarded as an
unidentified source by \citep{lou97}. \citet{gru09} consider this
object to be a (post-)AGB object, while conversely \citet{whi08}
classify this object as a high probability YSO. We agree with
\citet{gru09}, classifying it as {\tt C-PAGB}.

\emph{{SSTISAGEMC J053634.77$-$722658.6} (SSID151)}. This object
seems unknown in the literature, and will be the subject of a paper by
Hony et al., in prep. We classify it as {\tt GAL}.

\emph{{SSTISAGEMC J053655.60$-$681124.5} (SSID152)}. This star too
is relatively unstudied in the literature, but is classified as an
oxygen-rich AGB star by \citet{sri09} and \citet{sos09b}. We concur,
{\tt O-AGB}.

\emph{{RP 493} (SSID153)}. Identified by \citet{rei06} as a small,
elliptical PN with a high degree of confidence. These authors give a
diameter of $4\farcs0$\ in H$\alpha$\ and a velocity of $v_{\rm helio}
= 326.8$\,km s$^{-1}$. We classify this object as {\tt O-PN}.

\emph{{SSTISAGEMC J053730.59$-$674041.6} (SSID154)}. This object is a
Stage I YSO candidate according to \citet{whi08}, however we classify
it as {\tt GAL}.

\emph{{KDM 5345} (SSID155)}. This optically-identified
  carbon-star \citep{kon01} exhibits carbon-rich features in its IRS
  spectrum but also oxygen-rich features. We tentatively suggest that
  this object is a mixed-chemistry AGB star, and due to this
  uncertainty only classify it as {\tt UNK}.

\emph{{OGLE J053930.16$-$695755.8} (SSID156)}. This LPV has a
period of 291.6 days according to \citet{gro04}. It is an extreme AGB
star according to \citet{sri09} and \citet{vij09}. We classify it as
{\tt C-AGB}.

\emph{{HV 12631} (SSID157)}. \citet{alc98} identified this object
as an RV~Tau star, based on MACHO light curve (MACHO 14.9582.9), and
\citet{per03} confirmed this. This star has a period of 31 days. This
strengthens our classification of {\tt O-PAGB}. \citet{sri09} consider
it to be an O-AGB star.

\emph{{SSTISAGEMC J053942.45$-$711044.5} (SSID158)}. This could
be the planetary nebula RP618, listed by \citet{rei06} and
\citet{whi08} as a possible PN. However, \citet{vij09} and
\citet{gru09} believe this to be a YSO. We also classify it as a {\tt
  YSO}.

\emph{{SSTISAGEMC J053945.40$-$665809.4} (SSID159)}. This star has
a period of 125 days according to MACHO, and is classified as an
oxygen-rich AGB star by \citet{sri09}. We agree with this
classification, {\tt O-AGB}.

\emph{{SSTISAGEMC J053949.23$-$693747.0} (SSID160)}. The IRS
spectrum towards this object is very weak, and so we classify it as
{\tt UNK}. \citet{gru09} classify this object as a probable YSO or
star using Spitzer photometry.

\emph{{MACHO 81.9728.14} (SSID161)}.  This object is classified as
a possible RV~Tau star, based on MACHO light curve \citep{alc98}.
\citet{per03} re-analysed the light curve, and they identified this as
only a potential RV~Tau star, given the paucity of the MACHO
data. They derived a period of 47\,days, as did \citet{sos08}, who
classified it as an RV~Tau star. The IRS spectrum points to the same
conclusion, despite contamination by PAH emission, hence our
classification of {\tt O-PAGB}.

\emph{{MSX LMC 949} (SSID162)}. This object was not classified
by \citet{ega01} since it is rather faint in JHK$_S$.  It is an
extreme AGB star according to \citet{sri09} and a (post-)AGB star or
star according to \citet{gru09}. We classify it as {\tt O-PAGB}.

\emph{{RP 85} (SSID163)}. Listed by \citet{rei06} as a likely PN,
but as a YSO by \citet{gru09}. \Citet{vlo10} suggest that a faint and
flat dust continuum with weak emission lines at MIPS-SED wavelengths
is strongly suggestive that {RP 85} is a PN. We agree with
\citet{gru09}, classifying this object as a {\tt YSO}.

\emph{{SSTISAGEMC J054059.31$-$704402.5}
  (SSID164)}. {SSTISAGEMC J054059.31$-$704402.5} was successfully
modelled by \citet{whi08} and \citet{oli09} as a YSO, and classified
as such by \citet{gru09}, thus supporting our classification, {\tt
  YSO}.

\emph{{MSX LMC 947} (SSID165)}. Unanimous agreement between
\citet{ega01}, \citet{vij09}, \citet{sri09} and ourselves that this
object is an {\tt O-AGB}. It has a period of $\approx$700 days
\citep{sos09b}.

\emph{{SSTISAGEMC J054114.56$-$713236.0} (SSID166)}. We classify
this object as {\tt O-AGB}, as do \citet{sri09}. It has a period of 583
days according to the MACHO database.

\emph{{IRAS 05416$-$6906 } (SSID167)}. This object is variously
classed as YSO \citep{whi08, vij09}, H{\sc ii} region \citep{ega01},
and extreme AGB star \citep{sri09}. We classify it as {\tt C-AGB}.

\emph{{IRAS 05421$-$7116} (SSID168)}. {IRAS 05421-7116} was
successfully modelled by \citet{whi08} and \citet{oli09} as a Stage I
YSO, and classified as such by \citet{gru09}, thus supporting our
classification of {\tt YSO}.

\emph{{W61 6-24} (SSID169)}. A spectral type of M was assigned by
\citet{san77}, and with a \BVcol\ value of 2.7 magnitudes this object
does appear to be of early M-type. This star is also a member of the
very young cluster {NGC 2100}, which has an age of 6--10\,Myr
\citep{wol07,hod83,san95}. Thus our classification of {\tt RSG} is
reinforced.

\emph{{NGC 2100 Robb 4} (SSID170)}. This object has V=13.58 and
\BVcol\ equal to 2.04 \citep{reb83}, which implies that it is a late
M-type star.  The CO band index (in the 2.2\,$\mu$m window) measured
by \citet{eli85} matches the values for various early M-type
supergiants observed in that paper.  Reference is made to HST WFPC2
images of the {NGC 2100} cluster to get an estimated V
magnitude for the star, from which a \VKcol\ value is given by
\citet{kel99}.  The JHK values given in that paper do not agree with
the 2MASS photometry (potentially due to variability), but the
\JKcol\ values in either case are relatively large and suggest that
the star is of late type. Estimates of cluster age are on the order of
10--30\,Myr \citep{wol07}, and the chemistry is thought to be
oxygen-rich by \citet{vlo05b}, upholding our classification of {\tt
  RSG}.

\emph{{2MASS J05420676$-$6912312} (SSID171)}. UBV photometry is
given in the PhD thesis of \citet{bra01}.  The \BVcol\ value of 1.97
implies that the star is of late M-type.  The same is true of the
J$-$K value from 2MASS or DENIS, which affirms our classification of
{\tt RSG}.

\emph{{W61 6-57} (SSID172)}. This star also has \BVcol\ values from
the \citet{bra01} thesis and infrared data from DENIS and 2MASS.  The
values \BVcol\ = 2.16 and \JKcol\ = 1.2 suggest that the star is of
late M-type.  The K-band CO index from \citet{eli85} is consistent
with an early M spectral type. A medium optical spectrum was taken of
this star, as reported by \citet{jas94}.  They do not report a
spectral type but give fitted values T$_\mathrm{eff}$ = 4\,032\,K and
log($g$) of 1.0 derived from the spectrum.  That would seem to match a
K-type giant/supergiant rather than an M-type supergiant. However,
both types are consistent with our classification of {\tt RSG}.

\emph{{WOH G 494} (SSID173)}. This star is classified as an M
giant by \citet{wes81} and as a carbon-rich AGB star by OGLE-III
\citep[although the data quality is far from
  optimal;][]{sos09b}. Assigned periods range from 170--333 days
\citep{sos09b,fra05,fra08}. \citet{sri09} classify it as O-AGB, and
this agrees with our classification, {\tt O-AGB}.

\emph{{LM 2-42} (SSID174)}. This object has been classified as
an H$\alpha$\ emission-line object by \citet{lin63} since no other
nebular lines except that of H$\alpha$\ were detected on the analyzed
plates. The nature of this object as a PN has been recognized by
\citet{san78}. This PN could be member of the LMC cluster {KMHK
  1280}/{HS 398} \citep{kon96}. It is compact and elliptical in
H$\alpha$\ images \citep[{RP 10}][]{rei06} and is resolved in
HST images \citep[][0\farcs61$\times$0\farcs45]{sha06}. Its chemical
composition has been determined several times with the most recent
estimation by \citet{lei06}. We classify it as {\tt O-PN}.

\emph{{LH$\alpha$\ 120-N 178} (SSID175)}. This object also has
been recognized as a point-like emission nebula on objective prism
photographs of the LMC at H$\alpha$\ by \citet{hen56}, and tentatively
classified as a PN with strong H$\alpha$\ emission by
\citet{lin63}. The object has been recognised as a PN by \citet{wes64}
and classified as a PN of medium excitation class by \citet{san78}. It
is resolved and has an elliptical shape (0\farcs51$\times$0\farcs45)
with possible internal structure on HST images \citep{sha06}. Its
chemical composition has been determined several times with the most
recent estimation by \citet{lei06}. \citet{gru09} also classify this
object as PN, and we specify carbon chemistry, {\tt C-PN}.

\emph{{SSTISAGEMC J054254.38$-$700807.4} (SSID176)}. \citet{sos09b}
and \citet{sri09} both classify this object as O-AGB, as do we, {\tt
  O-AGB}.

\emph{{SSTISAGE1C J054310.86$-$672728.0} (SSID177)}. This object
appears very close to the position of {IRAS F05432$-$6728},
which has F$_{12}$=88 and F$_{25}$=136\,mJy, in reasonable agreement
with the \emph{Spitzer} IRS spectrum. However, F$_{60}$=1\,916 and
F$_{100}$=3\,873\,mJy for this object, which would be unusual for an
oxygen-rich post-AGB object, and does not fit with the long-wavelength
part of the IRS spectrum. AKARI recently detected an object in a
nearby position with F$_{9}$=75 and F$_{18}$=133 \citep{ish10}, also
in excellent agreement with the IRS spectrum. \citet{gru09} suggest
that this object is in the (post-)AGB phase, and that is our
conclusion from the IRS spectrum ({\tt O-PAGB}). SSID177 is probably
not related to the faint IRAS source, but potentially to the AKARI
source, 0543108-672730.

\emph{{SSTISAGEMC J054314.12-703835.1} (SSID178)}. This object is
located close to the 130--380\,Myr \citep{fra88} cluster {NGC 2107},
but there are no indications that this object is a part of it. The
MACHO and OGLE-III catalogs list a period of $\sim$160 days, and
OGLE-III classifies this object as a C-rich AGB star
\citep{sos09b}. However, an SiO feature is clearly visible in the IRS
spectrum, which leads us to classify this object as {\tt
  O-AGB}. \citet{sri09} agree.

\emph{{KDM 5841} (SSID179)}. This object shows near-infrared C$_2$
bands \citep{kon01} and has \emph{Spitzer} colours similar to that of a
carbon-rich AGB star \citep{sri09}, and thus is confirmed as {\tt
  C-AGB}.

\emph{{SSTISAGEMC J054406.01$-$683753.6} (SSID180)}. \citet{sri09}
class this as an O-rich AGB star, and it has periods in the range
239--554 days \citep{fra05,fra08,sos09b}. We also classify it as {\tt
  O-AGB}.

\emph{{SSTISAGEMC J054437.87$-$673657.7} (SSID181)}. According to
\citet{sri09} and \citet{vij09} this object is an extreme AGB star. We
classify it as {\tt C-AGB}.

\emph{{SSTISAGEMC J054440.11$-$691149.0} (SSID182)}. We classify
this object as {\tt O-AGB}, in agreement with \citet{sri09} and the
99-day period of \citet{sos09b}.

\emph{{IRAS 05452$-$6924} (SSID183)}. The SED of {IRAS
  05452-6924} was fitted by \citet{whi08} with a YSO model, and the
ice features in this object's IRS spectrum were modelled by
\citet{oli09}. \citet{gru09} are also in agreement. This confirms our
classification of {\tt YSO}.

\emph{{SSTISAGEMC J054524.23$-$683041.4} (SSID184)}. This is a YSO
candidate according to \citet{whi08}, whilst \citet{gru09} classify it
as a galaxy. We, too, classify it as {\tt GAL}.

\emph{{SSTISAGEMC J054546.32$-$673239.4} (SSID185)}. This is a YSO
candidate according to \citet{whi08}, but we classify it as {\tt
  O-AGB}.

\emph{{LH$\alpha$\ 120$-$N 170} (SSID186)}. This object also
has been recognized as a point-like emission nebula on objective prism
photographs of the LMC at H$\alpha$\ by \citet{hen56}, and tentatively
classified as PN with strong H$\alpha$\ emission by \citet{lin63}. The
object has been recognized as a PN by \citet{wes64} and classified as
a PN of high excitation class by \citet{san78}. This PN could be a
member of LMC cluster KMHK 1364 \citep{kon96}. It is resolved and has
an elliptical shape (0\farcs62$\times$0\farcs54) with internal
structure on HST images \citep{sha06}. Its chemical composition has
been determined several times with the most recent estimation by
\citet{lei06}. \citet{gru09} also agree that this object is a PN. This
corroborates our classification of {\tt O-PN}.

\emph{{SSTISAGEMC J054745.79$-$680734.1} (SSID187)}. This object
has an extremely long period of 2\,599 days according to
\citet{fra05,fra08}. It is a YSO candidate according to \citet{whi08},
and this agrees with our classification, {\tt YSO}.

\emph{{KDM 6247} (SSID188)}. This object appears in the carbon star
catalog of \citet{kon01}, based on the detection of Swan C$_2$
bands. \citet{san77} also include it in their catalog of carbon
stars. \citet{sri09}, however, class {KDM 6247} as an O-AGB. No
distinguishing features are seen in the IRS spectrum, and so we class
this object as {\tt STAR}.

\emph{{NGC 2121 LE 6} (SSID189)}. This object is a member of the
cluster {NGC 2121}, which has an age of $\sim$3.2\,Gyr
\citep{mac03}, and is thought to host carbon stars \citep{vlo05b}. An
extreme AGB according to \citet{sri09}, we class it as {\tt C-AGB}.

\emph{{IRAS 05495$-$7034} (SSID190)}. \citet{whi08} consider this
object to be a high-probability YSO, whereas \citet{gru08} class it as
an ERO. Our classification is in alignment with \citet{gru08}, {\tt
  C-AGB}.

\emph{{KDM 6486} (SSID191)}. This object is a part of the carbon
star catalog of \citet{kon01}, and part of the C-AGB grouping of
\citet{sri09} which fortifies our classification of {\tt C-AGB}.

\emph{{HV 2862} (SSID192)}.  This object is known to be a variable
star \citep{lin74}, with a period of 34 days
\citep{fra05,fra08}. \citet{sos08} have classified this star as an
RV~Tau star, which supports our classification of {\tt
  O-PAGB}. \citet{gru09}, however, class this as a star.

\emph{{SSTISAGEMC J055143.27$-$684543.0} (SSID193)}. According to
\citet{whi08}, this is a high-probability Stage I YSO. However, we
disagree and classify this object as {\tt GAL}. There is a radio
source located within 3$\arcmin$ \citep[MDM 111;][]{mar97}.

\emph{{PMP 337} (SSID194)}. This chemically peculiar star is a
member of {NGC 2136/7}, which has an age of 100\,Myr
\citep{pau06}. \citet{sri09} class this as a C-AGB, and we agree with
that classification, {\tt C-AGB}.

\emph{{PMP 133} (SSID195)}. This is a chemically peculiar star
included in the catalog of \citet{pau06}. As such, this makes it a
low- to intermediate-mass star. This agrees with our classification of
{\tt STAR}.

\emph{{IRAS 05537$-$7015} (SSID196)}. Mentioned in \citet{lou97} as
a good candidate for an evolved star, \citet{gru09} label this as
(post-)AGB. We agree, classifying this object as {\tt C-PAGB}.

\emph{{SSTISAGEMC J060053.62$-$680038.8} (SSID197)}. This star has
periods of 281 and 887 days \citep{fra05,fra08} and is classified as
an oxygen-rich AGB star by \citet{sri09}. We agree, with our
classification of {\tt O-AGB}.

\bsp

\label{lastpage}

\end{document}

The complete Table 3 follows.

\begin{deluxetable}{cccclllcc}
\rotate
\tabletypesize{\footnotesize}
\tablewidth{675pt}
\tablenum{3}
\tablecaption{Classification of point sources targeted in IRS staring mode as part of the \SSp\ program. \label{tab:psclass}}
\tablehead{ & & \colhead{Observed} & \colhead{SSTISAGEMC} & \colhead{MACHO} & \colhead{MACHO\tablenotemark{a}} & \colhead{OGLE-LMC} & \colhead{OGLE\tablenotemark{b}}\\
\colhead{SSID}&\colhead{Name\tablenotemark{c}} & \colhead{R.~A. \& Dec. (J2000)} &\colhead{designation\tablenotemark{d}} &\colhead{designation} & \colhead{period (d)} & \colhead{designation} & \colhead{period (d)} & \colhead{Classification}}
\startdata
1	&	 NGC 1651 SAGE IRS 1	&	  04$^\mathrm{h}$37$^\mathrm{m}$21.15$^\mathrm{s}$	 $-$70\degr34\arcmin44.57\arcsec 	&	J043721.15$-$703444.7 	&	\ldots	&	\ldots	&	LPV-00749	&	101	&	O-AGB	\\
2	&	\ldots	&	  04$^\mathrm{h}$37$^\mathrm{m}$27.69$^\mathrm{s}$	 $-$67\degr54\arcmin34.94\arcsec 	&	J043727.61$-$675435.1 	&	\ldots	&	\ldots	&	\ldots	&	\ldots	&	GAL	\\
3	&	\ldots	&	  04$^\mathrm{h}$46$^\mathrm{m}$27.15$^\mathrm{s}$	 $-$68\degr47\arcmin46.83\arcsec 	&	J044627.10$-$684747.0 	&	45.909.2	&	386--397	&	LPV-02759	&	389	&	C-AGB	\\
4	&	\ldots	&	  04$^\mathrm{h}$47$^\mathrm{m}$18.63$^\mathrm{s}$	 $-$69\degr42\arcmin20.53\arcsec 	&	J044718.63$-$694220.6 	&	\ldots	&	\ldots	&	\ldots	&	\ldots	&	RSG	\\
5	&	\ldots	&	  04$^\mathrm{h}$48$^\mathrm{m}$37.75$^\mathrm{s}$	 $-$69\degr23\arcmin36.85\arcsec 	&	J044837.77$-$692337.0 	&	\ldots	&	\ldots	&	\ldots	&	\ldots	&	YSO-2	\\
6	&	\ldots	&	  04$^\mathrm{h}$49$^\mathrm{m}$34.38$^\mathrm{s}$	 $-$69\degr05\arcmin49.17\arcsec 	&	J044934.31$-$690549.3 	&	\ldots	&	\ldots	&	\ldots	&	\ldots	&	O-AGB	\\
7	&	 MSX LMC 1128	&	  04$^\mathrm{h}$50$^\mathrm{m}$40.57$^\mathrm{s}$	 $-$68\degr58\arcmin18.76\arcsec 	&	J045040.52$-$685819.0 	&	45.1632.952	&	437--445	&	LPV-04314	&	441	&	C-AGB	\\
8	&	\ldots	&	  04$^\mathrm{h}$51$^\mathrm{m}$28.58$^\mathrm{s}$	 $-$69\degr55\arcmin49.92\arcsec 	&	J045128.58$-$695550.1 	&	44.1739.170	&	907--911	&	LPV-04685	&	884	&	O-AGB	\\
9	&	 IRAS 04518$-$6852	&	  04$^\mathrm{h}$51$^\mathrm{m}$40.63$^\mathrm{s}$	 $-$68\degr47\arcmin34.82\arcsec 	&	J045140.57$-$684734.6 	&	\ldots	&	\ldots	&	\ldots	&	\ldots	&	C-AGB	\\
10	&	\ldots	&	  04$^\mathrm{h}$52$^\mathrm{m}$00.38$^\mathrm{s}$	 $-$69\degr18\arcmin05.53\arcsec 	&	J045200.36$-$691805.6 	&	\ldots	&	\ldots	&	\ldots	&	\ldots	&	YSO-2	\\
11	&	\ldots	&	  04$^\mathrm{h}$52$^\mathrm{m}$28.66$^\mathrm{s}$	 $-$68\degr54\arcmin51.09\arcsec 	&	J045228.68$-$685451.3 	&	\ldots	&	\ldots	&	\ldots	&	\ldots	&	YSO-1	\\
12	&	 KDM 764	&	  04$^\mathrm{h}$52$^\mathrm{m}$32.49$^\mathrm{s}$	 $-$67\degr02\arcmin59.30\arcsec 	&	J045232.54$-$670259.2 	&	48.2024.11	&	250--873	&	LPV-05238	&	445	&	C-AGB	\\
13	&	\ldots	&	  04$^\mathrm{h}$53$^\mathrm{m}$09.54$^\mathrm{s}$	 $-$68\degr17\arcmin10.11\arcsec 	&	J045309.39$-$681710.8 	&	47.2127.16	&	207--1122	&	LPV-05538	&	915	&	O-AGB	\\
14	&	 IRAS F04532$-$6709	&	  04$^\mathrm{h}$53$^\mathrm{m}$11.03$^\mathrm{s}$	 $-$67\degr03\arcmin55.96\arcsec 	&	MA J045311.08$-$670355.6 	&	\ldots	&	\ldots	&	\ldots	&	\ldots	&	YSO-1	\\
15	&	\ldots	&	  04$^\mathrm{h}$53$^\mathrm{m}$28.71$^\mathrm{s}$	 $-$66\degr03\arcmin34.76\arcsec 	&	J045328.70$-$660334.4 	&	54.2160.204	&	2710	&	\ldots	&	\ldots	&	C-AGB	\\
16	&	GV 60	&	  04$^\mathrm{h}$53$^\mathrm{m}$30.86$^\mathrm{s}$	 $-$69\degr17\arcmin49.85\arcsec 	&	J045330.88$-$691749.7 	&	\ldots	&	\ldots	&	\ldots	&	\ldots	&	RSG	\\
17	&	LH$\alpha$ 120$-$N 82	&	  04$^\mathrm{h}$53$^\mathrm{m}$30.86$^\mathrm{s}$	 $-$69\degr17\arcmin49.85\arcsec 	&	2C J045330.15$-$691749.5	&	\ldots	&	\ldots	&	\ldots	&	\ldots	& OTHER (W-R)	\\
18	&	\ldots	&	  04$^\mathrm{h}$53$^\mathrm{m}$44.28$^\mathrm{s}$	 $-$66\degr11\arcmin45.76\arcsec 	&	J045344.24$-$661146.0 	&	\ldots	&	\ldots	&	\ldots	&	\ldots	&	C-AGB	\\
19	&	\ldots	&	  04$^\mathrm{h}$54$^\mathrm{m}$22.88$^\mathrm{s}$	 $-$70\degr26\arcmin56.64\arcsec 	&	J045422.82$-$702657.0 	&	\ldots	&	\ldots	&	\ldots	&	\ldots	&	STAR	\\
20	&	\ldots	&	  04$^\mathrm{h}$55$^\mathrm{m}$26.76$^\mathrm{s}$	 $-$68\degr25\arcmin07.93\arcsec 	&	J045526.69$-$682508.4 	&	\ldots	&	\ldots	&	\ldots	&	\ldots	&	YSO-3	\\
21	&	\ldots	&	  04$^\mathrm{h}$55$^\mathrm{m}$34.06$^\mathrm{s}$	 $-$65\degr57\arcmin00.92\arcsec 	&	J045534.07$-$655701.3 	&	\ldots	&	\ldots	&	\ldots	&	\ldots	&	YSO-4	\\
22	&	\ldots	&	  04$^\mathrm{h}$56$^\mathrm{m}$23.27$^\mathrm{s}$	 $-$69\degr27\arcmin48.05\arcsec 	&	J045623.21$-$692749.0 	&	17.2593.214	&	undef.	&	\ldots	&	\ldots	&	O-AGB	\\
23	&	 KDM 1238	&	  04$^\mathrm{h}$58$^\mathrm{m}$55.03$^\mathrm{s}$	 $-$69\degr11\arcmin18.21\arcsec 	&	J045855.02$-$691118.7 	&	18.2960.23	&	376--543	&	LPV-09687	&	133	&	C-AGB	\\
24	&	 RP 1805	&	  04$^\mathrm{h}$58$^\mathrm{m}$55.29$^\mathrm{s}$	 $-$68\degr50\arcmin36.13\arcsec 	&	---	&	\ldots	&	\ldots	&	\ldots	&	\ldots	&	UNK	\\
25	&	\ldots	&	  05$^\mathrm{h}$00$^\mathrm{m}$32.59$^\mathrm{s}$	 $-$66\degr21\arcmin12.60\arcsec 	&	J050032.61$-$662113.0 	&	\ldots	&	\ldots	&	\ldots	&	\ldots	&	YSO-4	\\
26	&	 RP 1631	&	  05$^\mathrm{h}$00$^\mathrm{m}$34.69$^\mathrm{s}$	 $-$70\degr52\arcmin00.34\arcsec 	&	J050034.61$-$705200.4 	&	\ldots	&	\ldots	&	\ldots	&	\ldots	&	OTHER (RCrB)	\\
27	&	 MSX LMC 1271	&	  05$^\mathrm{h}$02$^\mathrm{m}$21.52$^\mathrm{s}$	 $-$66\degr06\arcmin37.98\arcsec 	&	J050221.46$-$660638.3 	&	\ldots	&	\ldots	&	\ldots	&	\ldots	&	RSG	\\
28	&                	&	  05$^\mathrm{h}$02$^\mathrm{m}$24.21$^\mathrm{s}$	 $-$66\degr06\arcmin37.46\arcsec 	&	J050224.17$-$660637.4 	&	\ldots	&	\ldots	&	\ldots	&	\ldots	&	O-PAGB	\\
29	&	 HV 2281	&	  05$^\mathrm{h}$03$^\mathrm{m}$04.98$^\mathrm{s}$	 $-$68\degr40\arcmin24.90\arcsec 	&	J050304.95$-$684024.7 	&	19.3694.19	&	32	&	T2CEP-029	&	31	&	O-PAGB (RV Tau)	\\
30	&	 KDM 1656	&	  05$^\mathrm{h}$03$^\mathrm{m}$16.59$^\mathrm{s}$	 $-$65\degr49\arcmin44.79\arcsec 	&	J050316.60$-$654945.1 	&	55.3736.12	&	1035	&	\ldots	&	\ldots	&	C-AGB	\\
31	&	 KDM 1691	&	  05$^\mathrm{h}$03$^\mathrm{m}$36.89$^\mathrm{s}$	 $-$68\degr33\arcmin38.71\arcsec 	&	J050336.92$-$683338.5 	&	19.3817.12	&	503--512	&	LPV-14408	&	519	&	C-AGB	\\
32	&	 LMC$-$BM 11$-$19	&	  05$^\mathrm{h}$03$^\mathrm{m}$42.54$^\mathrm{s}$	 $-$67\degr59\arcmin18.83\arcsec 	&	J050342.57$-$675919.2 	&	24.3825.2539	&	894--1082	&	LPV-14513	&		&	STAR	\\
33	&	 LMC$-$BM 12$-$14	&	  05$^\mathrm{h}$03$^\mathrm{m}$53.50$^\mathrm{s}$	 $-$70\degr27\arcmin47.53\arcsec 	&	J050353.40$-$702747.6 	&	23.3788.9	&	405--894	&	LPV-14750	&	220	&	C-AGB	\\
34	&	\ldots	&	  05$^\mathrm{h}$03$^\mathrm{m}$54.60$^\mathrm{s}$	 $-$67\degr18\arcmin47.69\arcsec 	&	J050354.55$-$671848.7 	&	\ldots	&	\ldots	&	\ldots	&	\ldots	&	YSO-1	\\
35	&	 NGC 1818 WBT 5	&	  05$^\mathrm{h}$04$^\mathrm{m}$07.38$^\mathrm{s}$	 $-$66\degr26\arcmin42.71\arcsec 	&	J050407.42$-$662643.0 	&	53.3848.14	&	360--1707	&	\ldots	&	\ldots	&	RSG	\\
36	&	 NGC 1818 WBT 64	&	  05$^\mathrm{h}$04$^\mathrm{m}$07.73$^\mathrm{s}$	 $-$66\degr25\arcmin05.48\arcsec 	&	J050407.72$-$662505.9 	&	53.3849.19	&	831--848	&	\ldots	&	\ldots	&	C-AGB	\\
37	&	 NGC 1818 WBT 3	&	  05$^\mathrm{h}$04$^\mathrm{m}$11.09$^\mathrm{s}$	 $-$66\degr26\arcmin16.70\arcsec 	&	J050411.04$-$662616.8 	&	53.3848.13	&	360	&	\ldots	&	\ldots	&	RSG	\\
38	&	 MSX LMC 61	&	  05$^\mathrm{h}$04$^\mathrm{m}$28.91$^\mathrm{s}$	 $-$67\degr41\arcmin23.43\arcsec 	&	J050428.91$-$674123.9 	&	25.3951.67	&	576--1585	&	LPV-15419	&	577	&	O-AGB	\\
39	&	 RP 1878	&	  05$^\mathrm{h}$04$^\mathrm{m}$34.17$^\mathrm{s}$	 $-$67\degr52\arcmin21.05\arcsec 	&	J050434.20$-$675221.8 	&	\ldots	&	\ldots	&	\ldots	&	\ldots	&	UNK	\\
40	&	 IRAS 05047$-$6642	&	  05$^\mathrm{h}$04$^\mathrm{m}$51.71$^\mathrm{s}$	 $-$66\degr38\arcmin07.41\arcsec 	&	J050451.70$-$663807.5 	&	53.3966.918	&	20	&	\ldots	&	\ldots	&	YSO-3	\\
41	&	\ldots	&	  05$^\mathrm{h}$05$^\mathrm{m}$03.22$^\mathrm{s}$	 $-$69\degr24\arcmin26.51\arcsec 	&	J050503.21$-$692426.5 	&	\ldots	&	\ldots	&	LPV-16169	&	217	&	C-AGB	\\
42	&	\ldots	&	  05$^\mathrm{h}$05$^\mathrm{m}$17.19$^\mathrm{s}$	 $-$69\degr21\arcmin57.12\arcsec 	&	J050517.08$-$692157.0 	&	\ldots	&	\ldots	&	\ldots	&	\ldots	&	GAL	\\
43	&	 LMC$-$BM 13$-$2	&	  05$^\mathrm{h}$05$^\mathrm{m}$55.74$^\mathrm{s}$	 $-$67\degr22\arcmin09.24\arcsec 	&	J050555.66$-$672210.0 	&	52.4197.4218	&	2545	&	LPV-17258	&	206	&	C-AGB	\\
44	&	\ldots	&	  05$^\mathrm{h}$05$^\mathrm{m}$58.26$^\mathrm{s}$	 $-$68\degr09\arcmin23.79\arcsec 	&	J050558.23$-$680923.6 	&	19.4186.889	&	undef.	&	\ldots	&	\ldots	&	YSO-4	\\
45	&	\ldots	&	  05$^\mathrm{h}$06$^\mathrm{m}$07.51$^\mathrm{s}$	 $-$71\degr41\arcmin48.12\arcsec 	&	J050607.50$-$714148.4 	&	38.4132.5	&	346	&	\ldots	&	\ldots	&	C-AGB	\\
46	&	 KDM 1961	&	  05$^\mathrm{h}$06$^\mathrm{m}$12.61$^\mathrm{s}$	 $-$64\degr55\arcmin37.23\arcsec 	&	J050612.59$-$645537.5 	&	\ldots	&	\ldots	&	\ldots	&	\ldots	&	C-AGB	\\
47	&	 KDM 1966	&	  05$^\mathrm{h}$06$^\mathrm{m}$18.95$^\mathrm{s}$	 $-$64\degr56\arcmin10.88\arcsec 	&	J050618.98$-$645610.2 	&	\ldots	&	\ldots	&	\ldots	&	\ldots	&	C-AGB	\\
48	&	\ldots	&	  05$^\mathrm{h}$06$^\mathrm{m}$20.13$^\mathrm{s}$	 $-$64\degr54\arcmin58.01\arcsec 	&	J050620.12$-$645458.6 	&	\ldots	&	\ldots	&	\ldots	&	\ldots	&	C-AGB	\\
49	&	\ldots	&	  05$^\mathrm{h}$06$^\mathrm{m}$29.62$^\mathrm{s}$	 $-$68\degr55\arcmin34.54\arcsec 	&	J050629.61$-$685534.9 	&	1.4174.40	&	261--267	&	LPV-17926	&	154	&	C-AGB	\\
50	&	\ldots	&	  05$^\mathrm{h}$06$^\mathrm{m}$39.24$^\mathrm{s}$	 $-$68\degr22\arcmin09.32\arcsec 	&	J050639.14$-$682209.3 	&	\ldots	&	\ldots	&	\ldots	&	\ldots	&	OTHER (BSG)	\\
51	&	 SHV 0507252$-$690238	&	  05$^\mathrm{h}$07$^\mathrm{m}$09.47$^\mathrm{s}$	 $-$68\degr58\arcmin50.18\arcsec 	&	J050707.81$-$685852.7	&	1.4294.22	&	undef.	&	LPV-18832	&	507	&	C-AGB	\\
52	&	\ldots	&	  05$^\mathrm{h}$07$^\mathrm{m}$14.00$^\mathrm{s}$	 $-$67\degr48\arcmin46.46\arcsec 	&	J050713.90$-$674846.7 	&	\ldots	&	\ldots	&	\ldots	&	\ldots	&	C-PAGB	\\
53	&	\ldots	&	  05$^\mathrm{h}$07$^\mathrm{m}$53.01$^\mathrm{s}$	 $-$68\degr12\arcmin46.38\arcsec 	&	J050752.93$-$681246.5 	&	19.4427.193	&	335--335	&	LPV-19934	&	302	&	C-AGB	\\
54	&	\ldots	&	  05$^\mathrm{h}$07$^\mathrm{m}$59.36$^\mathrm{s}$	 $-$68\degr39\arcmin25.71\arcsec 	&	J050759.35$-$683925.8 	&	19.4541.306	&	359--362	&	LPV-20082	&	150	&	O-AGB	\\
55	&	\ldots	&	  05$^\mathrm{h}$08$^\mathrm{m}$26.27$^\mathrm{s}$	 $-$68\degr31\arcmin15.01\arcsec 	&	J050826.35$-$683115.1 	&	19.4543.3947	&	undef.	&	LPV-20709	&	400	&	C-AGB	\\
56	&	\ldots	&	  05$^\mathrm{h}$08$^\mathrm{m}$30.62$^\mathrm{s}$	 $-$69\degr22\arcmin37.39\arcsec 	&	J050830.51$-$692237.4 	&	\ldots	&	\ldots	&	LPV-20807	&	115	&	O-PAGB	\\
57	&	 KDM 2187	&	  05$^\mathrm{h}$08$^\mathrm{m}$36.42$^\mathrm{s}$	 $-$69\degr43\arcmin15.11\arcsec 	&	J050836.39$-$694315.7 	&	5.4525.42	&	1229--1361	&	LPV-20969	&	758	&	C-AGB	\\
58	&	 BMB$-$BW 180	&	  05$^\mathrm{h}$09$^\mathrm{m}$26.44$^\mathrm{s}$	 $-$69\degr06\arcmin56.99\arcsec 	&	J050926.57$-$690656.3 	&	79.4655.3986	&	69--69	&	LPV-22320	&	68	&	O-AGB	\\
59	&	 NGC 1856 SAGE IRS 1	&	  05$^\mathrm{h}$09$^\mathrm{m}$29.54$^\mathrm{s}$	 $-$69\degr07\arcmin50.90\arcsec 	&	J050929.53$-$690750.3 	&	\ldots	&	\ldots	&	\ldots	&	\ldots	&	O-AGB	\\
60	&	\ldots	&	  05$^\mathrm{h}$10$^\mathrm{m}$28.38$^\mathrm{s}$	 $-$68\degr44\arcmin31.44\arcsec 	&	J051029.27$-$684431.2 	&	\ldots	&	\ldots	&	LPV-24027	&	465	&	C-AGB	\\
61	&	\ldots	&	  05$^\mathrm{h}$10$^\mathrm{m}$59.06$^\mathrm{s}$	 $-$68\degr56\arcmin13.82\arcsec 	&	J051059.07$-$685613.7 	&	79.5021.10	&	175--280	&	LPV-24921	&	169	&	O-AGB	\\
62	&	 MSX LMC 209	&	  05$^\mathrm{h}$12$^\mathrm{m}$09.19$^\mathrm{s}$	 $-$71\degr06\arcmin49.52\arcsec 	&	J051209.02$-$710649.7 	&	\ldots	&	\ldots	&	LPV-26931	&	28	&	YSO-4	\\
63	&	\ldots	&	  05$^\mathrm{h}$12$^\mathrm{m}$13.57$^\mathrm{s}$	 $-$68\degr39\arcmin22.47\arcsec 	&	J051213.54$-$683922.8 	&	2.5146.30	&	846--847	&	LPV-27065	&	111	&	O-AGB	\\
64	&	\ldots	&	  05$^\mathrm{h}$12$^\mathrm{m}$28.17$^\mathrm{s}$	 $-$69\degr07\arcmin56.15\arcsec 	&	J051228.19$-$690755.8 	&	79.5260.1862	&	undef.	&	\ldots	&	\ldots	&	C-PAGB	\\
65	&	 IRAS 05133$-$6937	&	  05$^\mathrm{h}$13$^\mathrm{m}$01.80$^\mathrm{s}$	 $-$69\degr33\arcmin51.21\arcsec 	&	J051301.75$-$693351.0 	&	\ldots	&	\ldots	&	\ldots	&	\ldots	&	C-AGB	\\
66	&	 OGLE J051306.52$-$690946.4	&	  05$^\mathrm{h}$13$^\mathrm{m}$06.43$^\mathrm{s}$	 $-$69\degr09\arcmin46.53\arcsec 	&	J051306.49$-$690955.3 	&	79.5259.938	&	undef.	&	LPV-28579	&	356	&	C-AGB	\\
67	&	\ldots	&	  05$^\mathrm{h}$13$^\mathrm{m}$39.87$^\mathrm{s}$	 $-$66\degr38\arcmin52.70\arcsec 	&	J051339.94$-$663852.5 	&	58.5418.23	&	831--843	&	LPV-29579	&	742	&	O-AGB	\\
68	&	 NGC 1866 Robb B136	&	  05$^\mathrm{h}$13$^\mathrm{m}$41.42$^\mathrm{s}$	 $-$65\degr28\arcmin27.91\arcsec 	&	J051341.40$-$652828.2 	&	59.5436.11	&	122--122	&	\ldots	&	\ldots	&	O-AGB	\\
69	&	 BSDL 923	&	  05$^\mathrm{h}$13$^\mathrm{m}$42.83$^\mathrm{s}$	 $-$67\degr24\arcmin10.44\arcsec 	&	J051342.63$-$672409.9 	&	\ldots	&	\ldots	&	\ldots	&	\ldots	&	OTHER (BSG)	\\
70	&	\ldots	&	  05$^\mathrm{h}$13$^\mathrm{m}$47.79$^\mathrm{s}$	 $-$69\degr35\arcmin05.06\arcsec 	&	J051347.72$-$693505.2 	&	\ldots	&	\ldots	&	\ldots	&	\ldots	&	YSO-1	\\
71	&	\ldots	&	  05$^\mathrm{h}$13$^\mathrm{m}$48.33$^\mathrm{s}$	 $-$67\degr05\arcmin26.87\arcsec 	&	J051348.38$-$670527.0 	&	\ldots	&	\ldots	&	\ldots	&	\ldots	&	HII	\\
72	&	\ldots	&	  05$^\mathrm{h}$14$^\mathrm{m}$12.32$^\mathrm{s}$	 $-$68\degr50\arcmin58.29\arcsec 	&	J051412.33$-$685058.0 	&	79.5506.20	&	694	&	LPV-30640	&	795	&	O-AGB	\\
73	&	 HV 915	&	  05$^\mathrm{h}$14$^\mathrm{m}$18.15$^\mathrm{s}$	 $-$69\degr12\arcmin35.06\arcsec 	&	J051418.09$-$691234.9 	&	79.5501.13	&	97	&	T2CEP-067	&	48	&	O-PAGB (RV Tau)	\\
74	&	\ldots	&	  05$^\mathrm{h}$14$^\mathrm{m}$49.41$^\mathrm{s}$	 $-$67\degr12\arcmin22.24\arcsec 	&	J051449.43$-$671221.4 	&	\ldots	&	\ldots	&	\ldots	&	\ldots	&	YSO-1	\\
75	&	\ldots	&	  05$^\mathrm{h}$14$^\mathrm{m}$53.12$^\mathrm{s}$	 $-$69\degr17\arcmin23.70\arcsec 	&	J051453.10$-$691723.5 	&	\ldots	&	\ldots	&	\ldots	&	\ldots	&	O-PAGB	\\
76	&	\ldots	&	  05$^\mathrm{h}$15$^\mathrm{m}$26.47$^\mathrm{s}$	 $-$67\degr51\arcmin26.91\arcsec 	&	J051526.44$-$675126.9 	&	16.5763.20	&	441	&	LPV-33019	&	107	&	STAR	\\
77	&	\ldots	&	  05$^\mathrm{h}$16$^\mathrm{m}$12.52$^\mathrm{s}$	 $-$70\degr49\arcmin30.18\arcsec 	&	J051612.42$-$704930.3 	&	13.5840.19	&	746--776	&	LPV-34633	&	768	&	O-AGB	\\
78	&	 IRAS 05170$-$7156	&	  05$^\mathrm{h}$16$^\mathrm{m}$18.69$^\mathrm{s}$	 $-$71\degr53\arcmin59.21\arcsec 	&	J051618.69$-$715359.0 	&	\ldots	&	\ldots	&	\ldots	&	\ldots	&	UNK	\\
79	&	\ldots	&	  05$^\mathrm{h}$17$^\mathrm{m}$47.16$^\mathrm{s}$	 $-$68\degr18\arcmin42.64\arcsec 	&	J051747.18$-$681842.6 	&	\ldots	&	\ldots	&	LPV-37798	&	80	&	O-AGB	\\
80	&	\ldots	&	  05$^\mathrm{h}$18$^\mathrm{m}$03.28$^\mathrm{s}$	 $-$68\degr49\arcmin50.29\arcsec 	&	1C J051803.28$-$684950.6 	&	\ldots	&	\ldots	&	LPV-38325	&	349	&	C-AGB	\\
81	&	 KDM 3196	&	  05$^\mathrm{h}$18$^\mathrm{m}$07.94$^\mathrm{s}$	 $-$71\degr51\arcmin53.66\arcsec 	&	J051807.93$-$715153.7 	&	37.6066.16	&	867--1198	&	LPV-38462	&	118	&	STAR	\\
82	&	 HV 5715	&	  05$^\mathrm{h}$18$^\mathrm{m}$11.05$^\mathrm{s}$	 $-$67\degr26\arcmin48.92\arcsec 	&	J051811.08$-$672648.5 	&	49.6132.10	&	416--886	&	\ldots	&	\ldots	&	O-AGB	\\
83	&	\ldots	&	  05$^\mathrm{h}$18$^\mathrm{m}$32.63$^\mathrm{s}$	 $-$69\degr25\arcmin25.59\arcsec 	&	J051832.64$-$692525.5 	&	78.6224.39	&	256--269	&	LPV-39261	&	171	&	C-AGB	\\
84	&	 IRAS F05192$-$7008	&	  05$^\mathrm{h}$18$^\mathrm{m}$45.27$^\mathrm{s}$	 $-$70\degr05\arcmin34.70\arcsec 	&	J051845.23$-$700534.5 	&	78.6214.72	&	undef.	&	LPV-39671	&	39	&	C-PAGB	\\
85	&	 HV 2444	&	  05$^\mathrm{h}$18$^\mathrm{m}$45.46$^\mathrm{s}$	 $-$69\degr03\arcmin21.65\arcsec 	&	J051845.47$-$690321.8 	&	\ldots	&	\ldots	&	T2CEP-091	&	36	&	O-PAGB (RV Tau)	\\
86	&	\ldots	&	  05$^\mathrm{h}$19$^\mathrm{m}$08.52$^\mathrm{s}$	 $-$69\degr23\arcmin14.44\arcsec 	&	J051908.46$-$692314.3 	&	80.6345.4422	&	351--398	&	LPV-40384	&	188	&	C-AGB	\\
87	&	 2MASS J05191049$-$6933453	&	  05$^\mathrm{h}$19$^\mathrm{m}$10.63$^\mathrm{s}$	 $-$69\degr33\arcmin46.51\arcsec 	&	J051910.49$-$693345.3 	&	78.6342.4688	&	undef.	&	LPV-40465	&	481	&	C-AGB	\\
88	&	 2MASS J05194483$-$6929594	&	  05$^\mathrm{h}$19$^\mathrm{m}$44.87$^\mathrm{s}$	 $-$69\degr29\arcmin59.84\arcsec 	&	J051944.81$-$692959.4 	&	78.6343.28	&	115--1081	&	LPV-41591	&	107	&	STAR	\\
89	&	\ldots	&	  05$^\mathrm{h}$20$^\mathrm{m}$14.31$^\mathrm{s}$	 $-$70\degr29\arcmin31.33\arcsec 	&	J052014.24$-$702931.0 	&	6.6450.20	&	423--624	&	LPV-42635	&	686	&	O-AGB	\\
90	&	\ldots	&	  05$^\mathrm{h}$20$^\mathrm{m}$23.97$^\mathrm{s}$	 $-$69\degr54\arcmin23.08\arcsec 	&	J052023.97$-$695423.2 	&	78.6458.125	&	undef.	&	\ldots	&	\ldots	&	YSO-4	\\
91	&	\ldots	&	  05$^\mathrm{h}$20$^\mathrm{m}$51.86$^\mathrm{s}$	 $-$69\degr34\arcmin08.04\arcsec 	&	J052051.83$-$693407.6 	&	78.6584.36	&	732--770	&	LPV-43982	&	779	&	O-AGB	\\
92	&	 LH$\alpha$ 120$-$N 125	&	  05$^\mathrm{h}$20$^\mathrm{m}$52.44$^\mathrm{s}$	 $-$70\degr09\arcmin35.60\arcsec 	&	J052052.42$-$700935.5 	&	\ldots	&	\ldots	&	\ldots	&	\ldots	&	C-PN	\\
93	&	\ldots	&	  05$^\mathrm{h}$21$^\mathrm{m}$01.71$^\mathrm{s}$	 $-$69\degr14\arcmin17.07\arcsec 	&	J052101.66$-$691417.5 	&	80.6589.15	&	130--217	&	LPV-44309	&	92	&	O-AGB	\\
94	&	 HV 942	&	  05$^\mathrm{h}$21$^\mathrm{m}$47.99$^\mathrm{s}$	 $-$70\degr09\arcmin57.22\arcsec 	&	J052147.95$-$700957.0 	&	6.6696.60	&	946	&	\ldots	&	\ldots	&	OTHER (RCrB)	\\
95	&	 MACHO 78.6698.38	&	  05$^\mathrm{h}$21$^\mathrm{m}$49.40$^\mathrm{s}$	 $-$70\degr04\arcmin35.26\arcsec 	&	J052149.11$-$700434.2 	&	78.6698.38	&	25	&	T2CEP-104	&	25	&	O-PAGB (RV Tau)	\\
96	&	\ldots	&	  05$^\mathrm{h}$22$^\mathrm{m}$06.91$^\mathrm{s}$	 $-$71\degr50\arcmin17.89\arcsec 	&	J052206.92$-$715017.7 	&	\ldots	&	\ldots	&	LPV-46603	&	81	&	O-AGB	\\
97	&	\ldots	&	  05$^\mathrm{h}$22$^\mathrm{m}$22.98$^\mathrm{s}$	 $-$68\degr41\arcmin00.72\arcsec 	&	J052222.95$-$684101.2 	&	\ldots	&	\ldots	&	\ldots	&	\ldots	&	YSO-3	\\
98	&	 OGLE 052242.09$-$691526.2	&	  05$^\mathrm{h}$22$^\mathrm{m}$42.01$^\mathrm{s}$	 $-$69\degr15\arcmin26.04\arcsec 	&	J052241.93$-$691526.2 	&	80.6831.630	&	411	&	LPV-47888	&	417	&	C-AGB	\\
99	&	 SHV 0523185$-$693932	&	  05$^\mathrm{h}$22$^\mathrm{m}$54.96$^\mathrm{s}$	 $-$69\degr36\arcmin52.34\arcsec 	&	J052254.97$-$693651.7 	&	78.6947.2742	&	209--209	&	LPV-48376	&	208	&	O-AGB	\\
100	&	 LH$\alpha$ 120$-$N 136	&	  05$^\mathrm{h}$23$^\mathrm{m}$31.25$^\mathrm{s}$	 $-$69\degr04\arcmin04.56\arcsec 	&	J052331.11$-$690404.6 	&	\ldots	&	\ldots	&	\ldots	&	\ldots	&	O-PN	\\
101	&	 IRAS 05240$-$6809	&	  05$^\mathrm{h}$23$^\mathrm{m}$51.14$^\mathrm{s}$	 $-$68\degr07\arcmin12.37\arcsec 	&	J052351.13$-$680712.2 	&	\ldots	&	\ldots	&	\ldots	&	\ldots	&	YSO-1	\\
102	&	 IRAS 05246$-$7137	&	  05$^\mathrm{h}$23$^\mathrm{m}$53.95$^\mathrm{s}$	 $-$71\degr34\arcmin43.97\arcsec 	&	J052353.92$-$713443.9 	&	\ldots	&	\ldots	&	\ldots	&	\ldots	&	YSO-2	\\
103	&	\ldots	&	  05$^\mathrm{h}$24$^\mathrm{m}$05.25$^\mathrm{s}$	 $-$68\degr18\arcmin01.99\arcsec 	&	J052405.31$-$681802.5 	&	4.7087.2529	&	304--387	&	LPV-50988	&	330	&	C-AGB	\\
104	&	 MSX LMC 464	&	  05$^\mathrm{h}$24$^\mathrm{m}$13.30$^\mathrm{s}$	 $-$68\degr29\arcmin58.98\arcsec 	&	J052413.36$-$682958.8 	&	3.7084.121	&	596--1100	&	\ldots	&	\ldots	&	HII	\\
105	&	 OGLE J052445.53$-$691605.6	&	  05$^\mathrm{h}$24$^\mathrm{m}$45.36$^\mathrm{s}$	 $-$69\degr16\arcmin05.53\arcsec 	&	J052445.38$-$691605.3 	&	80.7194.86	&	399--401	&	LPV-52434	&	403	&	C-AGB	\\
106	&	 LH$\alpha$ 120$-$S 33	&	  05$^\mathrm{h}$24$^\mathrm{m}$57.86$^\mathrm{s}$	 $-$67\degr24\arcmin58.43\arcsec 	&	J052457.85$-$672458.3 	&	\ldots	&	\ldots	&	\ldots	&	\ldots	&	YSO-4	\\
107	&	 HV 5829	&	  05$^\mathrm{h}$25$^\mathrm{m}$19.52$^\mathrm{s}$	 $-$70\degr54\arcmin09.84\arcsec 	&	J052519.48$-$705410.0 	&	\ldots	&	\ldots	&	T2CEP-119	&	34	&	O-PAGB (RV Tau)	\\
108	&	\ldots	&	  05$^\mathrm{h}$25$^\mathrm{m}$46.52$^\mathrm{s}$	 $-$66\degr14\arcmin11.30\arcsec 	&	J052546.51$-$661411.5 	&	\ldots	&	\ldots	&	\ldots	&	\ldots	&	YSO-1	\\
109	&	\ldots	&	  05$^\mathrm{h}$26$^\mathrm{m}$13.35$^\mathrm{s}$	 $-$68\degr47\arcmin15.24\arcsec 	&	J052613.39$-$684715.0 	&	\ldots	&	\ldots	&	\ldots	&	\ldots	&	YSO-4	\\
110	&	 OGLE J052620.25$-$693902.4	&	  05$^\mathrm{h}$26$^\mathrm{m}$20.15$^\mathrm{s}$	 $-$69\degr39\arcmin02.59\arcsec 	&	J052620.10$-$693902.1 	&	77.7430.51	&	820	&	LPV-55565	&	810	&	O-AGB	\\
111	&	 HV 2522	&	  05$^\mathrm{h}$26$^\mathrm{m}$27.19$^\mathrm{s}$	 $-$66\degr42\arcmin58.66\arcsec 	&	J052627.23$-$664258.7 	&	62.7474.1378	&	undefined	&	\ldots	&	\ldots	&	O-PAGB	\\
112	&	 RP 589	&	  05$^\mathrm{h}$26$^\mathrm{m}$37.58$^\mathrm{s}$	 $-$70\degr29\arcmin07.18\arcsec 	&	J052637.81$-$702906.7 	&	\ldots	&	\ldots	&	\ldots	&	\ldots	&	UNK	\\
113	&	\ldots	&	  05$^\mathrm{h}$27$^\mathrm{m}$07.14$^\mathrm{s}$	 $-$70\degr20\arcmin02.12\arcsec 	&	J052707.10$-$702001.9 	&	7.7541.31	&	undefined	&	\ldots	&	\ldots	&	O-PAGB	\\
114	&	\ldots	&	  05$^\mathrm{h}$27$^\mathrm{m}$23.24$^\mathrm{s}$	 $-$71\degr24\arcmin25.41\arcsec 	&	J052723.14$-$712426.3 	&	\ldots	&	\ldots	&	\ldots	&	\ldots	&	YSO-3	\\
115	&	 LH$\alpha$ 120$-$N 145	&	  05$^\mathrm{h}$27$^\mathrm{m}$35.64$^\mathrm{s}$	 $-$69\degr08\arcmin56.22\arcsec 	&	J052735.63$-$690856.3 	&	\ldots	&	\ldots	&	\ldots	&	\ldots	&	O-PAGB	\\
116	&	 HV 2551	&	  05$^\mathrm{h}$27$^\mathrm{m}$38.76$^\mathrm{s}$	 $-$69\degr28\arcmin45.57\arcsec 	&	J052738.58$-$692843.9 	&	\ldots	&	\ldots	&	\ldots	&	\ldots	&	RSG	\\
117	&	 W61 11$-$16	&	  05$^\mathrm{h}$27$^\mathrm{m}$39.62$^\mathrm{s}$	 $-$69\degr09\arcmin01.57\arcsec 	&	J052739.63$-$690901.4 	&	\ldots	&	\ldots	&	\ldots	&	\ldots	&	RSG	\\
118	&	\ldots	&	  05$^\mathrm{h}$27$^\mathrm{m}$47.60$^\mathrm{s}$	 $-$71\degr48\arcmin52.75\arcsec 	&	J052747.62$-$714852.8 	&	\ldots	&	\ldots	&	\ldots	&	\ldots	&	O-PAGB	\\
119	&	 SHV 0528350$-$701014	&	  05$^\mathrm{h}$28$^\mathrm{m}$05.91$^\mathrm{s}$	 $-$70\degr07\arcmin54.03\arcsec 	&	J052805.91$-$700753.4 	&	\ldots	&	\ldots	&	LPV-59143	&	476	&	C-AGB	\\
120	&	 OGLE J052825.96$-$694647.4	&	  05$^\mathrm{h}$28$^\mathrm{m}$25.86$^\mathrm{s}$	 $-$69\degr46\arcmin47.45\arcsec 	&	J052825.81$-$694647.3 	&	\ldots	&	\ldots	&	LPV-59817	&	133	&	C-AGB	\\
121	&	 IRAS 05298$-$6957	&	  05$^\mathrm{h}$29$^\mathrm{m}$24.61$^\mathrm{s}$	 $-$69\degr55\arcmin14.19\arcsec 	&	J052924.85$-$695519.2 	&	\ldots	&	\ldots	&	\ldots	&	\ldots	&	O-AGB	\\
122	&	 HV 5879	&	  05$^\mathrm{h}$29$^\mathrm{m}$54.80$^\mathrm{s}$	 $-$69\degr04\arcmin15.73\arcsec 	&	J052954.73$-$690415.7 	&	82.8044.1115	&	122--1566	&	\ldots	&	\ldots	&	RSG	\\
123	&	 SP77 46$-$50	&	  05$^\mathrm{h}$30$^\mathrm{m}$04.67$^\mathrm{s}$	 $-$68\degr47\arcmin29.08\arcsec 	&	J053004.56$-$684728.8 	&	\ldots	&	\ldots	&	\ldots	&	\ldots	&	RSG	\\
124	&	 SHV 0530472$-$690607	&	  05$^\mathrm{h}$30$^\mathrm{m}$27.56$^\mathrm{s}$	 $-$69\degr03\arcmin59.04\arcsec 	&	J053027.49$-$690358.3 	&	82.8165.1133	&	212--212	&	LPV-63816	&	214	&	O-AGB	\\
125	&	 IRAS 05315$-$7145	&	  05$^\mathrm{h}$30$^\mathrm{m}$44.72$^\mathrm{s}$	 $-$71\degr42\arcmin59.62\arcsec 	&	J053044.10$-$714300.6 	&	\ldots	&	\ldots	&	\ldots	&	\ldots	&	C-AGB	\\
126	&	 KDM 4554	&	  05$^\mathrm{h}$30$^\mathrm{m}$45.03$^\mathrm{s}$	 $-$68\degr21\arcmin29.11\arcsec 	&	J053044.55$-$682119.2 	&	8.8176.2019	&	467--612	&	LPV-64345	&	426	&	C-AGB	\\
127	&	 NGC 2004 Robb B45	&	  05$^\mathrm{h}$30$^\mathrm{m}$46.74$^\mathrm{s}$	 $-$67\degr16\arcmin56.92\arcsec 	&	J053046.81$-$671657.2 	&	61.8192.23	&	381	&	\ldots	&	\ldots	&	RSG	\\
128	&	 NGC 2004 Wes 18$-$13	&	  05$^\mathrm{h}$30$^\mathrm{m}$48.40$^\mathrm{s}$	 $-$67\degr16\arcmin45.88\arcsec 	&	J053048.42$-$671645.8 	&	\ldots	&	\ldots	&	\ldots	&	\ldots	&	RSG	\\
129	&	 NGC 2004 Wes 6$-$14	&	  05$^\mathrm{h}$30$^\mathrm{m}$52.25$^\mathrm{s}$	 $-$67\degr17\arcmin34.22\arcsec 	&	J053052.28$-$671734.4 	&	\ldots	&	\ldots	&	\ldots	&	\ldots	&	RSG	\\
130	&	\ldots	&	  05$^\mathrm{h}$31$^\mathrm{m}$28.43$^\mathrm{s}$	 $-$70\degr10\arcmin27.65\arcsec 	&	J053128.44$-$701027.1 	&	7.8269.63	&	401--405	&	LPV-65660	&	393	&	O-AGB	\\
131	&	 MACHO 82.8405.15	&	  05$^\mathrm{h}$31$^\mathrm{m}$51.01$^\mathrm{s}$	 $-$69\degr11\arcmin46.56\arcsec 	&	J053150.98$-$691146.4 	&	82.8405.15	&	93	&	T2CEP-147	&	47	&	O-PAGB (RV Tau)	\\
132	&	 KDM 4665	&	  05$^\mathrm{h}$31$^\mathrm{m}$58.96$^\mathrm{s}$	 $-$72\degr44\arcmin36.35\arcsec 	&	J053158.92$-$724436.0 	&	35.8352.3	&	118--212	&	\ldots	&	\ldots	&	C-AGB	\\
133	&	\ldots	&	  05$^\mathrm{h}$32$^\mathrm{m}$06.64$^\mathrm{s}$	 $-$70\degr10\arcmin25.34\arcsec 	&	J053206.70$-$701024.8 	&	7.8390.36	&	115--206	&	LPV-66766	&	124	&	STAR	\\
134	&	\ldots	&	  05$^\mathrm{h}$32$^\mathrm{m}$18.66$^\mathrm{s}$	 $-$67\degr31\arcmin46.16\arcsec 	&	J053218.64$-$673145.9 	&	\ldots	&	\ldots	&	\ldots	&	\ldots	&	RSG	\\
135	&	 NGC 2011 SAGE IRS 1	&	  05$^\mathrm{h}$32$^\mathrm{m}$19.31$^\mathrm{s}$	 $-$67\degr31\arcmin20.34\arcsec 	&	J053219.33$-$673120.5 	&	\ldots	&	\ldots	&	\ldots	&	\ldots	&	RSG	\\
136	&	 KDM 4718	&	  05$^\mathrm{h}$32$^\mathrm{m}$26.52$^\mathrm{s}$	 $-$73\degr10\arcmin06.99\arcsec 	&	J053226.51$-$731006.8 	&	\ldots	&	\ldots	&	\ldots	&	\ldots	&	C-AGB	\\
137	&	 RP 774	&	  05$^\mathrm{h}$32$^\mathrm{m}$39.71$^\mathrm{s}$	 $-$69\degr30\arcmin49.25\arcsec 	&	J053239.68$-$693049.4 	&	81.8521.506	&	undef.	&	\ldots	&	\ldots	&	YSO-4	\\
138	&	\ldots	&	  05$^\mathrm{h}$32$^\mathrm{m}$53.35$^\mathrm{s}$	 $-$66\degr07\arcmin27.17\arcsec 	&	J053253.36$-$660727.8 	&	64.8572.519	&	undef.	&	\ldots	&	\ldots	&	GAL	\\
139	&	 KDM 4774	&	  05$^\mathrm{h}$32$^\mathrm{m}$54.98$^\mathrm{s}$	 $-$67\degr36\arcmin47.10\arcsec 	&	J053254.99$-$673647.2 	&	\ldots	&	\ldots	&	\ldots	&	\ldots	&	C-AGB	\\
140	&	 MSX LMC 736	&	  05$^\mathrm{h}$33$^\mathrm{m}$06.86$^\mathrm{s}$	 $-$70\degr30\arcmin34.22\arcsec 	&	J053306.86$-$703033.9 	&	\ldots	&	\ldots	&	\ldots	&	\ldots	&	C-AGB	\\
141	&	\ldots	&	  05$^\mathrm{h}$33$^\mathrm{m}$18.61$^\mathrm{s}$	 $-$66\degr00\arcmin39.91\arcsec 	&	J053318.58$-$660040.2 	&	64.8695.450	&	undef.	&	\ldots	&	\ldots	&	C-AGB	\\
142	&	\ldots	&	  05$^\mathrm{h}$33$^\mathrm{m}$43.18$^\mathrm{s}$	 $-$70\degr59\arcmin21.16\arcsec 	&	J053343.27$-$705921.1 	&	14.8620.13	&	151--182	&	LPV-69535	&	106	&	O-AGB	\\
143	&	\ldots	&	  05$^\mathrm{h}$33$^\mathrm{m}$44.00$^\mathrm{s}$	 $-$70\degr59\arcmin01.14\arcsec 	&	J053343.98$-$705901.9 	&	14.8620.21	&	146--1080	&	LPV-69554	&	148	&	O-AGB	\\
144	&	 LH$\alpha$ 120$-$N 151	&	  05$^\mathrm{h}$33$^\mathrm{m}$46.97$^\mathrm{s}$	 $-$68\degr36\arcmin44.08\arcsec 	&	J053346.97$-$683644.2 	&	\ldots	&	\ldots	&	\ldots	&	\ldots	&	C-PN	\\
145	&	\ldots	&	  05$^\mathrm{h}$34$^\mathrm{m}$41.46$^\mathrm{s}$	 $-$69\degr26\arcmin30.74\arcsec 	&	J053441.40$-$692630.6 	&	\ldots	&	\ldots	&	LPV-71117	&	503	&	C-AGB	\\
146	&	 SHP LMC 256	&	  05$^\mathrm{h}$34$^\mathrm{m}$44.20$^\mathrm{s}$	 $-$67\degr37\arcmin50.82\arcsec 	&	J053444.17$-$673750.1 	&	\ldots	&	\ldots	&	\ldots	&	\ldots	&	UNK	\\
147	&	 HV 2700	&	  05$^\mathrm{h}$35$^\mathrm{m}$19.01$^\mathrm{s}$	 $-$67\degr02\arcmin19.50\arcsec 	&	J053518.91$-$670219.5 	&	\ldots	&	\ldots	&	\ldots	&	\ldots	&	RSG	\\
148	&	\ldots	&	  05$^\mathrm{h}$35$^\mathrm{m}$48.02$^\mathrm{s}$	 $-$70\degr31\arcmin46.92\arcsec 	&	J053548.07$-$703146.6 	&	11.8990.10	&	965--981	&	LPV-72897	&	953	&	O-AGB	\\
149	&	\ldots	&	  05$^\mathrm{h}$36$^\mathrm{m}$02.42$^\mathrm{s}$	 $-$67\degr45\arcmin17.41\arcsec 	&	J053602.36$-$674517.3 	&	8.9032.176	&	1137	&	\ldots	&	\ldots	&	YSO-4	\\
150	&	 IRAS 05370$-$7019	&	  05$^\mathrm{h}$36$^\mathrm{m}$32.48$^\mathrm{s}$	 $-$70\degr17\arcmin38.81\arcsec 	&	J053632.56$-$701738.4 	&	\ldots	&	\ldots	&	\ldots	&	\ldots	&	C-PAGB	\\
151	&	\ldots	&	  05$^\mathrm{h}$36$^\mathrm{m}$34.82$^\mathrm{s}$	 $-$72\degr26\arcmin58.67\arcsec 	&	J053634.77$-$722658.6 	&	\ldots	&	\ldots	&	\ldots	&	\ldots	&	GAL	\\
152	&	\ldots	&	  05$^\mathrm{h}$36$^\mathrm{m}$55.68$^\mathrm{s}$	 $-$68\degr11\arcmin24.31\arcsec 	&	J053655.60$-$681124.6 	&	8.9146.17	&	752--810	&	LPV-74445	&	748	&	O-AGB	\\
153	&	 RP 493	&	  05$^\mathrm{h}$37$^\mathrm{m}$10.12$^\mathrm{s}$	 $-$71\degr23\arcmin14.06\arcsec 	&	J053710.26$-$712314.3 	&	\ldots	&	\ldots	&	\ldots	&	\ldots	&	O-PN	\\
154	&	\ldots	&	  05$^\mathrm{h}$37$^\mathrm{m}$30.63$^\mathrm{s}$	 $-$67\degr40\arcmin41.19\arcsec 	&	J053730.59$-$674041.6 	&	\ldots	&	\ldots	&	\ldots	&	\ldots	&	GAL	\\
155	&	 KDM 5345	&	  05$^\mathrm{h}$38$^\mathrm{m}$23.66$^\mathrm{s}$	 $-$66\degr09\arcmin00.91\arcsec 	&	J053823.58$-$660900.3 	&	66.9419.10	&	112	&	\ldots	&	\ldots	&	UNK	\\
156	&	 OGLE J053930.16$-$695755.8	&	  05$^\mathrm{h}$39$^\mathrm{m}$29.94$^\mathrm{s}$	 $-$69\degr57\arcmin56.27\arcsec 	&	J053930.06$-$695800.9 	&	81.9603.512	&	405	&	LPV-77631	&	399	&	C-AGB	\\
157	&	 HV 12631	&	  05$^\mathrm{h}$39$^\mathrm{m}$33.17$^\mathrm{s}$	 $-$71\degr21\arcmin55.45\arcsec 	&	J053932.92$-$712200.6 	&	14.9582.9	&	31	&	T2CEP-169	&	31	&	O-PAGB (RV Tau)	\\
158	&	\ldots	&	  05$^\mathrm{h}$39$^\mathrm{m}$42.37$^\mathrm{s}$	 $-$71\degr10\arcmin45.03\arcsec 	&	J053942.45$-$711044.5 	&	\ldots	&	\ldots	&	\ldots	&	\ldots	&	YSO-3	\\
159	&	\ldots	&	  05$^\mathrm{h}$39$^\mathrm{m}$45.46$^\mathrm{s}$	 $-$66\degr58\arcmin09.75\arcsec 	&	J053945.40$-$665809.4 	&	67.9648.7	&	125--126	&	\ldots	&	\ldots	&	O-AGB	\\
160	&	\ldots	&	  05$^\mathrm{h}$39$^\mathrm{m}$49.22$^\mathrm{s}$	 $-$69\degr37\arcmin47.03\arcsec 	&	J053949.23$-$693747.0 	&	\ldots	&	\ldots	&	\ldots	&	\ldots	&	UNK	\\
161	&	 MACHO 81.9728.14	&	  05$^\mathrm{h}$40$^\mathrm{m}$00.47$^\mathrm{s}$	 $-$69\degr42\arcmin14.85\arcsec 	&	J054000.52$-$694214.6 	&	81.9728.14	&	47	&	T2CEP-174	&	47	&	O-PAGB (RV Tau)	\\
162	&	 MSX LMC 949	&	  05$^\mathrm{h}$40$^\mathrm{m}$14.78$^\mathrm{s}$	 $-$69\degr28\arcmin49.33\arcsec 	&	J054014.83$-$692849.1 	&	76.9732.1316	&	403--566	&	\ldots	&	\ldots	&	O-PAGB	\\
163	&	 RP 85	&	  05$^\mathrm{h}$40$^\mathrm{m}$33.51$^\mathrm{s}$	 $-$70\degr32\arcmin41.06\arcsec 	&	J054033.54$-$703240.8 	&	\ldots	&	\ldots	&	\ldots	&	\ldots	&	YSO-4	\\
164	&	\ldots	&	  05$^\mathrm{h}$40$^\mathrm{m}$59.28$^\mathrm{s}$	 $-$70\degr44\arcmin02.82\arcsec 	&	J054059.31$-$704402.5 	&	\ldots	&	\ldots	&	\ldots	&	\ldots	&	YSO-3	\\
165	&	 MSX LMC 947	&	  05$^\mathrm{h}$41$^\mathrm{m}$02.04$^\mathrm{s}$	 $-$70\degr43\arcmin10.55\arcsec 	&	J054103.55$-$704307.6 	&	11.9834.92	&	694--703	&	LPV-79341	&	709	&	O-AGB	\\
166	&	\ldots	&	  05$^\mathrm{h}$41$^\mathrm{m}$14.58$^\mathrm{s}$	 $-$71\degr32\arcmin36.01\arcsec 	&	J054114.56$-$713236.0 	&	15.9822.4	&	583	&	\ldots	&	\ldots	&	O-AGB	\\
167	&	 IRAS 05416$-$6906	&	  05$^\mathrm{h}$41$^\mathrm{m}$20.69$^\mathrm{s}$	 $-$69\degr04\arcmin44.46\arcsec 	&	J054121.18$-$690438.7 	&	\ldots	&	\ldots	&	\ldots	&	\ldots	&	C-AGB	\\
168	&	 IRAS 05421$-$7116	&	  05$^\mathrm{h}$41$^\mathrm{m}$25.08$^\mathrm{s}$	 $-$71\degr15\arcmin32.74\arcsec 	&	J054125.09$-$711532.6 	&	\ldots	&	\ldots	&	\ldots	&	\ldots	&	YSO-1	\\
169	&	 W61 6$-$24	&	  05$^\mathrm{h}$41$^\mathrm{m}$57.43$^\mathrm{s}$	 $-$69\degr12\arcmin18.61\arcsec 	&	J054157.40$-$691218.1 	&	\ldots	&	\ldots	&	\ldots	&	\ldots	&	RSG	\\
170	&	 NGC 2100 Robb 4	&	  05$^\mathrm{h}$42$^\mathrm{m}$03.91$^\mathrm{s}$	 $-$69\degr13\arcmin07.64\arcsec 	&	J054206.15$-$691307.7 	&	\ldots	&	\ldots	&	\ldots	&	\ldots	&	RSG	\\
171	&	 2MASS J05420676$-$6912312	&	  05$^\mathrm{h}$42$^\mathrm{m}$06.76$^\mathrm{s}$	 $-$69\degr12\arcmin31.35\arcsec 	&	J054206.76$-$691231.1 	&	\ldots	&	\ldots	&	\ldots	&	\ldots	&	RSG	\\
172	&	 W61 6$-$57	&	  05$^\mathrm{h}$42$^\mathrm{m}$09.98$^\mathrm{s}$	 $-$69\degr13\arcmin28.76\arcsec 	&	J054209.95$-$691328.7 	&	\ldots	&	\ldots	&	\ldots	&	\ldots	&	RSG	\\
173	&	 WOH G 494	&	  05$^\mathrm{h}$42$^\mathrm{m}$30.51$^\mathrm{s}$	 $-$69\degr48\arcmin57.47\arcsec 	&	J054230.55$-$694857.3 	&	76.10090.12	&	183--333	&	LPV-80890	&	170	&	O-AGB	\\
174	&	 LM 2$-$42	&	  05$^\mathrm{h}$42$^\mathrm{m}$33.35$^\mathrm{s}$	 $-$70\degr29\arcmin24.18\arcsec 	&	J054233.17$-$702924.1 	&	\ldots	&	\ldots	&	\ldots	&	\ldots	&	O-PN	\\
175	&	 LH$\alpha$ 120$-$N 178	&	  05$^\mathrm{h}$42$^\mathrm{m}$36.63$^\mathrm{s}$	 $-$70\degr09\arcmin32.50\arcsec 	&	J054236.65$-$700932.0 	&	\ldots	&	\ldots	&	\ldots	&	\ldots	&	C-PN	\\
176	&	\ldots	&	  05$^\mathrm{h}$42$^\mathrm{m}$54.62$^\mathrm{s}$	 $-$70\degr08\arcmin07.74\arcsec 	&	J054254.38$-$700807.4 	&	76.10206.25	&	166--274	&	LPV-81294	&	278	&	O-AGB	\\
177	&	\ldots	&	  05$^\mathrm{h}$43$^\mathrm{m}$10.87$^\mathrm{s}$	 $-$67\degr27\arcmin28.42\arcsec 	&	J054310.86$-$672728.0 	&	\ldots	&	\ldots	&	\ldots	&	\ldots	&	O-PAGB	\\
178	&	\ldots	&	  05$^\mathrm{h}$43$^\mathrm{m}$14.23$^\mathrm{s}$	 $-$70\degr38\arcmin35.12\arcsec 	&	J054314.12$-$703835.1 	&	12.10198.12	&	156--467	&	LPV-81615	&	162	&	O-AGB	\\
179	&	 KDM 5841	&	  05$^\mathrm{h}$43$^\mathrm{m}$28.77$^\mathrm{s}$	 $-$69\degr42\arcmin43.88\arcsec 	&	J054328.84$-$694243.7 	&	76.10212.28	&	417--918	&	LPV-81866	&	191	&	C-AGB	\\
180	&	\ldots	&	  05$^\mathrm{h}$44$^\mathrm{m}$06.10$^\mathrm{s}$	 $-$68\degr37\arcmin53.68\arcsec 	&	J054406.01$-$683753.6 	&	33.10349.11	&	470--554	&	LPV-82411	&	239	&	O-AGB	\\
181	&	\ldots	&	  05$^\mathrm{h}$44$^\mathrm{m}$37.92$^\mathrm{s}$	 $-$67\degr36\arcmin58.17\arcsec 	&	J054437.87$-$673657.7 	&	68.10486.408	&	undef.	&	\ldots	&	\ldots	&	C-AGB	\\
182	&	\ldots	&	  05$^\mathrm{h}$44$^\mathrm{m}$40.08$^\mathrm{s}$	 $-$69\degr11\arcmin49.11\arcsec 	&	J054440.11$-$691149.0 	&	\ldots	&	\ldots	&	LPV-82889	&	99	&	O-AGB	\\
183	&	 IRAS 05452$-$6924	&	  05$^\mathrm{h}$44$^\mathrm{m}$50.23$^\mathrm{s}$	 $-$69\degr23\arcmin04.72\arcsec 	&	J054450.23$-$692304.2 	&	\ldots	&	\ldots	&	\ldots	&	\ldots	&	YSO-1	\\
184	&	\ldots	&	  05$^\mathrm{h}$45$^\mathrm{m}$24.23$^\mathrm{s}$	 $-$68\degr30\arcmin41.61\arcsec 	&	J054524.23$-$683041.4 	&	\ldots	&	\ldots	&	\ldots	&	\ldots	&	GAL	\\
185	&	\ldots	&	  05$^\mathrm{h}$45$^\mathrm{m}$46.35$^\mathrm{s}$	 $-$67\degr32\arcmin39.16\arcsec 	&	J054546.32$-$673239.4 	&	68.10608.258	&	484--491	&	\ldots	&	\ldots	&	O-AGB	\\
186	&	 LH$\alpha$ 120$-$N 170	&	  05$^\mathrm{h}$47$^\mathrm{m}$04.62$^\mathrm{s}$	 $-$69\degr27\arcmin34.24\arcsec 	&	J054704.54$-$692733.9 	&	31.10821.779	&	undef.	&	LPV-84717	&	362	&	O-PN	\\
187	&	\ldots	&	  05$^\mathrm{h}$47$^\mathrm{m}$45.80$^\mathrm{s}$	 $-$68\degr07\arcmin34.26\arcsec 	&	J054745.79$-$680734.1 	&	32.10962.261	&	2599	&	\ldots	&	\ldots	&	YSO-4	\\
188	&	 KDM 6247	&	  05$^\mathrm{h}$47$^\mathrm{m}$57.25$^\mathrm{s}$	 $-$68\degr14\arcmin56.90\arcsec 	&	J054757.37$-$681457.0 	&	\ldots	&	\ldots	&	\ldots	&	\ldots	&	STAR	\\
189	&	 NGC 2121 LE 6	&	  05$^\mathrm{h}$48$^\mathrm{m}$16.84$^\mathrm{s}$	 $-$71\degr28\arcmin39.64\arcsec 	&	J054816.79$-$712839.3 	&	28.11033.2559	&	387--387	&	LPV-85475	&	385	&	C-AGB	\\
190	&	 IRAS 05495$-$7034	&	  05$^\mathrm{h}$49$^\mathrm{m}$00.12$^\mathrm{s}$	 $-$70\degr33\arcmin22.51\arcsec 	&	J054900.01$-$703322.5 	&	\ldots	&	\ldots	&	\ldots	&	\ldots	&	C-AGB	\\
191	&	 KDM 6486	&	  05$^\mathrm{h}$50$^\mathrm{m}$36.62$^\mathrm{s}$	 $-$68\degr28\arcmin52.27\arcsec 	&	J055036.69$-$682852.4 	&	32.11441.5	&	336--643	&	LPV-86831	&	663	&	C-AGB	\\
192	&	 HV 2862	&	  05$^\mathrm{h}$51$^\mathrm{m}$22.58$^\mathrm{s}$	 $-$69\degr53\arcmin51.05\arcsec 	&	J055122.52$-$695351.4 	&	30.11540.16	&	34	&	T2CEP-191	&	34	&	O-PAGB (RV Tau)	\\
193	&	\ldots	&	  05$^\mathrm{h}$51$^\mathrm{m}$43.25$^\mathrm{s}$	 $-$68\degr45\arcmin42.79\arcsec 	&	J055143.27$-$684543.0 	&	72.11557.1055	&	642	&	\ldots	&	\ldots	&	GAL	\\
194	&	 PMP 337	&	  05$^\mathrm{h}$52$^\mathrm{m}$51.07$^\mathrm{s}$	 $-$69\degr28\arcmin39.09\arcsec 	&	J055251.05$-$692839.6 	&	31.11789.29	&	269--1015	&	LPV-87711	&	151	&	C-AGB	\\
195	&	 PMP 133	&	  05$^\mathrm{h}$52$^\mathrm{m}$52.51$^\mathrm{s}$	 $-$69\degr30\arcmin35.54\arcsec 	&	J055252.50$-$693035.6 	&	31.11788.10	&	25	&	\ldots	&	\ldots	&	STAR	\\
196	&	 IRAS 05537$-$7015	&	  05$^\mathrm{h}$53$^\mathrm{m}$12.04$^\mathrm{s}$	 $-$70\degr15\arcmin22.71\arcsec 	&	J055311.98$-$701522.5 	&	\ldots	&	\ldots	&	\ldots	&	\ldots	&	C-PAGB	\\
197	&	\ldots	&	  06$^\mathrm{h}$00$^\mathrm{m}$53.60$^\mathrm{s}$	 $-$68\degr00\arcmin39.08\arcsec 	&	J060053.62$-$680038.8 	&	74.13142.12	&	281--887	&	\ldots	&	\ldots	&	O-AGB	\\
\enddata
%\tablenotetext{a}{Periods are taken from \citet{alc98} and \citet{fra05,fra08}.}
%\tablenotetext{b}{Periods are taken from the online
%  OGLE-III catalog, as of 2009 October 21 \citep{sos08,sos09b}}
%\tablenotetext{c}{We employ commonly-used abbreviations, e.g., RP for [RP2006] \citep{rei06}, KDM for [KDM2001] \citep{kon01}}
%\tablenotetext{d}{\emph{SAGE-LMC} identifier \citep{mei06}}
\end{deluxetable}